\documentclass[11pt]{article}

\usepackage{verbatim,amsmath,amssymb}
\usepackage{epsfig,float,color}
\usepackage{amsthm,mathrsfs,amsfonts,dsfont} 
\usepackage{hyperref}
\usepackage{a4wide}
\usepackage{graphicx}
\usepackage{setspace}
\usepackage{float}
\usepackage{longtable}
\usepackage{listings}
\usepackage{cite}
\usepackage{epstopdf}
\usepackage{subfig}
\usepackage{multirow}
\usepackage{makeidx}
\usepackage[bottom]{footmisc}
\usepackage{varwidth}
\usepackage{geometry}
\usepackage{setspace}
\usepackage{natbib}
\usepackage[section]{placeins}

\definecolor{darkblue}{rgb}{0,0,1}
\hypersetup{pdftex=true, colorlinks=true, breaklinks=true, linkcolor=darkblue, menucolor=darkblue, pagecolor=darkblue, citecolor=darkblue, urlcolor=darkblue}

\numberwithin{equation}{section}
\numberwithin{table}{section}
\numberwithin{figure}{section}
\makeindex
\begin{document}
\pagenumbering{roman}
\newcommand {\eqb}[1]{\begin{equation}\begin{array}{#1}}
\newcommand {\eqe}{\end{array}\end{equation}}
\newcommand {\eqbb}[1]{\begin{equation}\boxed{\begin{array}{#1}}
\newcommand {\eqeb}{\end{array}}\end{equation}}
\newcommand {\esb}[1]{\begin{equation*}\begin{array}{#1}}
\newcommand {\ese}{\end{array}\end{equation*}}
\newcommand {\ds}{\displaystyle}
\newcommand {\ts}{\textstyle}
\newcommand {\df}[2]{\frac{\ds#1}{\ds#2}}
\newcommand {\pa}[2]{\frac{\partial{#1}}{\partial{#2}}}
\newcommand {\paq}[2]{\frac{\partial^2{#1}}{\partial{#2}^2}}
\newcommand {\paqq}[3]{\frac{\partial^2{#1}}{\partial{#2}\,\partial{#3}}}
\newcommand {\back}{\! \! \!}
\newcommand {\is}{\back &=& \back}
\newcommand {\dis}{\back &:=& \back}
\newcommand {\isd}{\back &=:& \back}
\newcommand {\ais}{\back &\approx& \back}
\newcommand {\isn}{\back &\neq& \back}
\newcommand {\plus}{\back &+& \back}
\newcommand {\mi}{\back &-& \back}
\newcommand {\isl}{\back &<& \back}
\newcommand {\isleq}{\back &\leq& \back}
\newcommand {\isg}{\back &>& \back}
\newcommand {\isgeq}{\back &\geq& \back}
\newcommand {\isin}{\back &\in& \back}
\newcommand {\with}{\quad,$with:$\quad}
\newcommand {\s}{~\!}

\newcommand {\mc}[3]{\multicolumn{#1}{#2}{#3}}
\newcommand {\norm}[1]{\|#1\|}
\newcommand {\tr}{\mathrm{tr}\,}
\newcommand {\spur}{\mathrm{sp}\,}
\newcommand {\grad}{\mathrm{grad}\,}
\newcommand {\divz}{\mathrm{div}\,}
\newcommand {\rot}{\mathrm{rot}\,}
\newcommand {\Grad}{\mathrm{Grad}\,}
\newcommand {\Divz}{\mathrm{Div}\,}

\newcommand {\dif}{\mathrm{d}}

\newcommand {\ave}[1]{\langle#1\rangle}
\newcommand {\avo}[1]{\langle #1 \rangle_{\Omega}}
\newcommand {\ava}[1]{\langle #1 \rangle_{\Omega_e}}
\newcommand {\avb}[1]{\langle #1 \rangle_{\Omega/\Omega_e}}
\newcommand {\avI}[1]{\langle #1 \rangle_{\Omega_I}}
\newcommand {\avE}[1]{\langle #1 \rangle_{\Omega_E}}
\newcommand {\iom}{\int_\Omega}
\newcommand {\pom}{\int_{\partial\Omega}}

\newcommand {\II}{{I\kern-.3em I}}
\newcommand {\III}{{I\kern-.3em I\kern-.3em I}}

\newcommand {\into}{\int_\Omega}
\newcommand {\intoo}{\int_{\Omega_0}}
\newcommand {\intI}{\int_{\Omega_I}}
\newcommand {\inta}{\int_{\Omega_1}}
\newcommand {\intb}{\int_{\Omega_2}}
\newcommand {\intc}{\int_{\Omega_3}}
\newcommand {\inti}{\int_{\Omega_i}}
\newcommand {\intj}{\int_{\Omega_j}}
\newcommand {\intk}{\int_{\Omega_k}}
\newcommand {\intl}{\int_{\Omega_\ell}}
\newcommand {\intab}{\int_{\bar{\Omega}_1}}
\newcommand {\intbb}{\int_{\bar{\Omega}_2}}
\newcommand {\intcb}{\int_{\bar{\Omega}_3}}
\newcommand {\intaob}{\int_{\bar{\Omega}_{10}}}
\newcommand {\intbob}{\int_{\bar{\Omega}_{20}}}
\newcommand {\intaobb}{\int_{\bar{\bar{\Omega}}_{10}}}
\newcommand {\intbobb}{\int_{\bar{\bar{\Omega}}_{20}}}
\newcommand {\intib}{\int_{\bar{\Omega}_i}}
\newcommand {\intjb}{\int_{\bar{\Omega}_j}}
\newcommand {\intkb}{\int_{\bar\Omega_k}}
\newcommand {\intlb}{\int_{\bar\Omega_\ell}}
\newcommand {\intie}{\int_{\Omega_i^e}}
\newcommand {\intje}{\int_{\Omega_j^e}}
\newcommand {\intpa}{\int_{\partial\bar{\Omega}_1}}
\newcommand {\intpb}{\int_{\partial\bar{\Omega}_2}}
\newcommand {\intpi}{\int_{\Gamma_i^e}}
\newcommand {\intpj}{\int_{\Gamma_j^e}}
\newcommand {\intpk}{\int_{\Gamma_k^e}}
\newcommand {\intpio}{\int_{\Gamma_{i0}^{e}}}
\newcommand {\intpjo}{\int_{\Gamma_{j0}^{e}}}
\newcommand {\intpko}{\int_{\Gamma_{k0}^{e}}}

\newcommand {\mra}{\mathrm{a}}
\newcommand {\mrb}{\mathrm{b}}
\newcommand {\mrc}{\mathrm{c}}
\newcommand {\mrd}{\mathrm{d}}
\newcommand {\mre}{\mathrm{e}}
\newcommand {\mrf}{\mathrm{f}}
\newcommand {\mrg}{\mathrm{g}}
\newcommand {\mrh}{\mathrm{h}}
\newcommand {\mri}{\mathrm{i}}
\newcommand {\mrj}{\mathrm{j}}
\newcommand {\mrk}{\mathrm{k}}
\newcommand {\mrl}{\mathrm{l}}
\newcommand {\mrm}{\mathrm{m}}
\newcommand {\mrn}{\mathrm{n}}
\newcommand {\mro}{\mathrm{o}}
\newcommand {\mrp}{\mathrm{p}}
\newcommand {\mrq}{\mathrm{q}}
\newcommand {\mrr}{\mathrm{r}}
\newcommand {\mrs}{\mathrm{s}}
\newcommand {\mrt}{\mathrm{t}}
\newcommand {\mru}{\mathrm{u}}
\newcommand {\mrv}{\mathrm{v}}
\newcommand {\mrw}{\mathrm{w}}
\newcommand {\mrx}{\mathrm{x}}
\newcommand {\mry}{\mathrm{y}}
\newcommand {\mrz}{\mathrm{z}}

\newcommand {\ma}{\mathbf{a}}
\newcommand {\mcc}{\mathbf{c}}
\newcommand {\md}{\mathbf{d}}
\newcommand {\me}{\mathbf{e}}
\newcommand {\mf}{\mathbf{f}}
\newcommand {\mg}{\mathbf{g}}
\newcommand {\mh}{\mathbf{h}}
\newcommand {\mii}{\mathbf{i}}
\newcommand {\mj}{\mathbf{j}}
\newcommand {\mk}{\mathbf{k}}
\newcommand {\ml}{\mathbf{l}}
\newcommand {\mm}{\mathbf{m}}
\newcommand {\mn}{\mathbf{n}}
\newcommand {\mo}{\mathbf{o}}
\newcommand {\mpp}{\mathbf{p}}
\newcommand {\mq}{\mathbf{q}}
\newcommand {\mr}{\mathbf{r}}
\newcommand {\mt}{\mathbf{t}}
\newcommand {\muu}{\mathbf{u}}
\newcommand {\mv}{\mathbf{v}}
\newcommand {\mw}{\mathbf{w}}
\newcommand {\mx}{\mathbf{x}}
\newcommand {\my}{\mathbf{y}}
\newcommand {\mz}{\mathbf{z}}

\newcommand {\ba}{\boldsymbol{a}}
\newcommand {\bb}{\boldsymbol{b}}
\newcommand {\bc}{\boldsymbol{c}}
\newcommand {\bd}{\boldsymbol{d}}
\newcommand {\be}{\boldsymbol{e}}
\newcommand {\bff}{\boldsymbol{f}}
\newcommand {\bg}{\boldsymbol{g}}
\newcommand {\bh}{\boldsymbol{h}}
\newcommand {\bi}{\boldsymbol{i}}
\newcommand {\bj}{\boldsymbol{j}}
\newcommand {\bk}{\boldsymbol{k}}
\newcommand {\bl}{\boldsymbol{l}}
\newcommand {\bell}{\boldsymbol{\ell}}
\newcommand {\bm}{\boldsymbol{m}}
\newcommand {\bn}{\boldsymbol{n}}
\newcommand {\bo}{\boldsymbol{o}}
\newcommand {\bp}{\boldsymbol{p}}
\newcommand {\bq}{\boldsymbol{q}}
\newcommand {\br}{\boldsymbol{r}}
\newcommand {\bs}{\boldsymbol{s}}
\newcommand {\bt}{\boldsymbol{t}}
\newcommand {\bu}{\boldsymbol{u}}
\newcommand {\bv}{\boldsymbol{v}}
\newcommand {\bw}{\boldsymbol{w}}
\newcommand {\bx}{\boldsymbol{x}}
\newcommand {\by}{\boldsymbol{y}}
\newcommand {\bz}{\boldsymbol{z}}
\newcommand {\balp}{\mbox{\boldmath$\alpha$}}
\newcommand {\bbet}{\mbox{\boldmath$\beta$}}
\newcommand {\bdel}{\mbox{\boldmath$\delta$}}
\newcommand {\blam}{\mbox{\boldmath$\lambda$}}
\newcommand {\bnu}{\mbox{\boldmath$\nu$}}
\newcommand {\bphi}{\mbox{\boldmath$\phi$}}
\newcommand {\bpsi}{\mbox{\boldmath$\psi$}}
\newcommand {\bvphi}{\mbox{\boldmath$\varphi$}}
\newcommand {\brho}{\mbox{\boldmath$\rho$}}
\newcommand {\bxi}{\mbox{\boldmath$\xi$}}
\newcommand {\bet}{\mbox{\boldmath$\eta$}}
\newcommand {\btheta}{\mbox{\boldmath$\theta$}}
\newcommand {\bome}{\mbox{\boldmath$\omega$}}

\newcommand {\mA}{\mathbf{A}}
\newcommand {\mB}{\mathbf{B}}
\newcommand {\mC}{\mathbf{C}}
\newcommand {\mD}{\mathbf{D}}
\newcommand {\mE}{\mathbf{E}}
\newcommand {\mF}{\mathbf{F}}
\newcommand {\mG}{\mathbf{G}}
\newcommand {\mH}{\mathbf{H}}
\newcommand {\mI}{\mathbf{I}}
\newcommand {\mJ}{\mathbf{J}}
\newcommand {\mK}{\mathbf{K}}
\newcommand {\mL}{\mathbf{L}}
\newcommand {\mM}{\mathbf{M}}
\newcommand {\mN}{\mathbf{N}}
\newcommand {\mO}{\mathbf{O}}
\newcommand {\mP}{\mathbf{P}}
\newcommand {\mQ}{\mathbf{Q}}
\newcommand {\mR}{\mathbf{R}}
\newcommand {\mS}{\mathbf{S}}
\newcommand {\mT}{\mathbf{T}}
\newcommand {\mU}{\mathbf{U}}
\newcommand {\mV}{\mathbf{V}}
\newcommand {\mW}{\mathbf{W}}
\newcommand {\mX}{\mathbf{X}}
\newcommand {\mY}{\mathbf{Y}}
\newcommand {\mZ}{\mathbf{Z}}

\newcommand {\bA}{\boldsymbol{A}}
\newcommand {\bB}{\boldsymbol{B}}
\newcommand {\bC}{\boldsymbol{C}}
\newcommand {\bD}{\boldsymbol{D}}
\newcommand {\bE}{\boldsymbol{E}}
\newcommand {\bF}{\boldsymbol{F}}
\newcommand {\bG}{\boldsymbol{G}}
\newcommand {\bH}{\boldsymbol{H}}
\newcommand {\bI}{\boldsymbol{I}}
\newcommand {\bJ}{\boldsymbol{J}}
\newcommand {\bK}{\boldsymbol{K}}
\newcommand {\bL}{\boldsymbol{L}}
\newcommand {\bM}{\boldsymbol{M}}
\newcommand {\bN}{\boldsymbol{N}}
\newcommand {\bO}{\boldsymbol{O}}
\newcommand {\bP}{\boldsymbol{P}}
\newcommand {\bQ}{\boldsymbol{Q}}
\newcommand {\bR}{\boldsymbol{R}}
\newcommand {\bS}{\boldsymbol{S}}
\newcommand {\bT}{\boldsymbol{T}}
\newcommand {\bU}{\boldsymbol{U}}
\newcommand {\bV}{\boldsymbol{V}}
\newcommand {\bW}{\boldsymbol{W}}
\newcommand {\bX}{\boldsymbol{X}}
\newcommand {\bY}{\boldsymbol{Y}}
\newcommand {\bZ}{\boldsymbol{Z}}

\newcommand {\eps}{\varepsilon}
\newcommand {\sig}{\sigma}
\newcommand {\vphi}{\varphi}
\newcommand {\beps}{\mbox{\boldmath$\varepsilon$}}
\newcommand {\bLam}{\mbox{\boldmath$\Lambda$}}
\newcommand {\bsig}{\mbox{\boldmath$\sigma$}}
\newcommand {\bSig}{\mbox{\boldmath$\Sigma$}}
\newcommand {\btau}{\mbox{\boldmath$\tau$}}
\newcommand {\bbphi}{\mbox{\boldmath$\Phi$}}
\newcommand {\bom}{\mbox{\boldmath$\omega$}}
\newcommand {\bOme}{\mbox{\boldmath$\Omega$}}
\newcommand {\bThe}{\mbox{\boldmath$\Theta$}}
\newcommand {\bXi}{\mbox{\boldmath$\Xi$}}
\newcommand {\kd}[1]{\delta_{#1}}
\newcommand {\bone}{\mathbf{1}}

\newcommand {\bbA}{\mathbb{A}}
\newcommand {\bbB}{\mathbb{B}}
\newcommand {\bbC}{\mathbb{C}}
\newcommand {\bbD}{\mathbb{D}}
\newcommand {\bbE}{\mathbb{E}}
\newcommand {\bbF}{\mathbb{F}}
\newcommand {\bbG}{\mathbb{G}}
\newcommand {\bbH}{\mathbb{H}}
\newcommand {\bbI}{\mathbb{I}}
\newcommand {\bbJ}{\mathbb{J}}
\newcommand {\bbK}{\mathbb{K}}
\newcommand {\bbL}{\mathbb{L}}
\newcommand {\bbM}{\mathbb{M}}
\newcommand {\bbN}{\mathbb{N}}
\newcommand {\bbO}{\mathbb{O}}
\newcommand {\bbP}{\mathbb{P}}
\newcommand {\bbQ}{\mathbb{Q}}
\newcommand {\bbR}{\mathbb{R}}
\newcommand {\bbS}{\mathbb{S}}
\newcommand {\bbT}{\mathbb{T}}
\newcommand {\bbU}{\mathbb{U}}
\newcommand {\bbV}{\mathbb{V}}
\newcommand {\bbW}{\mathbb{W}}
\newcommand {\bbX}{\mathbb{X}}
\newcommand {\bbY}{\mathbb{Y}}
\newcommand {\bbZ}{\mathbb{Z}}

\newcommand {\IR}{{\rm\kern.24em
   \vrule width.02em height1.53ex depth-.05ex
   \kern-.3em R}}
\newcommand {\ic}{{\rm\kern.20em
   \vrule width.02em height1.0ex depth-.05ex
   \kern-.22em c}}
\newcommand {\ia}{{\rm\kern.20em
   \vrule width.02em height1.05ex depth-.0ex
   \kern-.25em a}}
\newcommand {\IC}{{\rm\kern.24em
   \vrule width.02em height1.4ex depth-.05ex
   \kern-.26em C}}
\newcommand {\ID}{{\rm\kern.34em
   \vrule width.02em height1.5ex depth-.05ex
   \kern-.36em D}}
\newcommand {\IS}{{\rm\kern.24em
   \vrule width.02em height1.6ex depth.05ex
   \kern-.26em S}}
\newcommand {\IT}{{\rm\kern.50em
   \vrule width.02em height1.55ex depth-.05ex
   \kern-.52em T}}
\newcommand {\Sa}{\ds\frac{1}{2(1-\nu)}}
\newcommand {\Sb}{\ds\frac{3-4\nu}{4(1-\nu)}}
\newcommand {\Sc}{\ds\frac{1}{4(1-\nu)(3-4\nu)}}
\newcommand {\SC}{\ds\frac{1}{2(1-\nu)(3-4\nu)}}
\newcommand {\Sco}{\ds\frac{1}{4(1-\nu)(3-4\nu)R^2_0}}
\newcommand {\IE}{{\rm\kern.24em
   \vrule width.02em height1.55ex depth-.05ex
   \kern-.33em E}}
\newcommand {\IEa}{{\rm\kern.24em
   \vrule width.02em height1.55ex depth-.05ex
   \kern-.33em E}^{1}_{ijkl}}
\newcommand {\IEb}{{\rm\kern.24em
   \vrule width.02em height1.55ex depth-.05ex
   \kern-.33em E}^{2}_{ijkl}}

\newcommand {\sA}{\mathcal{A}}
\newcommand {\sB}{\mathcal{B}}
\newcommand {\sC}{\mathcal{C}}
\newcommand {\sD}{\mathcal{D}}
\newcommand {\sE}{\mathcal{E}}
\newcommand {\sF}{\mathcal{F}}
\newcommand {\sG}{\mathcal{G}}
\newcommand {\sH}{\mathcal{H}}
\newcommand {\sI}{\mathcal{I}}
\newcommand {\sJ}{\mathcal{J}}
\newcommand {\sK}{\mathcal{K}}
\newcommand {\sL}{\mathcal{L}}
\newcommand {\sM}{\mathcal{M}}
\newcommand {\sN}{\mathcal{N}}
\newcommand {\sO}{\mathcal{O}}
\newcommand {\sP}{\mathcal{P}}
\newcommand {\sQ}{\mathcal{Q}}
\newcommand {\sR}{\mathcal{R}}
\newcommand {\sS}{\mathcal{S}}
\newcommand {\sT}{\mathcal{T}}
\newcommand {\sU}{\mathcal{U}}
\newcommand {\sV}{\mathcal{V}}
\newcommand {\sW}{\mathcal{W}}
\newcommand {\sX}{\mathcal{X}}
\newcommand {\sY}{\mathcal{Y}}
\newcommand {\sZ}{\mathcal{Z}}

\newcommand{\Pint}[1]{\Pi_{\mathrm{int}#1}}
\newcommand{\Pext}[1]{\Pi_{\mathrm{ext}#1}}
\newcommand{\Pic}{\Pi_{\mathrm{c}}}
\newcommand{\fint}[1]{\mf_{\mathrm{int}#1}}
\newcommand{\fext}[1]{\mf_{\mathrm{ext}#1}}
\newcommand{\fc}[1]{\mf_{\mathrm{c}#1}}
\newcommand{\kint}[1]{\mk_{\mathrm{int}#1}}
\newcommand{\kc}[1]{\mk_{\mathrm{c}#1}}

\newcommand{\mhx}{{\mz}}
\newcommand{\mhX}{{\mZ}}

\newcommand {\dphi}{\delta\bvphi}

\newcommand {\Ass}[2]{\kern 0.9ex \vrule width0.45em height0.2ex depth0ex \kern -2.1ex \bigwedge_{#1}^{#2}}
\newcommand {\ASS}[2]{\kern 1.45ex \vrule width0.5em height0.2ex depth0ex \kern -2.65ex \bigwedge_{#1}^{#2}}


		\begin{center}
		\begin{spacing}{1.5}
			\textbf{\Large Adaptive local surface refinement based on LR NURBS and its application to contact}\\
		\end{spacing}
		\vspace{3mm}
		Christopher Zimmermann, Roger A. Sauer\footnote{corresponding author, email: sauer@aices.rwth-aachen.de} \\
		\vspace{3mm}
		\textit{Aachen Institute for Advanced Study in Computational Engineering Science (AICES),\\ RWTH Aachen University, Templergraben 55, 52062 Aachen, Germany}
		
		\vspace{5mm}
		

Published\footnote{This pdf is the personal version of an article whose final publication is available at \href{http://dx.doi.org/10.1007/s00466-017-1455-7}{http:/\!/link.springer.com}} 
in \textit{Computational Mechanics}, 
\href{http://dx.doi.org/10.1007/s00466-017-1455-7}{DOI: 10.1007/s00466-017-1455-7} \\
Submitted on 18.~April 2017, Revised on 13.~July 2017, Accepted on 21.~July 2017 

	\end{center}

\pagenumbering{arabic}

\section*{Abstract}
\addcontentsline{toc}{section}{Abstract}

\hrule
A novel adaptive local surface refinement technique based on Locally Refined Non-Uniform Rational B-Splines (LR NURBS) is presented. LR NURBS can model complex geometries exactly and are the rational extension of LR B-splines. The local representation of the parameter space overcomes the drawback of non-existent local refinement in standard NURBS-based isogeometric analysis. For a convenient embedding into general finite element code, the B\'ezier extraction operator for LR NURBS is formulated. An automatic remeshing technique is presented that allows adaptive local refinement and coarsening of LR NURBS. \\
In this work, LR NURBS are applied to contact computations of 3D solids and membranes. For solids, LR NURBS-enriched finite elements are used to discretize the contact surfaces with LR NURBS finite elements, while the rest of the body is discretized by linear Lagrange finite elements. For membranes, the entire surface is discretized by LR NURBS. Various numerical examples are shown, and they demonstrate the benefit of using LR NURBS: Compared to uniform refinement, LR NURBS can achieve high accuracy at lower computational cost.\\
\textbf{Keywords:} Adaptive local refinement, computational contact mechanics, isogeometric analysis, LR B-splines, nonlinear finite element analysis\\
\hrule

\section{Introduction}
\pagestyle{plain}
A wide range of engineering applications, especially those that are governed by local surface effects, necessitate an accurate surface description. An example are contact problems. Due to their strong nonlinear behavior, those are usually only solvable by numerical methods. For this, isogeometric finite element discretizations are advantageous, since they offer at least $C^1$-continuity across element boundaries. Standard NURBS based discretizations lack local refinement, but their ability to discretize complex geometries exactly makes them powerful. \\
Isogeometric Analysis (IGA) was developed by \cite{hughes05} and the work of \cite{cottrell} summarizes the concept of IGA. \cite{borden11} introduced the B\'ezier extraction operator, which allows a suitable embedding of isogeometric analysis into existing finite element codes. The work of \cite{scott11} introduces the extension of the B\'ezier extraction operator to T-splines. This offers a local representation of the B\'ezier extraction operator. B-splines and NURBS are the most widespread element types in IGA. Those element types are only globally refinable. Hierarchical B-splines were introduced by \cite{Forsey1988-1} and allow local refinement. An analysis-suitable approach that bases on hierarchichal NURBS discretization can be found in, e.g. \cite{schill12}. The work of \cite{Sederberg2003-1} introduces T-splines as an approach to discretize surfaces more efficiently than hierarchical B-splines. \cite{scott2012} developed a local refinement approach based on analysis-suitable T-splines. \\
\textit{Locally Refined B-splines} (LR B-splines) are developed as a new approach to allow local refinement in IGA. LR B-splines were introduced by \cite{dokken13} and further advanced by \cite{Bressan2013-1} and \cite{johannessen14}. 
\cite{johannessen2015-2} studied the similarities and differences of LR B-spline, classical hierarchical and truncated hierarchical discretizations. The classical hierarchical approach leads to a much denser sparsity pattern of matrices than LR B-spline or truncated hierarchical discretizations. Whether an LR B-spline or a truncated hierarchical discretization is beneficial is depending on the ratio of the locally refined domain w.r.t. the entire mesh. 
In the IGA community, T-splines are more widespread than LR splines and commercial software for the automatic mesh generation is available, e.g. Rhinoceros \citep{mcneel2012} with a T-splines plug-in \citep{auto2015}.
In contrast to T-splines, the refinement by LR B-splines is directly performed in the parameter domain, which reflects the piecewise polynomial structure. This is more convenient than the multiple vertex grids of T-splines. With an LR B-spline description new possibilites are achieved on how the computational mesh can be generated. This makes them interesting in many fields of engineering. In the works of \cite{nortoft14} and \cite{johannessen15} the LR B-splines are successfully applied to the computation of Navier-Stokes flows. To the best of the authors' knowledge \textit{rational} basis functions in the context of LR splines have not been used so far. In this work, LR B-splines are extended to LR NURBS to combine the advantages of local refinement and the ability of modeling complex geometries exactly.\\
The numerical examples presented here consider 3D solids and membranes using a classical finite element formulation for solids \citep{wriggers-fee} and the membrane finite element formulation of \cite{membrane}. This work is based on computational contact mechanics \citep{laursen, wriggers-contact} using isogeometric finite elements \citep{lorenzis11, lu11, temizer11, lorenzis12} and an unbiased contact algorithm \citep{Sauer2013-1,Sauer2014-2}. 
In the literature several approaches can be found that account for local refinement and adaptive meshing.
\cite{Lee1994} present an adaptive method for $h$-$p$ refined meshes in the context of frictional contact and non-isogeometric finite elements.
The work of \cite{hager12} presents a non-conforming domain decomposition method. 
The domain decomposition consists of a global coarse mesh and an overlapping fine mesh for the contact computation. 
In the context of IGA and frictionless contact a recent work based on analysis-suitable T-splines is presented in \cite{dimitri14}. The work by \cite{dimitri2017} is based on analysis-suitable T-splines and investigates frictional contact and mixed mode debonding problems. In the work of \cite{temizer16} hierarchical NURBS are used to allow local refinement.
In this work, LR NURBS are applied to frictionless and frictional sliding contact. \\
LR B-splines have been successfully used in 2D, since they are linearly independent in 2D. In 3D, the linear independence for arbitrary, locally refined meshes has not been proven yet and requires further study. To overcome this issue, this work uses 2D LR NURBS and combines them with an enrichment technique \citep{Corbett2014-1, nece2} along the third direction. This combines the high accuracy, which is achieved by isogeometric analysis and the efficiency of standard finite elements. This formulation is adapted for LR NURBS-enrichment to gain the possibility of local refinement. In this work we present a novel technique for adaptive local surface refinement and coarsening that
\begin{itemize}
\item considers the extension of LR B-spline finite elements to LR NURBS finite elements
\item uses a local formulation of the B\'ezier extraction operator
\item is automatically controlled by a proposed refinement criterion
\item is extended to LR NURBS-enriched finite elements for 3D solids
\item is applied to frictionless and frictional sliding contact.
\end{itemize}
In Sec.~\ref{sec:FE} a brief summary of the nonlinear finite element formulation for membranes and 3D solids is given. A short introduction to computational contact mechanics follows in Sec.~\ref{sec:cont}. The fundamentals of LR B-splines are presented in Sec.~\ref{sec:LR_B-splines}. The geometric modeling using LR NURBS and the formulation of the B\'ezier extraction operator is presented in Sec.~\ref{sec:LR_NURBS}. The technique of adaptive local refinement and coarsening using LR NURBS discretizations is presented in Sec.~\ref{sec:ALR}. Numerical results are shown in Sec.~\ref{sec:num_res}. The performance of LR NURBS within the proposed local refinement technique is compared with uniform meshes using NURBS discretizations. A conclusion is given in Sec.~\ref{sec:conclusion}.

\section{Preliminaries}
\label{sec:prelim}

This section presents an overview of the fundamental mathematical formulations that are used within this work. A general finite element formulation for membranes and 3D solids in the context of contact computations is discussed in Sec.~\ref{sec:FE} and Sec.~\ref{sec:cont}. The fundamentals of LR B-splines are given in Sec.~\ref{sec:LR_B-splines}.

\subsection{Finite element formulation}
\label{sec:FE}
The finite element formulation for nonlinear membranes and 3D solids is similar. Their governing equations (e.g. equilibrium) are different but obtaining their discretized weak form follows the same approach. The quasi-static weak form of a system of two deformable bodies in contact is described by
\begin{equation}
\sum^\mathrm{II}_{k=\mathrm{I}} \left[\, \delta \Pi^k_{\mathrm{int}} - \delta \Pi^k_{\mathrm{c}} - \delta \Pi^k_{\mathrm{ext}}\, \right] = 0~,
\label{eq:FE}
\end{equation}
considering two objects $(k=\mathrm{I},\mathrm{II})$\footnote{the index $k$ is used as super- and subscript but the position has no special meaning}. This describes the equilibrium between the internal virtual work $\delta \Pi^k_{\mathrm{int}}$, the virtual contact work $\delta \Pi^k_{\mathrm{c}}$ and the external virtual work. In our examples $\delta \Pi^k_{\mathrm{ext}}~=~0$ is considered. The computational formulation for nonlinear membranes in $\IR^3$ in the framework of curvilinear coordinates is taken from \citet{membrane}. A general constitutive setting and finite element formulation for solids in $\IR^3$ can be found in \citet{wriggers-fee}. The internal virtual work for membranes $\mathcal{S}_k$ in $\IR^3$ and solids $\mathcal{B}_k$ in $\IR^3$ is
\eqb{lll}
\delta \Pi^k_{\mathrm{int}}=\ds \begin{cases}
\ds\int_{\mathcal{S}_k} \delta\boldsymbol{\varphi}^k_{,\alpha}\,\cdot\, {\sigma_k^{\alpha\beta}} \,\ba^k_{\beta} \, \text{d}a_k~,\quad\forall\,\delta\boldsymbol{\varphi}_k \in \mathcal{V}_k~, & \text{for} \quad \mathcal{S}_k~,\\[4mm]
\ds\int_{\mathcal{B}_k} \text{grad}(\delta\boldsymbol{\varphi}_k)\,:\, \boldsymbol{\sigma}_k \, \text{d}v_k~,\quad\forall\,\delta\boldsymbol{\varphi}_k \in \mathcal{V}_k~, \ds & \text{for} \quad\mathcal{B}_k~.
\end{cases}
\eqe
Here, $\sigma^{\alpha\beta}_k$ are the contra-variant components of the Cauchy stress tensor $\bsig$ with $\alpha = 1,2$ and $\beta = 1,2$. The parametric derivative of the virtual displacement field is denoted by $\delta\boldsymbol{\varphi}^k_{,\alpha}$ and $\ba^k_{\alpha}$ are the co-variant tangent vectors. The virtual contact work for the contact surfaces $\Gamma^k_{\mathrm{c}}$ is expressed as
\eqb{lll}
\label{eq:cf}
\delta \Pi^k_{\mathrm{c}} = \ds\int_{\Gamma^k_{\mathrm{c}}}  \delta\boldsymbol{\varphi}_k\cdot\boldsymbol{t}^k_{\mathrm{c}}\, \text{d}a_k~,\quad\forall\,\delta\boldsymbol{\varphi}_k \in \mathcal{V}_k~,
\eqe
where $\bt^k_{\mathrm{c}}$ denotes the contact traction on the surface $\Gamma^k_{\mathrm{c}}\subset \sS_k$. The computation of the contact integrals is performed using the two-half pass algorithm of \citet{Sauer2013-1}. This algorithm evaluates Eq.~\eqref{eq:cf} on each body equivalently. System~\eqref{eq:FE} is solved for the unknown deformation field $\boldsymbol{\varphi}_k \in \mathcal{V}_k$, with $\mathcal{V}_k$ as a suitable space of kinematically admissible variations. The membrane and the contact surface are discretized into a set of membrane elements $\Gamma^{k}_e$ and a set of contact elements $\Gamma^{k}_{\mathrm{c}\,e}$ such that $\mathcal{S}_k \approx \mathcal{S}_k^h = \bigcup_e \Gamma^{k}_e$ and $\Gamma^k_{\mathrm{c}} \approx \Gamma^{k\,h}_{\mathrm{c}} = \bigcup_e \Gamma^{k}_{\mathrm{c}\,e}$, respectively. Here, the element domains $\Gamma^{k}_e$ and $\Gamma^{k}_{\mathrm{c}\,e}$ are considered equal and have the same finite element discretization. Similarly, the solids are discretized into a set of volume elements $\Omega^k_e$ and a set of surface contact elements $\Gamma^k_{\mathrm{c}\,e}$ such that $\mathcal{B}_k \approx \mathcal{B}_k^h = \bigcup_e \Omega^k_e$ and $\Gamma^k_{\mathrm{c}} \approx \Gamma^{k\,h}_{\mathrm{c}} = \bigcup_e \Gamma^k_{\mathrm{c}\,e}$, respectively. The superscript $h$ denotes the approximated discrete setting. With a set of $n_e$ basis functions $\mathbf{N} = \left[N_1 \bI,...,N_{n_e}\bI \right]$ with identity matrix $\bI$ in $\IR^3$ and discrete points $\mathbf{x}_e$, a point $\bx \in \mathcal{B}_k$ is interpolated within each element as 
\begin{equation}
\bx\approx\bx^h=\mathbf{N}\,\mathbf{x}_e~.
\end{equation}
The same holds for the reference configuration $\bX\approx\bX^h$, the displacement field $\bu\approx\bu^h$ and the virtual displacement field $\delta\boldsymbol{\varphi}\approx\boldsymbol{v}^h$. In here we drop the index $k$ for convenience. In standard finite elements the discrete points $\mathbf{x}$ are known as nodes, while in IGA they are called control points. For standard finite element basis functions Lagrange polynomials are used, while for IGA the basis functions will be discussed in Sec.~\ref{sec:LR_B-splines}. As a consequence of the above definitions, the dicretized weak form can be written as
\begin{equation}
\mathbf{v}^{\mT}\left[\mathbf{f}_{\mathrm{int}}-\mathbf{f}_{\mathrm{c}} \right] = \boldsymbol{0},\,\,\,\,\,\,\,\, \forall\, \mathbf{v} \in \mathcal{V}^h~,
\end{equation}
with the internal forces
\eqb{lll}
\mathbf{f}_\mathrm{int}=\ds \begin{cases}
\ds\int_{\mathcal{S}_k}\mathbf{N}^\mathrm{T}_{,\alpha}\,\sigma^{\alpha\beta}\,\mathbf{N}_{,\beta}\,\text{d}a_k~,  & \text{for} \quad \mathcal{S}_k~,\\[4mm]
\ds\int_{\mathcal{B}_k}\mathbf{N}_{,\bx}\,\bsig\,\text{d}v_k~, \ds & \text{for} \quad\mathcal{B}_k~,
\end{cases}
\eqe
and the contact forces
\begin{equation}
\mathbf{f}_\mathrm{c} = \int_{\Gamma^k_\mathrm{c}}\mathbf{N}^\mathrm{T}\,\bt_\mathrm{c}\,\text{d}a_k~.
\end{equation}
A brief overview of computational contact mechanics is given in the next section.

\subsection{Computational contact formulation}
\label{sec:cont}
This section gives a brief summary of computational contact mechanics. For more detailed information, the monograph by \cite{wriggers-contact} is recommended. Here, the penalty method is used in order to enforce contact. The gap vector $\bg$ between two points $\boldsymbol{x}_\mathrm{I}$ and $\boldsymbol{x}_\mathrm{II}$ on the surfaces $\sS_\mathrm{I}$ and $\sS_\mathrm{II}$ is given by
\begin{equation}
\boldsymbol{g} = \boldsymbol{x}_\mathrm{I}-\boldsymbol{x}_\mathrm{II}~.
\end{equation}
The closest projection point $\boldsymbol{x}_\mrp$ of $\boldsymbol{x}_\mathrm{I}$ is obtained by orthogonal projection of $\boldsymbol{x}_\mathrm{I}$ onto $\sS_\mathrm{II}$. The contact traction is expressed as $\boldsymbol{t}_\mathrm{c} = \bt_\mathrm{n}+\bt_\mathrm{t}$, i.e. it is decomposed into its normal and tangential components. According to the penalty formulation, the normal traction
\begin{equation}
\boldsymbol{t}_\mathrm{n}(\boldsymbol{x}_k)=
\begin{cases}
-\eps_\mathrm{n}\, g_\mathrm{n}\, \boldsymbol{n}_\mathrm{p}~,
& g_\mathrm{n}<0~,\\
0~,
& g_\mathrm{n}\geq0~,
\end{cases}
\end{equation}
is applied at each contact point. The traction is proportional to the normal gap $g_\mathrm{n} = (\boldsymbol{x}_\mathrm{I} - \boldsymbol{x}_\mathrm{p})\cdot \boldsymbol{n}_\mathrm{p}$, the surface outward normal $\bn_\mrp$ at $\bx_\mrp$ and the constant penalty parameter $\varepsilon_\mathrm{n}$. $k$ is the index for the body under consideration. The tangential contact traction $\bt_\mrt$ during frictional sliding is given by Coulomb's law
\begin{equation}
\boldsymbol{t}_\mathrm{t} = -\mu\, p\, \frac{\boldsymbol{\dot{g}}_\mathrm{t}}{\boldsymbol{||\dot{g}}_\mathrm{t}||}~.
\end{equation}
Here, we find the friction coefficient $\mu$, the contact pressure $p=||\bt_\mathrm{n}||$ and the relative tangential sliding velocity $\boldsymbol{\dot{g}}_\mathrm{t}$. The tangential contact slip 
\begin{equation}
\boldsymbol{g}_\mathrm{t} = \Delta \boldsymbol{g}_\mathrm{e} + \boldsymbol{g}_\mathrm{s}~,
\end{equation}
consists of the reversible $\Delta\bg_\mathrm{e}$ and irreversible $\bg_\mathrm{s}$ part. The slip criterion
\begin{equation}
f_\mathrm{s} =\, ||\bt_\mathrm{t}||-\mu\, p~,
\end{equation}
is used to distinguish between sticking and slipping. Sliding occurs for $f_\mathrm{s} =0$ and sticking for $f_\mathrm{s} < 0$. The examples presented below deal with frictionless and frictional contact. For frictionless contact, the tangential traction $\boldsymbol{t}_\mathrm{t}$ vanishes and the contact traction is given by $\boldsymbol{t}_\mathrm{c} = \boldsymbol{t}_\mathrm{n}$. The implementation and formulation of frictional contact is taken from \citet{Sauer2014-2}.
\subsection{LR B-spline fundamentals}
\label{sec:LR_B-splines}
The core idea of LR B-splines is to disband the tensor-product mesh structure of classical B-splines in order to obtain a local representation of the parameter domain. As introduced in the B-spline theory, a knot vector $\Xi$ of size $n+p+1$ generates $n$ linearly independent basis functions of order\footnote{also known as degree} $p$. The local representation of the parameter space and the geometric discretization are obtained by splitting the global knot vector $\Xi = [\xi_1,...,\xi_{n+p+1}]$ into local knot vectors $\Xi_i = [\xi_i,...,\xi_{i+p+1}]$ each constructing a single basis function. With the local knot vector representation, the domain of the basis function is minimal, i.e. the basis has \textit{minimal support}. The basis functions are defined by the Cox-de Boor recursion formula, which depends on the entries of the local knot vector $\Xi_i$ and desired polynomial order $p$, see \cite{Cox1971-1} and \cite{DeBoor1972-1}. For $p=0$
\begin{equation}
\label{eq:cox1}
N_{i}^0(\xi)=
\begin{cases}
1~,
& $if $ \xi_{i}\leq\xi<\xi_{i+1}~,\\
0~,
&$otherwise~,$
\end{cases}
\end{equation}
and for $p>0$
\begin{equation}
\label{eq:cox2}
N_{i}^p(\xi)=
\frac{\xi-\xi_i}{\xi_{i+p}-\xi_i}\,N_{i}^{p-1}(\xi_i,...,\xi_{p+i})+
\frac{\xi_{i+p+1}-\xi}{\xi_{i+p+1}-\xi_{i+1}}\,N_{i+1}^{p-1}(\xi_{i+1},...,\xi_{p+i+1})~.
\end{equation}
The LR B-spline basis  $B^{p\,q}_{i\,j}(\xi,\eta)$ of order $p$ and $q$ in 2D is defined as a separable function $B : \IR^{2}\rightarrow \IR$ 
\begin{equation}
B^{p\,q}_{i\,j}(\xi,\eta) = N_{i}^p(\xi)\,M_{j}^q(\eta)~. 
\end{equation}
$N_{i}^p(\xi)$ and $M_{j}^q(\eta)$ are the shape functions in each parametric direction. The properties of LR B-splines follow directly from standard isogeometric analysis. The basis is non-negative and forms a partition of unity, and the resulting geometry lies within the convex hull of the control points. The continuity across the element boundaries is ${C}^{p-{m_v}}$, i.e. it is defined by the polynomial order $p$ and the multiplicity of the knot vector entry $m_v$. The LR B-spline surface is then constructed by a set of control points $\mathbf{x}_{i\,j}$ and the local basis functions $B^{p\,q}_{i\,j}(\xi,\eta)$, as
\begin{equation}
\bx(\xi,\eta)=\sum_{i=1}^n\sum_{j=1}^m B_{i\,j}^{p\,q}(\xi,\eta)\,\mathbf{x}_{i\,j}~,
\end{equation}
with $n,m$ as the number of single basis functions in each parametric dimension. To ensure that LR B-splines keep the partition of unity property during local refinement the basis functions are multiplied by a scaling factor $\gamma_{i\,j}\in (0,1]$
\begin{equation}
\bx(\xi,\eta)=\sum_{i=1}^n\sum_{j=1}^m B_{i\,j}^{p\,q}(\xi,\eta)\,\mathbf{x}_{i\,j}\,\gamma_{i\,j}~.
\end{equation}
With the above formulation, the global B-spline representation is split into a local representation consisting of a set of locally refinable B-splines. 
In the following we use the terminology LR B-spline for both \textit{locally refinable} and \textit{locally refined} B-splines.
It is important to note that the global B-splines do not exist anymore. Global B-splines do not allow local refinement. The ability is only given by the LR B-splines. Locally refined meshes and the local refinement procedure are discussed in the following sections.

\subsubsection{Locally refined meshes}

\begin{figure}[h]
\subfloat[LR mesh]{\includegraphics[width=0.32\linewidth, trim = 0 0 0 0,clip]{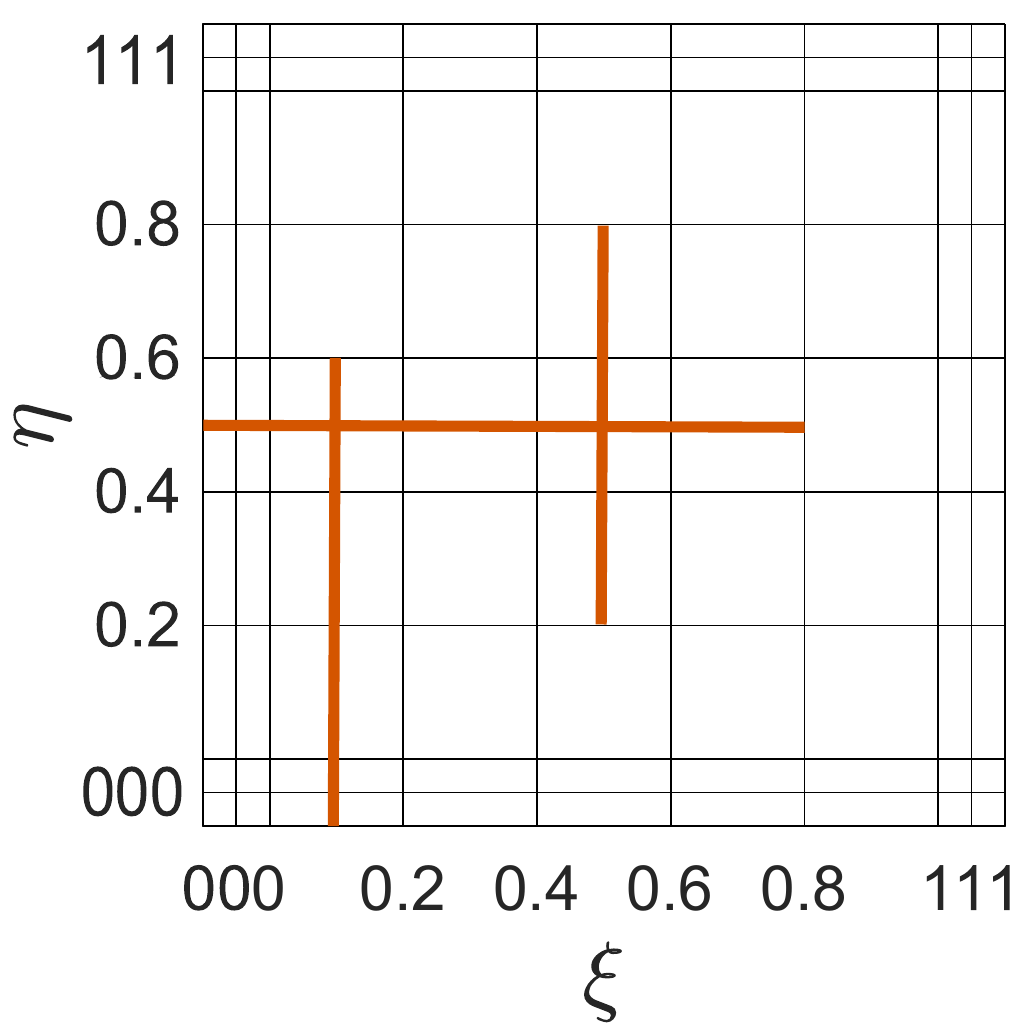}
\label{fig:LRm}}
\subfloat[Minimal support]{\includegraphics[width=0.32\linewidth]{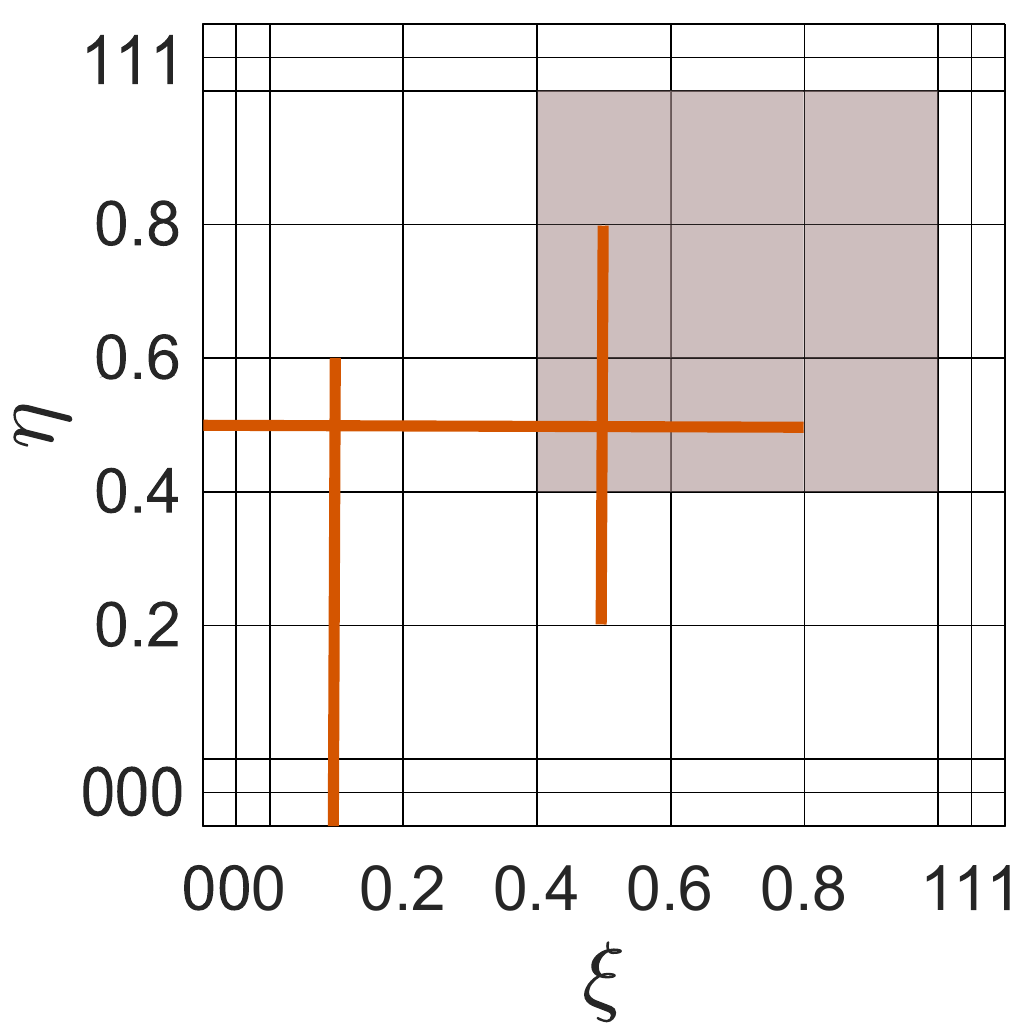}
\label{fig:minsupp}}
\subfloat[No minimal support]{\includegraphics[width=0.32\linewidth]{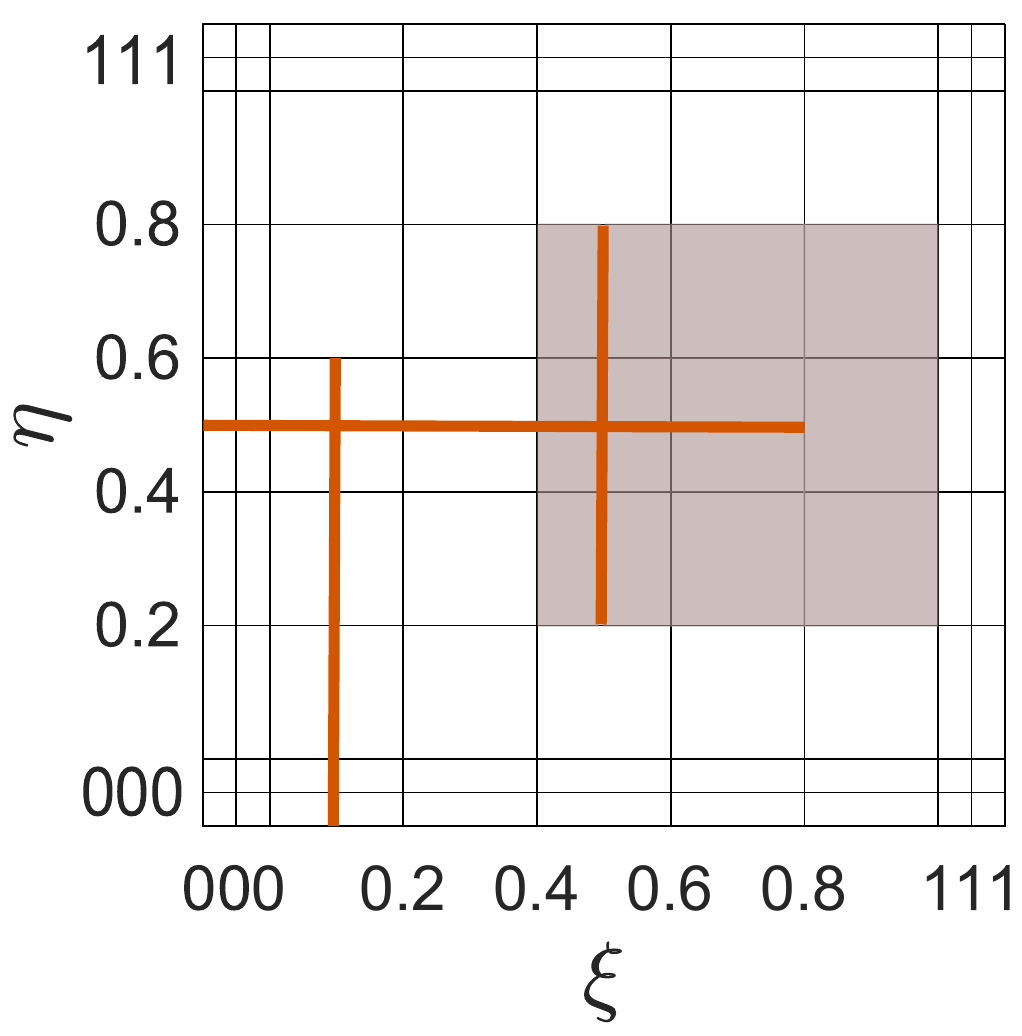}
\label{fig:notminsupp1}}
\caption{LR meshes contructed by meshline insertion (marked by thick lines) into a tensor mesh of a bivariate parameter domain. (a)~Example of an LR mesh. (b)~Support domain of the LR B-spline basis, marked in gray, which has minimal support. (c)~ Support domain of the LR B-spline basis, marked in gray, that does \textit{not} have minimal support. The meshline extension, which is spanned by $\eta=[0.2,0.8]$ at $\xi=0.5$ fully crosses the support domain.}
\label{fig:notminsupp}
\end{figure}

The local representation of the parametric domain leads to new possibilities on how the finite element mesh can be constructed. The ability of local refinement is the major advantage in comparison to classical tensor meshes. The LR mesh is the result of a series of meshline insertions into an initial tensor mesh. Fig.~\ref{fig:LRm} shows an example of an LR mesh. The meshlines never stop in the center of an element (knot span). As a knot vector of an LR B-spline basis function has size of $p+2$ knot vector entries, the meshlines cross at least $p+2$ knots. A meshline insertion can be either a new meshline, an elongation of an existing one, a joining of two meshlines or the increase of the multiplicity of a meshline. Increasing the multiplicity decreases the continuity of the LR B-spline. In the bi-variate case, horizontal and vertical meshlines can be inserted. Any type of meshline insertion implies that an LR B-spline loses the property of minimal support. An LR B-spline has minimal support if the support domain of the basis functions is not fully crossed by any other meshline. The basis function domain of the LR B-spline in Figure~\ref{fig:minsupp}, marked in gray, has minimal support because its support domain is not fully crossed by any meshline insertion completely. The LR B-spline basis marked in the Fig.~\ref{fig:notminsupp1} does not have minimal support. Its basis function domain is crossed by the vertical meshline, which is spanned by $\eta=[0.2,0.8]$ at $\xi=0.5$. If an LR B-spline loses the property of minimal support refinement is performed as discussed in the next section.

\subsubsection{Procedure of local refinement}

\begin{figure}[h]
\centering
\includegraphics[width=0.7\linewidth]{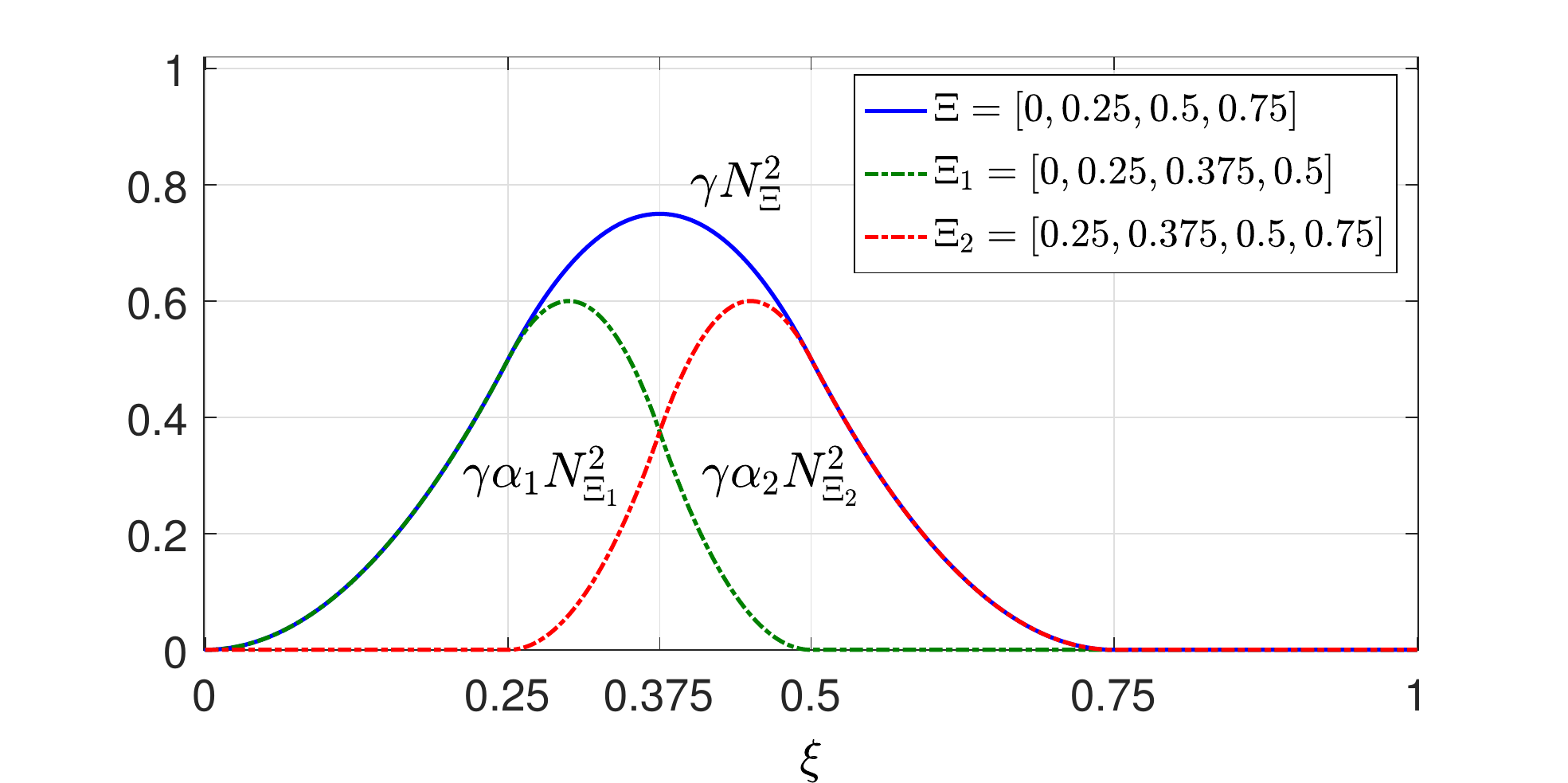}
\caption{The original basis function of the knot vector $\Xi$ is split by knot insertion at $\xi = 0.375$ into two local basis functions characterized by the knot vectors $\Xi_1$ and $\Xi_2$.}
\label{fig:sep_basisfun}
\end{figure}

Local refinement is perfomed if an LR B-spline loses the property of minimal support due to meshline extensions. In general, the procedure of local refinement is realized by single knot insertion. From the classical theory it is known that this enriches the basis while the geometry remains unchanged. The insertion of a single knot $\hat{\xi}$ into the knot span $\bxi_i = [\xi_{i-1},\xi_i]$ of a local knot vector $\Xi$ of size $p+2$ leads to a knot vector of size $p+3$. Consequently, two locally refined B-splines are generated by splitting the enlarged knot vector $\Xi=\left[\xi_1,...,\xi_{i-1},\hat{\xi},\xi_i,...,\xi_{p+2}\right]$ into the local knot vectors of size $p+2$
\eqb{lll}
\Xi_1&=\left[\xi_1,...,\xi_{i-1},{\hat{\xi}},\xi_i,...,\xi_{p+1}\right]~,\\[2mm]
\Xi_2&=\left[\xi_2,...,\xi_{i-1},{\hat{\xi}},\xi_i,...,\xi_{p+2}\right]~.
\eqe
The relation for an LR B-spline basis in one parametric direction on the LR mesh is then given by
\begin{equation}
\gamma N^p_{\Xi}(\xi)=\gamma_1\,N^p_{\Xi_1}(\xi)+\gamma_2\,N^p_{\Xi_2}(\xi)~.
\end{equation}
with
\eqb{lll}
\label{eq:gamma}
\gamma_1 &= \alpha_1\,\gamma~,\\
\gamma_2 &= \alpha_2\,\gamma~.
\eqe
The associated $\alpha_{1}$ and $\alpha_{2}$ are determined by
\eqb{lll}
\label{eq:calc_alpha}
\alpha_1=
\begin{cases}
1~,
&\xi_{p+1}\leq\hat{\xi}<\xi_{p+2}~,\\
\ds\frac{\hat{\xi}-\xi_1}{\xi_{p+1}-\xi_1}~,
&\xi_{1}<\hat{\xi}<\xi_{p+1}~,\\[4mm]
\end{cases}\\
\alpha_2=
\begin{cases}
\ds\frac{\xi_{p+2}-\hat{\xi}}{\xi_{p+2}-\xi_2}~,
&\xi_{2}<\hat{\xi}<\xi_{p+2}~,\\
1~,
&\xi_{1}<\hat{\xi}\leq\xi_{2}~.
\end{cases}
\eqe
Consider the knot $\hat{\xi} = 0.375$ is inserted into the knot vector $\Xi = [0, 0.25, 0.5, 0.75]$. The knot vector entries construct the quadratic basis function $N^2_{\Xi}$. As illustrated in Fig.~\ref{fig:sep_basisfun}, the original basis function is split into two new functions each describing the LR B-spline basis functions $\gamma\,\alpha_{1}\,N^2_{\Xi_1}$ and $\gamma\,\alpha_{2}\,N^2_{\Xi_2}$. 
The refinement process for bivariate functions is performed in one parametric direction at a time, i.e. first all horizontal meshlines and then all vertical meshlines are inserted (or vice versa). Consider two local knot vectors $\Xi$ and $\mathcal{H}$, one in each parametric direction. For a weighted, bivariate LR B-spline basis ${\gamma}\,B^{p\,q}_{\boldsymbol{\Xi}}(\xi,\eta)$ with $\mathbf{\Xi} = [\Xi,\mathcal{H}]$ the following relation is given
\begin{flalign}
\label{eq:weight_spline}
\begin{split}
\gamma \,B^{p\,q}_{\boldsymbol{\Xi}}(\xi,\eta) &=\gamma \,N^p_{\Xi}(\xi)\,M^q_{\mathcal{H}}(\eta)~,\\
&=\gamma\,(\alpha_1\,N^p_{\Xi_1}(\xi)+\alpha_2\,N^p_{\Xi_2}(\xi))\,M^q_{\mathcal{H}}(\eta)~, \\
&=\gamma_1\,B_{\mathbf{\Xi}_1}^{p\,q}(\xi,\eta)+\gamma_2\,B_{\mathbf{\Xi}_2}^{p\,q}(\xi,\eta)~,\\
\end{split}
\end{flalign}
for the case that $\Xi$ is split into $\Xi_{1}$ and $\Xi_{2}$. The refinement process is given by two steps. At first all LR B-spline basis functions whose support domain is crossed by meshline extensions will be split. The second step is to check every newly created basis function if it has minimal support. In the case that a newly created LR B-spline does not have minimal support an additional splitting is performed. By splitting $\gamma \,B^{p\,q}_{\boldsymbol{\Xi}}(\xi,\eta)$ into ${\gamma_j}\, B^{p\,q}_{\boldsymbol{\Xi}_j}(\xi,\eta)$, with $j=1,2$, the following cases can occur:
\begin{enumerate}
\item The LR B-spline does not exist and a new LR B-spline is created. In this case the new control point $\mathbf{x}_j$ is a copy of the control point $\mathbf{x}$ of the parent LR B-spline $\mathbf{x}_j = \mathbf{x}$. The weight $\gamma_j$ is set by Eq.~\eqref{eq:gamma}.
\item If $B^{p\,q}_{\boldsymbol{\Xi}_j}(\xi,\eta)$ already exists, the control point and the weight are set to
\begin{equation}
\label{eq:cp_update}
\mathbf{x}_{j} = \frac{\mathbf{x}_{j} \,\gamma_{j}+\mathbf{x}\,\gamma\,\alpha_{j}}{\gamma_{j}+\alpha_{j}\,\gamma}~,
\end{equation}
and
\begin{equation}
\label{eq:gamma_up}
\gamma_{j} = \gamma_{j}+\gamma\alpha_{j}~.
\end{equation}
\end{enumerate}
After splitting the former LR B-spline, $\gamma \,B^{p\,q}_{\boldsymbol{\Xi}}(\xi,\eta)$ is deleted in both cases. The algorithm proceeds with the second step and checks if the support domain of the new local basis functions is fully crossed by any existing meshline. If one does not have minimal support the first step is performed again. Note that at every step of the refinement process the partition of unity is maintained and the geometric mapping is left unchanged. For the use of LR B-splines in IGA it is required to ensure that the resulting spline space is linearly independent. In the bivariate case this can be guaranteed by only using primitive meshline extensions\footnote{A primitive meshline extension is (a) a meshline spanning $p+1$ elements, (b) elongating a meshline by one element or (c) raising the multiplicity of a meshline (length of $p+1$ elements).}, see \cite{Mourrain2014-1} and \cite{johannessen14}. In this work, all meshline extensions are formulated as primitives.

\section{Geometric modeling using LR NURBS}
\label{sec:LR_NURBS}

So far, non-rational LR B-splines have been discussed. In this section, their extension to Locally Refined Non-Uniform $Rational$ B-Splines (LR NURBS) is introduced. With a NURBS representation of objects one gains the ability to describe many geometries, which cannot be represented by polynomials, exactly. Especially conic sections, ellipsoids, spheres, cylinders, etc. can be constructed by a projective transformation of piecewise quadratic curves exactly, see \cite{Farin1992-1}. This is one of the defining features of NURBS.

\subsection{LR NURBS}
The fundamentals of standard NURBS in isogeometric analysis can be found in, e.g. \cite{hughes05}. Combining NURBS with the LR B-spline theory from Sec.~\ref{sec:LR_B-splines}, the extension of LR B-splines to LR NURBS follows. An LR NURBS object in $\IR^d$ is constructed by the projective transformation of an LR B-spline entity in $\IR^{d+1}$. The first $d$ entries of a projective control point $\boldsymbol{\mathbf{x}}_i^\mrw$ represent the spatial coordinates and the $d+1$ entry the weight, e.g. $\boldsymbol{\mathbf{x}}_i^\mrw= [x_i,y_i,z_i,\mrw_i]$ for $d=3$. The control points of the LR NURBS object result from the projective transformation
\begin{equation}
\label{eq:NURBS_trans}
(\mathbf{x}_i)_k=\frac{(\mathbf{x}_i^\mrw)_k}{\mrw_i}~, \quad k=1,...,d~,\quad \text{where}\quad \mrw_i=(\mathbf{x}_i^\mrw)_{d+1}~,
\end{equation}
%
with $(\mathbf{x}_i)_k$ as the $k^{th}$ component of the vector $\mathbf{x}_i$ and $\mrw_i$ as the $i^{th}$ weight. To generalize this relation the weighting function for the bivariate case is introduced by
\begin{equation}
W(\xi,\eta)=\sum_{i=0}^n\sum_{j=0}^m B^{p\,q}_{i\,j}(\xi,\eta)\, \mrw_{i\,j}~.
\end{equation}
$B^{p\,q}_{i\,j}(\xi,\eta)$ are the standard LR B-spline basis functions and the transformation is applied by dividing every point of the curve by $W(\xi,\eta)$. Each element of surface ${\mathcal{S}}$ is a polynomial, which is divided by another polynomial of the same order. The resulting LR NURBS basis is then defined as
\begin{equation}
R^{p\,q}_{i\,j}(\xi,\eta)= \frac{B^{p\,q}_{i\,j}(\xi,\eta)\, \mrw_{i\,j}}{W(\xi,\eta)} = \frac{B^{p\,q}_{i\,j}(\xi,\eta)\, \mrw_{i\,j}}{\sum_{\hat{i}=0}^n \sum_{\hat{j}=0}^m B^{p\,q}_{\hat{i}\,\hat{j}}(\xi,\eta)\,\mrw_{\hat{i}\,\hat{j}}}~.
\end{equation}
With a set of control points $\mathbf{x}_{i\,j}$ and the scaling factors $\gamma_{i\,j}$ the LR NURBS surface is then defined as
\begin{equation}
\label{eq:NURBS_surf}
\bx(\xi,\eta)=\sum_{i=0}^n\sum_{j=0}^m R_{i\,j}^{p\,q}(\xi,\eta)\,\mathbf{x}_{i\,j}\,\gamma_{i\,j}~.
\end{equation}
The properties of the LR NURBS basis, such as its continuity and support, follow directly from the knot vectors as before. The basis is still non-negative and it still forms a partition of unity. This leads to the strong convex hull property of the LR NURBS. Note that the weights are separated from any explicit geometric representation. The weights are each associated with a specific control point and a manipulation of them leads to a change of the resulting geometry. In the case that all weights are equal, the surface is again a polynomial and consequently $R^{p\,q}_{i\,j}(\xi,\eta)=B^{p\,q}_{i\,j}(\xi,\eta)$. Therefore, LR B-splines are a special case of LR NURBS. In order to locally refine an LR NURBS object it is necessary to take special care of the weights. Before the refinement procedure starts, the projective control points $\mathbf{x}_{i\,j}^\mrw$ of the LR NURBS are computed by Eq.~\eqref{eq:NURBS_trans}. The weights $\mrw_{i\,j}$ are included in vector $\mathbf{x}_{i\,j}^\mrw$ and treated as the fourth part of the control points. This results in 
\begin{equation}
\mathbf{x}_{i\,j}^\mrw= [x_{i\,j}\,\mrw_{i\,j},\,y_{i\,j}\,\mrw_{i\,j},\,z_{i\,j}\,\mrw_{i\,j},\,\mrw_{i\,j}]~.
\end{equation}
After the refinement process, the control points are transformed back by dividing the spatial coordinates by their associated weight. Note that while the refinement is performed, $\mrw_{i\,j}$ is treated in the same way as the control points. The linear independence of LR NURBS spaces follow directly from the LR B-splines.

\subsection{B\'ezier extraction operator for LR NURBS}
\label{sec:BEO_LR}

\begin{figure}[h]
\centering
\includegraphics[width=0.99\linewidth]{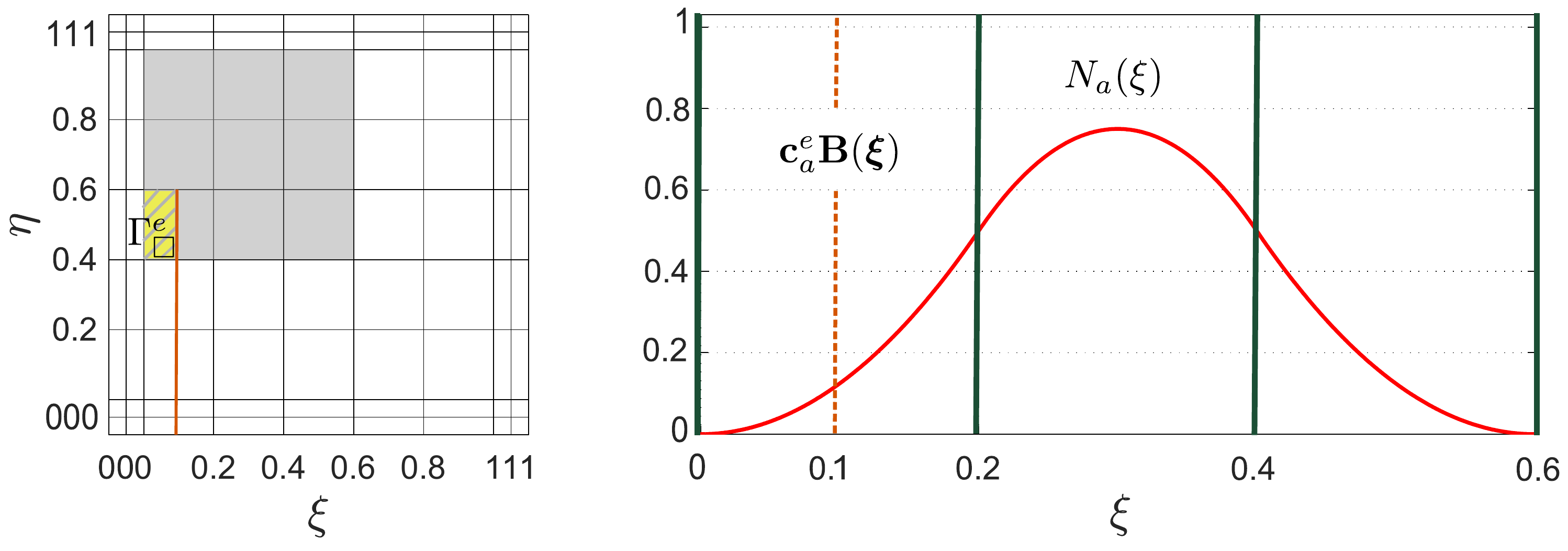}
\caption{Bivariate LR mesh~(left) and the local basis function of a quadratic LR NURBS in one parametric direction~(right). The support width of $N_a^e(\xi)$ is represented by ${\mathbf{c}}_a^{e}\, \mathbf{B}({\bxi})$ in the knot span ${\bxi}=[0,0.2]$ exceeding the actual domain of $\Gamma^e_\square$.}
\label{fig:BEO_supp_1}
\end{figure}

This section discusses the B\'ezier extraction of LR NURBS. The B\'ezier extraction is an advantageous technique for the decomposition of splines into their B\'ezier elements. It allows for the embedding of isogeometric analysis into a standard finite element framework. The B\'ezier extraction of NURBS is introduced by \cite{borden11} and advanced to T-splines by \cite{scott11}. 
The B\'ezier extraction of LR NURBS shares similarities with the formulation of the B\'ezier extraction of T-splines. Both element types have a local parametric representation. The B\'ezier decomposition of each local basis function, that is nonzero within an element, is performed separately.
The basis function in one parametric direction of an LR NURBS element $\Gamma^e_\square$ within the knot span $\bxi$ can be expressed in terms of a set of Bernstein polynomials $\mathbf{B}({{\bxi}})$ and a linear operator $\mathbf{c}_a^{e}$.
Each local basis function ${N}_a^{e}({\bxi}) $ of an element $\Gamma^e_\square$ can then be expressed by 
\begin{equation}
{N}_a^{e}({\bxi}) = {\mathbf{c}}_a^{e}\, \mathbf{B}({\bxi})~.
\label{eq:beo}
\end{equation}
Here, $a=1,2,...,n_e$ where $n_e$ is the number of nonzero basis functions of $\Gamma^e_\square$. In matrix-vector form this becomes
\begin{equation}
{\mN}^{e}({\bxi}) = {\mathbf{C}}^{e} \,\mathbf{B}({\bxi})~.
\end{equation}
$\mC^e$ is the element-wise B\'ezier extraction operator. The linear operator ${\mathbf{c}}_a^{e}$ is of dimension $[1\times (p+1)]$ and represents a single row of $\mC^e$. The coefficients of ${\mathbf{c}}_a^{e}$ are calculated by an algorithm presented in \cite{scott11}. The algorithm inserts knots into the knot vector until all entries have multiplicity $p$. The knot insertion performs the spline decomposition into its B\'ezier elements. With ${\mathbf{c}}_a^{e}$ and the Bernstein polynomials the relation in Eq.~\ref{eq:beo} is obtained. Due to the unstructured LR meshes it is required to account for two aspects:
\begin{enumerate}
\item To determine ${\mathbf{c}}_a^{e}$ it is required that the knot vectors are \textit{open}, i.e. the first and last knot vector entry have multiplicity of $p+1$. The knot vectors of LR NURBS are in general \textit{not open}. For this, the local knot vectors are extended, which is simply done by adding knot vector entries. The extension of the knot vectors does not affect the basis of LR NURBS itself because it is only used for the B\'ezier decomposition.
\item The unstructured meshes of LR NURBS or T-splines necessitate a special consideration. An example is shown in Fig.~\ref{fig:BEO_supp_1}. Consider the element $\Gamma^e_\square$ with $\tilde\bxi=[0,0.1]$ marked in yellow in Fig.~\ref{fig:BEO_supp_1} on the left. The support domain of one bivariate local basis function is marked in gray. In $\xi$-direction, the basis function $N_a(\xi)$ is spanned by the local knot vector ${\Xi}=[0,0.2,0.4,0.6]$ as shown in Fig.~\ref{fig:BEO_supp_1} on the right. The basis function is non-zero within the domain of $\Gamma^e_\square$. From $\Xi$ follows that ${\mathbf{c}}_a^{e}$ is defined within the knot span ${\bxi}=[0,0.2]$ and exceeds $\tilde\bxi=[0,0.1]$ (the domain of $\Gamma^e_\square$). Those non-matching domains must be considered and treated. For T-splines an additional knot is inserted into the knot vector so that ${\mathbf{c}}_a^{e}$ can be determined for the correct parametric domain $\tilde\bxi=[0,0.1]$. In contrast to T-splines, no additional knot insertion is required for the formulation for LR NURBS. 
\end{enumerate}

For LR NURBS the linear operator ${\mathbf{c}}_a^{e}$ is mapped by using an additional linear operator $\mT_a^e$ to obtain a new $\tilde{\mathbf{c}}_a^{e}$ such that
\begin{equation}
{N}_a^{e}(\tilde{\bxi}) = \tilde{\mathbf{c}}_a^{e}\,\mathbf{B}({\bxi})~.
\end{equation}
With $\tilde{\mathbf{c}}_a^{e}={\mathbf{c}}_a^{e}\,\mT_a^e$
\begin{equation}
{N}_a^{e}(\tilde{\bxi}) = {\mathbf{c}}_a^{e}\,\mT_a^e \,\mathbf{B}({\bxi})~,
\end{equation}
follows. The matrix $\mT_a^e$ follows from the relation
\begin{equation}
\tilde{\mathbf{c}}_a^{e}\, \mathbf{B}(\tilde{\bxi})={\mathbf{c}}_a^{e}\, \mathbf{B}({\bxi})~.
\end{equation}
After some algebraic manipulations this equation is expressed as
\begin{equation}
\tilde{\mathbf{c}}_a^{e}={\mathbf{c}}_a^{e}\left(\left(\mathbf{B}(\tilde{\bxi})\right)^{-1}\mathbf{B}({\bxi})\right)^\mT~,
\end{equation}
such that
\begin{equation}
\mathbf{T}^{e}_a := \left(\left(\mathbf{B}(\tilde{\bxi})\right)^{-1}\mathbf{B}({\bxi})\right)^\mT~.
\end{equation}
If the knot spans coincide, i.e. $\tilde\bxi=\bxi$, then
\begin{equation}
\mathbf{B}(\tilde{\bxi})=\mathbf{B}({\bxi})~,
\end{equation}
\begin{equation}
\mT_a^e= \mathbf{I}~,
\end{equation}
and $\tilde{\mathbf{c}}_a^{e}={\mathbf{c}}_a^e$ follows. With this formulation, the basis functions having support in each element $\Gamma^e_\square$ are expressed by the Bernstein polynomials $\mathbf{B}({{\bxi}})$ and the linear operators $\tilde{\mathbf{c}}_a^{e}$.

\section{Adaptive local refinement and coarsening}
\label{sec:ALR}

This section presents a technique for adaptive local refinement and coarsening for LR NURBS discretizations. The technique is formulated in the context of frictional contact. This requires the mapping of contact variables from one mesh to another.

\subsection{Adaptive local refinement}
\label{sec:ALR2}
\begin{figure}[h]
\centering
\includegraphics[width=0.99\linewidth, trim = 0 0 0 0,clip]{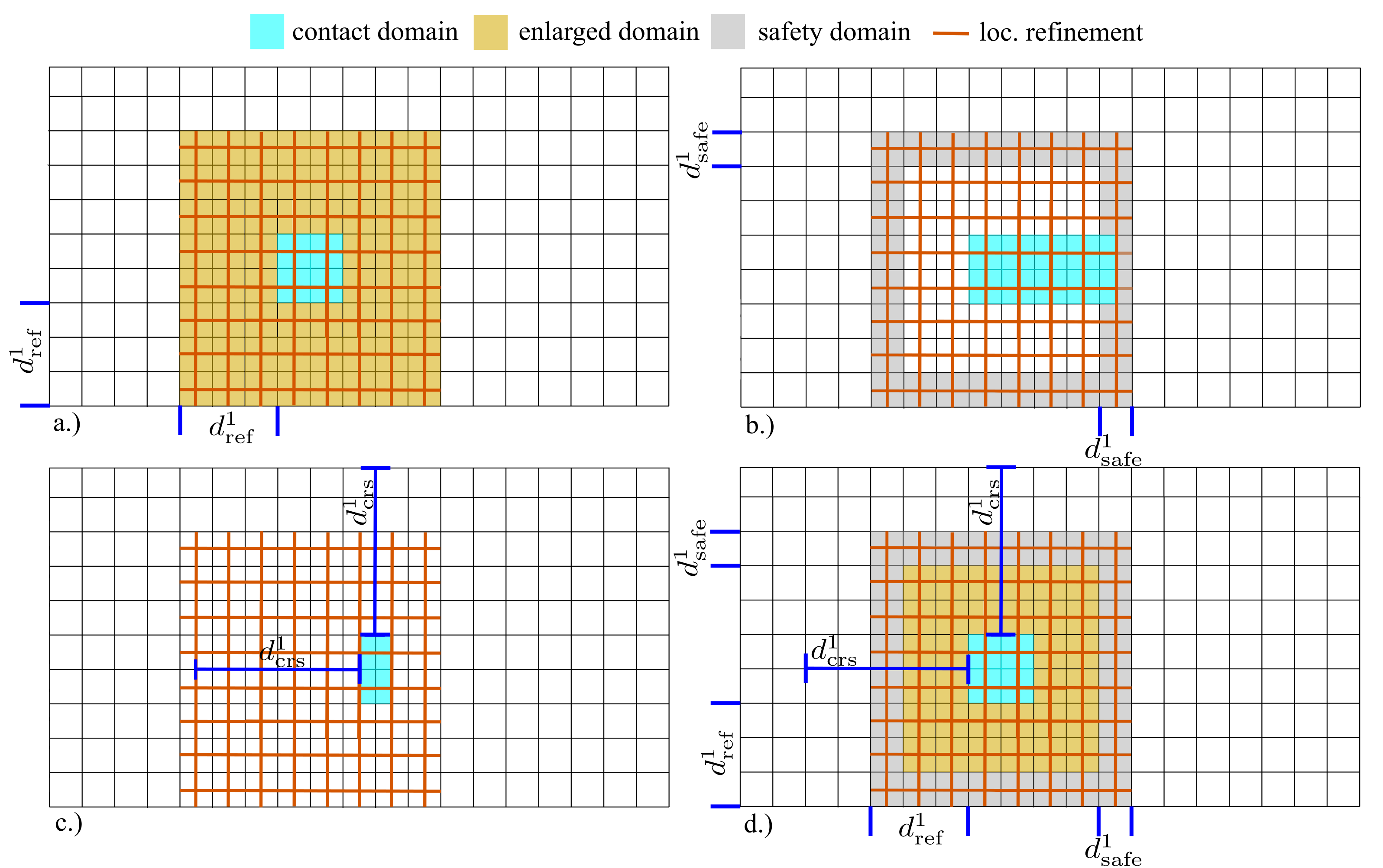}
\caption{Automatic control of the refinement and coarsening procedure based on the parameters $d_{\mathrm{ref}}^d$, $d_{\mathrm{safe}}^d$ and $d_{\mathrm{crs}}^d$. (a.)~Mesh refinement: During a refinement step, mesh refinement is considered within the enlarged contact domain. (b.)~Refinement detection: The need for mesh refinement is detected once the contact domain overlaps with the safety domain. Mesh refinement is then applied by redetermining the domain as shown in~(a.). (c.)~Coarsening detection: The need for mesh coarsening is detected when refined elements are detected beyond the distance $d_\mathrm{crs}^d$ from the contact domain. Mesh coarsening is then applied as described in Sec.~\ref{sec:alr3}. (d.)~Combined setup of the parameters $d_{\mathrm{ref}}^d$, $d_{\mathrm{safe}}^d$ and $d_{\mathrm{crs}}^d$.}
\label{fig:ALR1}
\end{figure}

An adaptive local refinement technique necessitates an indicator for refinement. 
Commonly used indicators use \textit{a posteriori} error estimation that provide a reliable error distribution, see e.g. \cite{Ainsworth1997-1}, \cite{johannessen14} and \cite{kumar15}.
In this work, a refinement indicator based on the contact state is used. If new contact is detected in an element, refinement is considered. If contact is lost in an element, coarsening is considered. After the contact domain is determined, the local refinement is performed by meshline extensions in the parameter domain. 
The refinement is performed until a prescribed refinement \textit{depth} or smallest element size, is obtained. 
For sliding contact it is desirable to automatically control the technique of adaptive local refinement and coarsening. A changing contact domain can lead to a high number of newly detected contacts and lost contacts. To reduce the number of refinement and coarsening events, three parameters are used.
In Fig.~\ref{fig:ALR1} four cases are sketched to illustrate the automatic control by using the parameters $d_{\mathrm{ref}}^d$, $d_{\mathrm{safe}}^d$ and $d_{\mathrm{crs}}^d$. 
The superscript $d$ denotes the current refinement depth that is performed.
\begin{itemize}
\item The parameter $d_{\mathrm{ref}}^d\geq 0$ enlarges the refinement domain by a number of unrefined elements (see Fig.~\ref{fig:ALR1}a). In this example, $d_{\mathrm{ref}}^d=3\,d_\mre^{1-d}$ in both directions. Here, 
\eqb{l}
d_\mre^{1-d} = \ds \frac{d^0_\mre}{2^{1-d}}~,
\eqe
is the minimum element length of the previous refinement depth and $d^0_\mre$ is the original element length.
The elements within the enlarged contact domain are flagged for refinement.
The LR mesh is the result of splitting all flagged elements equally.
\item The parameter $d_{\mathrm{safe}}^d\geq 0$ sets a safety domain at the boundary between coarse and refined elements. If contact is detected within this domain, local refinement is performed. An example with $d_{\mathrm{safe}}^d=d_\mre^{1-d}$ in both directions is shown in Fig.~\ref{fig:ALR1}b. This parameter ensures that at every time-step the contact domain is surrounded by refined elements.
$d_{\mathrm{safe}}^d$ detects a refinement event and does not specify a domain for refinement.
\item The parameter $d_{\mathrm{crs}}^d\geq 0$ controls the coarsening. An example with $d_{\mathrm{crs}}^d=5\,d_\mre^{1-d}$ in both directions is shown in Fig.~\ref{fig:ALR1}c. Coarsening is performed if refined elements are detected beyond this distance from the contact domain. $d_{\mathrm{crs}}^d$ detects a coarsening event and does not specify a domain for coarsening.
\end{itemize}
Fig.~\ref{fig:ALR1}d shows the three parameters combined. The parameters ensure that the contact domain and the neighboring elements are at every time-step represented in the desired discretization. The parameters $d_{\mathrm{ref}}^d$, $d_{\mathrm{safe}}^d$ and $d_{\mathrm{crs}}^d$ can be set arbitrarily and a good choice is depending on the specific problem setup. 
A good compromise of the size of the locally refined domain and the number of refinement/coarsening events is desirable. The coarsening is discussed in the next section.

\FloatBarrier

\subsection{Adaptive coarsening}
\label{sec:alr3}

\begin{figure}[h]
\centering
\includegraphics[width=1\linewidth, trim = 0 0 0 0,clip]{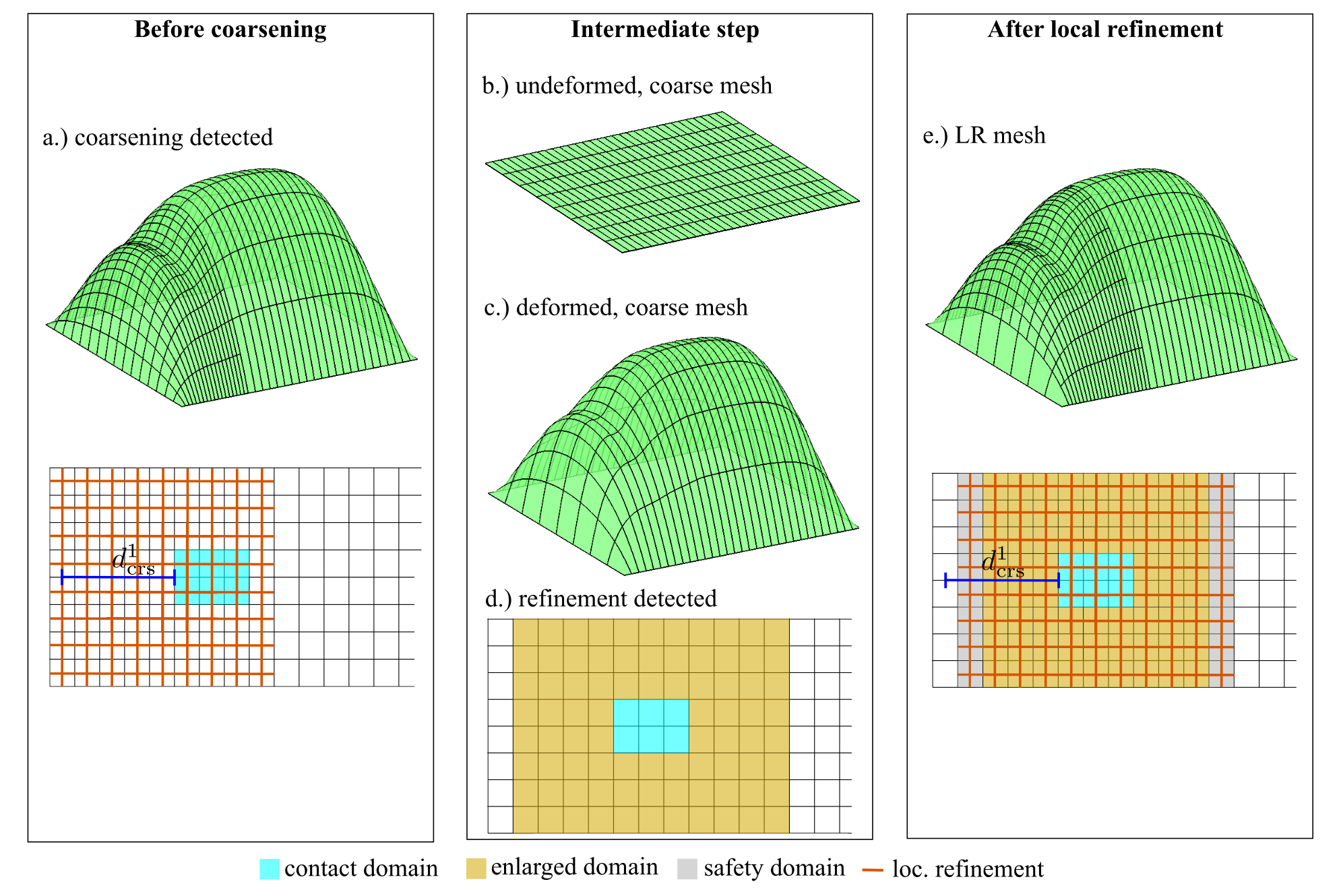}
\caption{Sketch of the proposed coarsening and refinement procedure. (a.)~LR mesh where coarsening is detected (shown are the deformed configuration and the corresponding parametric representation). (b.)~Undeformed, reference mesh in its coarse representation. (c.)~Deformed configuration in its coarse representation. (d.)~Local refinement detected. (e.)~Deformed, locally refined mesh and parametric representation of the mesh.}
\label{fig:ALR2}
\end{figure}

Coarsening LR meshes in the context of contact computations is challenging. The reasons are the unstructured parametric representation of LR NURBS and that frictional contact is history dependent. The contact variables need to be preserved during the process of coarsening and refinement. A technique for NURBS and T-spline coarsening is presented in \cite{Thomas2015-1}. This technique has not been advanced to LR NURBS yet. To overcome this here, the coarsening is combined with local refinement. Therefore, an intermediate step is introduced, in which the entire mesh is coarsened. 
Based on the coarse mesh, the local refinement is then performed to obtain the desired LR mesh. 
The coarsening and refinement process requires that the contact domain has the same discretization and identical contact variables before and after the process.
Due to the intermediate step, the contact domain is temporary coarsened. To preserve the contact variables a mapping is required.
The contact variables are stored before the mesh is coarsened, and then mapped to the new mesh after the refinement step. 
The technique consists of six main steps to perform a consistent adaptive coarsening within the context of contact computations. The technique is sketched in Fig.~\ref{fig:ALR2} and the steps are:
\begin{enumerate}
\item Coarsening is detected when refined elements are detected beyond the distance $d_\mathrm{crs}^d$ (Fig.~\ref{fig:ALR2}a). Store the current configuration and the contact variables for each element in contact. 
\item Recover the undeformed, reference mesh (Fig.~\ref{fig:ALR2}b). 
\item Interpolate the control points of the mesh in Fig.~\ref{fig:ALR2}a to obtain the deformed, coarse mesh (Fig.~\ref{fig:ALR2}c). The interpolation leads to a geometric approximation error, which is addressed in step~6.
\item Perform local refinement within the enlarged contact domain $d_\mathrm{ref}^d$ (Fig.~\ref{fig:ALR2}d) to obtain the desired LR mesh (Fig.~\ref{fig:ALR2}e).
\item Preserve the contact variables by mapping them to the current configuration.
\item Reduce the geometric approximation error by recovering control points from the configuration in Fig.~\ref{fig:ALR2}a: By comparing the meshes in Fig.~\ref{fig:ALR2}a and Fig.~\ref{fig:ALR2}e it turns out that their discretization only differ slightly. The control points of the domains, that remain unchanged in their discretization, are recovered.
\end{enumerate}
As Fig.~\ref{fig:ALR2} shows, the discretization of the contact domain, at a given time-step, is unaffected by the coarsening and refinement procedure. 
With the adaptive local refinement and coarsening technique, a refinement/coarsening event is not detected at every computational step. A refinement/coarsening event is performed only when it is detected according to the criteria of Fig.~\ref{fig:ALR1}. The results of this technique can be seen in the following examples, e.g. Fig.~\ref{fig:RSS_2}.

\FloatBarrier

\section{Numerical examples}
\label{sec:num_res}
\pagestyle{plain}

The performance of the adaptive local refinement and coarsening technique using LR NURBS elements is illustrated in this section by several numerical examples. The first example is used for validation. In the remaining examples, the performance of LR NURBS for frictionless and frictional contact is investigated. In the last example, LR NURBS-enriched contact elements are considered for 3D friction of two deformable solids. To investigate the benefit of LR NURBS discretization, comparisons to reference models with uniform NURBS discretization are made.

\subsection{Inflation of a hemisphere}
\label{sec:infl}

\begin{figure}[h]
\centering
\includegraphics[width=0.49\linewidth, trim = 180 10 365 80,clip]{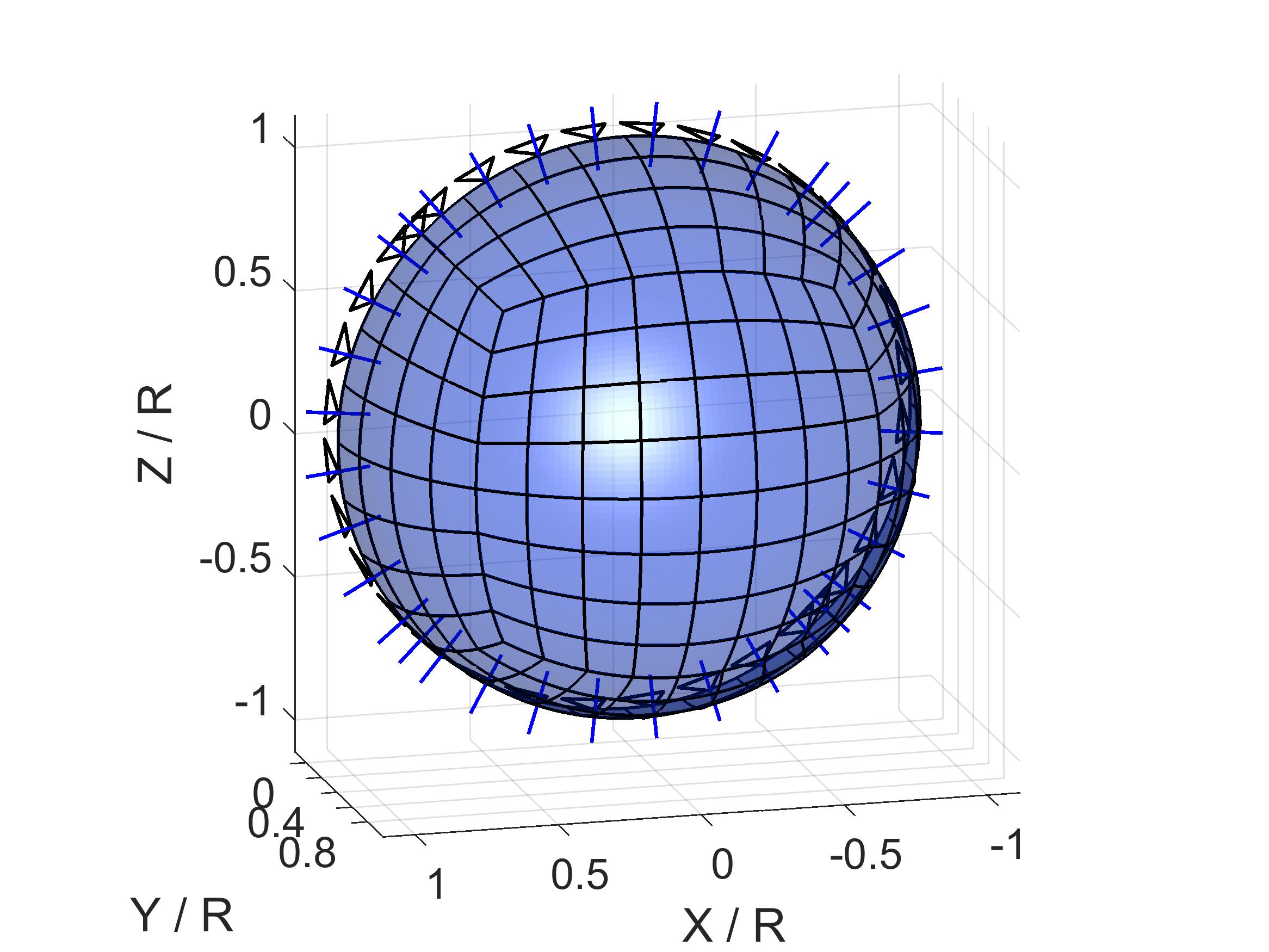}
\includegraphics[width=0.49\linewidth, trim = 180 10 365 80,clip]{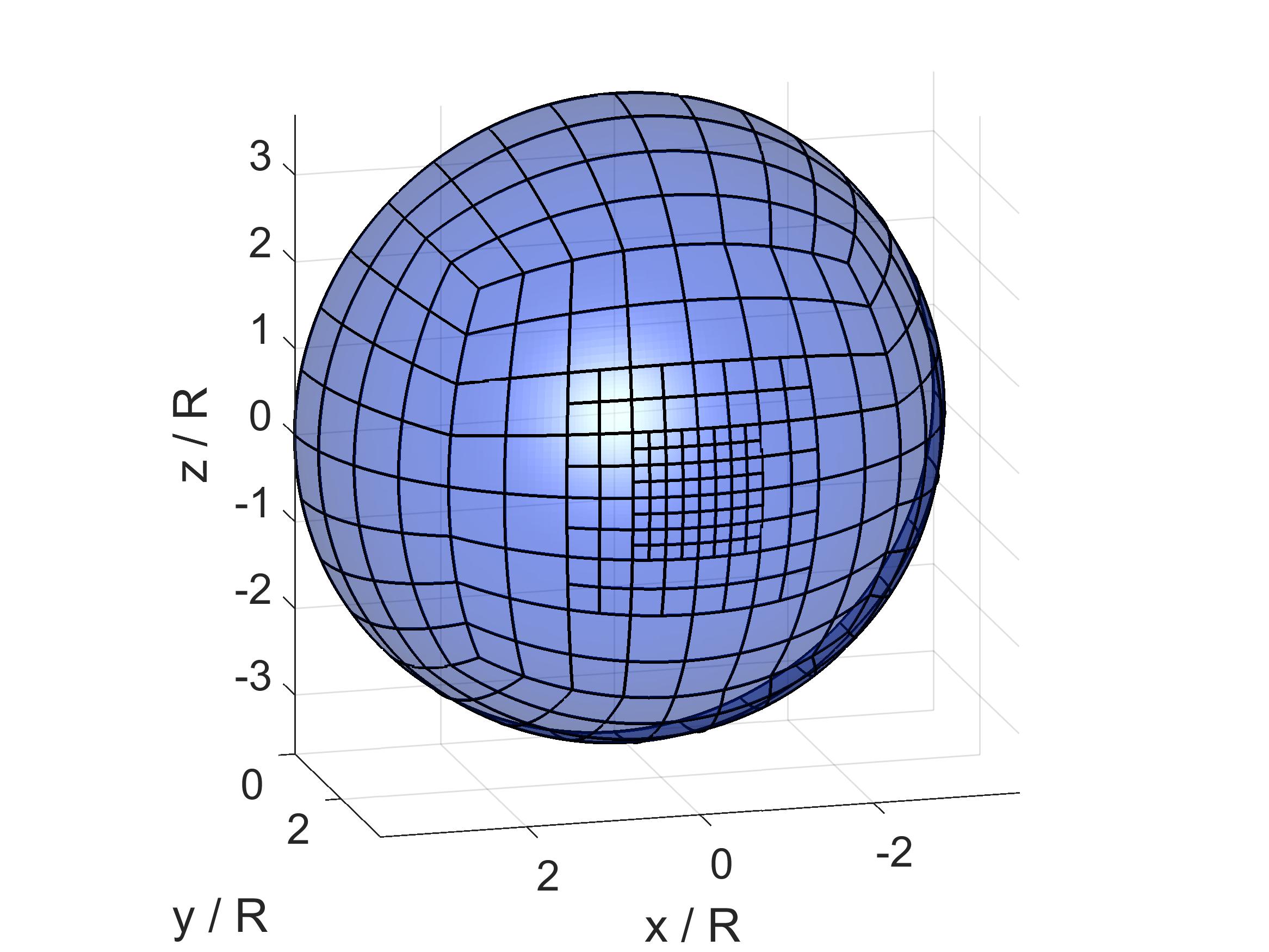}
\caption{Inflation of a hemispherical membrane: Model setup in the initial configuration~(left). Blue lines denote the allowed direction of deformation. Black supports denote the fixed directions. Current configuration at $V=50\,V_0$, two times locally refined~(right).}
\label{fig:inflate_1}
\end{figure}

\begin{figure}[h]
\centering
	\includegraphics[width=0.5\linewidth]{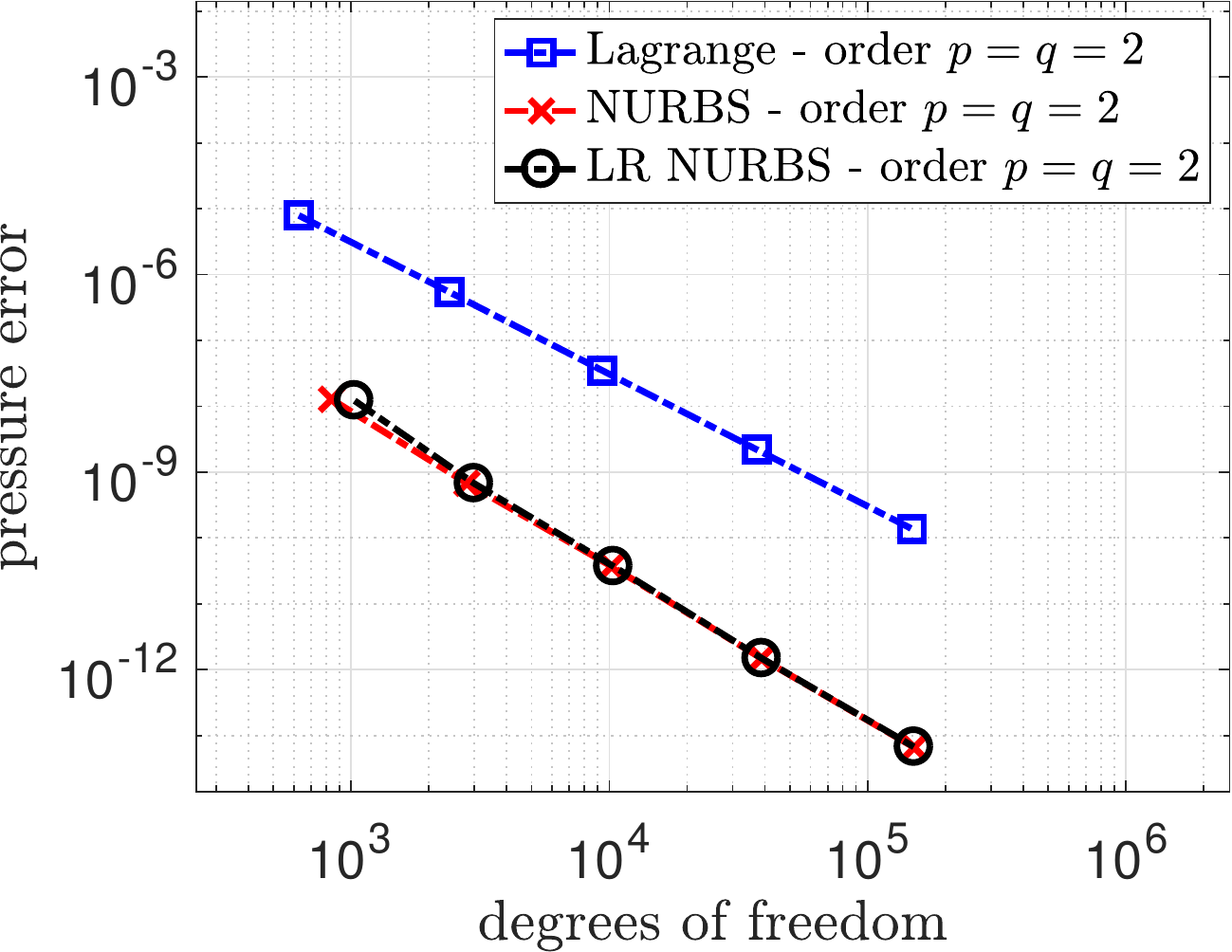}
	\caption{Inflation of a hemispherical membrane: Pressure error for meshes discretized by quadratic Lagrange, quadratic NURBS and quadratic LR NURBS elements in comparison to the analytical solution. There is no benefit in local refinement for a problem with a uniform solution and hence LR NURBS perform equally well than uniform NURBS.}
	\label{fig:inflate_2}
\end{figure}

The first example considers the inflation of a spherical membrane balloon. The inflation can be described by an analytical formula, which is used for validation. The system is modeled as a hemisphere and the boundary conditions are chosen such that the symmetry of the model is maintained during inflation, see the left side of Fig.~\ref{fig:inflate_1}. The discretized geometry consists of five patches and the NURBS weights differ for each. This chosen geometry is not perfectly spherical but still an approximation. The material is described by the incompressible Neo-Hookean membrane material model \citep{membrane}
\begin{equation}
\sig^{\alpha\beta}=\frac{\mu}{J}\left( A^{\alpha\beta} - \frac{a^{\alpha\beta}}{J^2} \right)~.
\end{equation}
Here, $\mu$ is the shear modulus, $J$ is the surface area change and $A^{\alpha\beta}$ and $a^{\alpha\beta}$ are the contra-variant components of the metric tensor in the reference and current configuration. 
The problem setup is similar to the one of \citet{membrane} in which the balloon is modeled as  $1/8$th of a sphere with NURBS and Lagrange discretizations. Initially, the balloon has the volume $V_0 = 4\pi R^3 /3$ with the radius $R$. The volume is increased step-wise until it reaches $V=50\,V_0$. The analytical pressure-volume relation is given by
\begin{equation}
\frac{p_\mathrm{int} R}{\mu} = 2\left(\left(\frac{V_0}{V}\right)^{\frac{1}{3}}-\left(\frac{V_0}{V}\right)^{\frac{7}{3}}\right)~,
\end{equation}
with the internal pressure $p_{\mathrm{int}}$.
We now investigate the behavior of the proposed local refinement procedure using quadratic LR NURBS elements. Five different uniform meshes with quadratic NURBS discretizations are taken each as the starting point. The number of Gaussian quadrature points is $n_{qp}=3\times 3$ for all elements. As an homogeneous hemisphere has no domain of major interest, the center is chosen to apply local refinement of depth 2, see the right side of Fig.~\ref{fig:inflate_1}. The performance of the five meshes is compared to quadratic NURBS and quadratic Lagrange discretizations examining the pressure error, see Fig.~\ref{fig:inflate_2}. The isogeometric elements behave as observed in \citet{membrane} and show better convergence behavior than standard Lagrange elements. The LR NURBS meshes behave equal well than the uniform NURBS meshes. Better behavior can not be expected in this example, since the hemisphere has a homogeneous behavior and local refinement does not improve the numerical results significantly. But with this first example we can conclude that the applied local refinement procedure works successful and the computations with LR NURBS are valid. In the following examples, the performance of adaptive local refinement with respect to the accuracy and the computational cost is investigated.
\FloatBarrier

\subsection{Rigid sphere in contact with a square membrane sheet}

\begin{figure}[h]
\centering
{\includegraphics[width=0.99\linewidth, height=0.4\linewidth, trim = 150 20 150 260,clip]{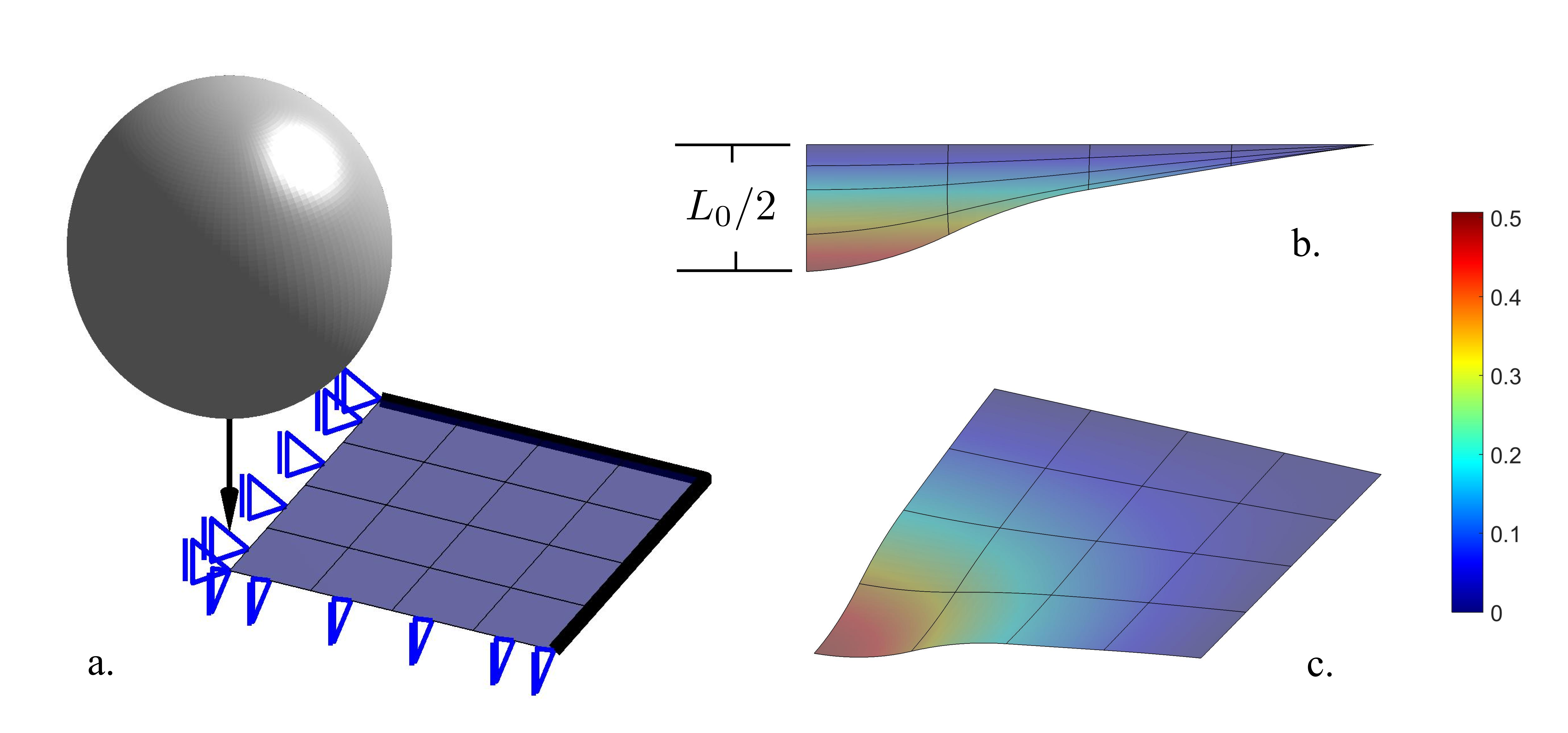}}
\caption{Rigid sphere in contact with a square membrane sheet: (a)~Initial membrane discretization, boundary conditions and rigid sphere. (b),~(c)~Deformed membrane surface from two different points of view. The coloring shows the downward displacement of the membrane.}
\label{fig:RSM_1}
\end{figure}

\begin{figure}[h]
\centering
\includegraphics[width=0.19\linewidth, trim = 93 150 65 60,clip]{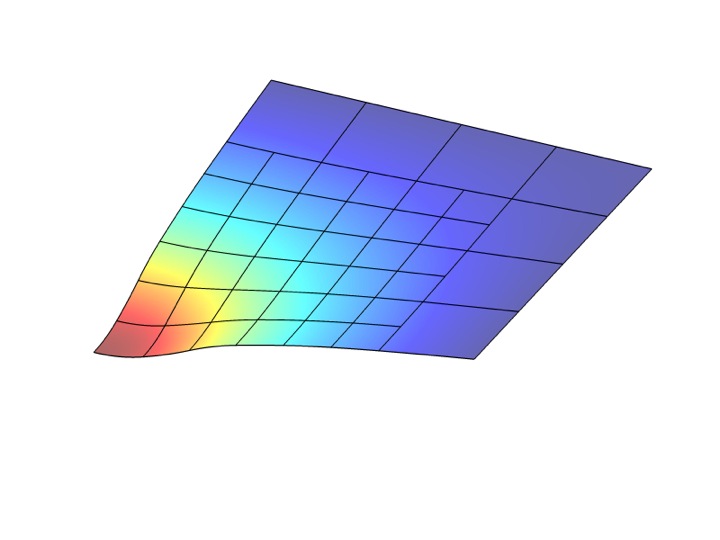}
\includegraphics[width=0.19\linewidth, trim = 93 150 65 60,clip]{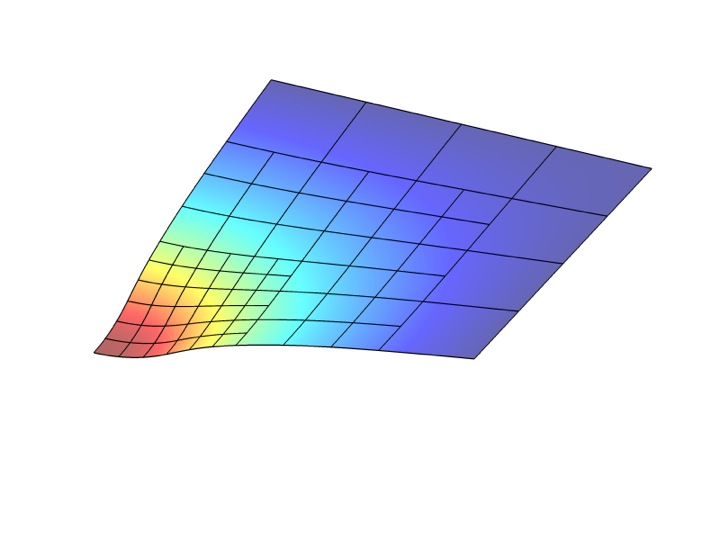}
\includegraphics[width=0.19\linewidth, trim = 93 150 65 60,clip]{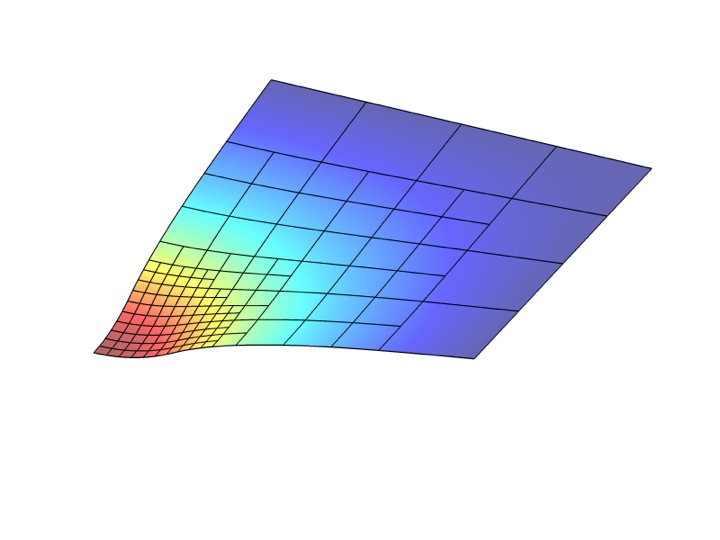}
\includegraphics[width=0.19\linewidth, trim = 93 150 65 60,clip]{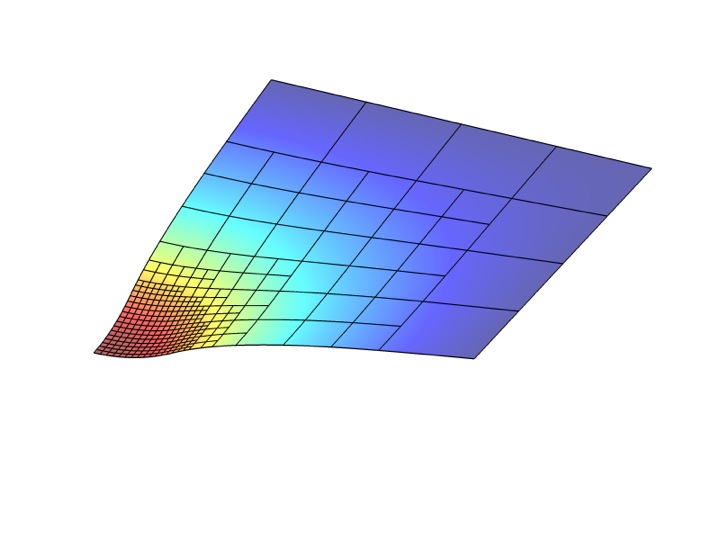}
\includegraphics[width=0.19\linewidth, trim = 93 150 65 60,clip]{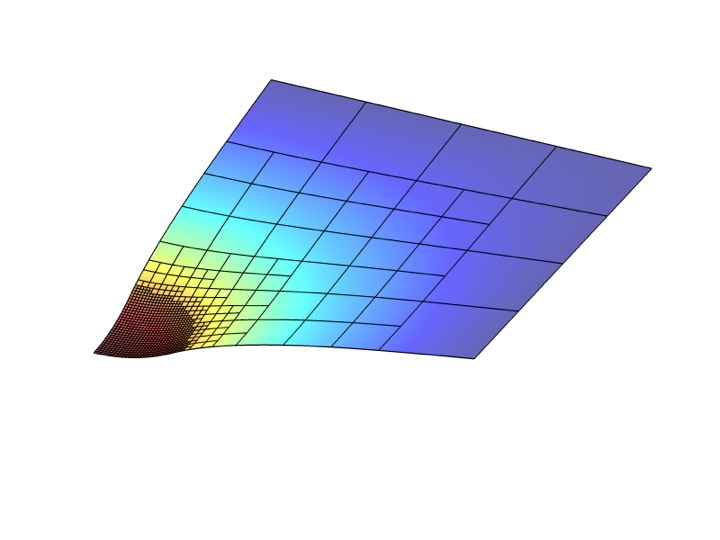}
\caption{Rigid sphere in contact with a square membrane sheet: Deformation series with a locally refined contact domain. A refinement depth from 1 to 5 is applied. The coloring shows the displacement of the membrane.}
\label{fig:RSM_2}
\end{figure}

\begin{figure}[h]
{\includegraphics[width=0.5\linewidth]{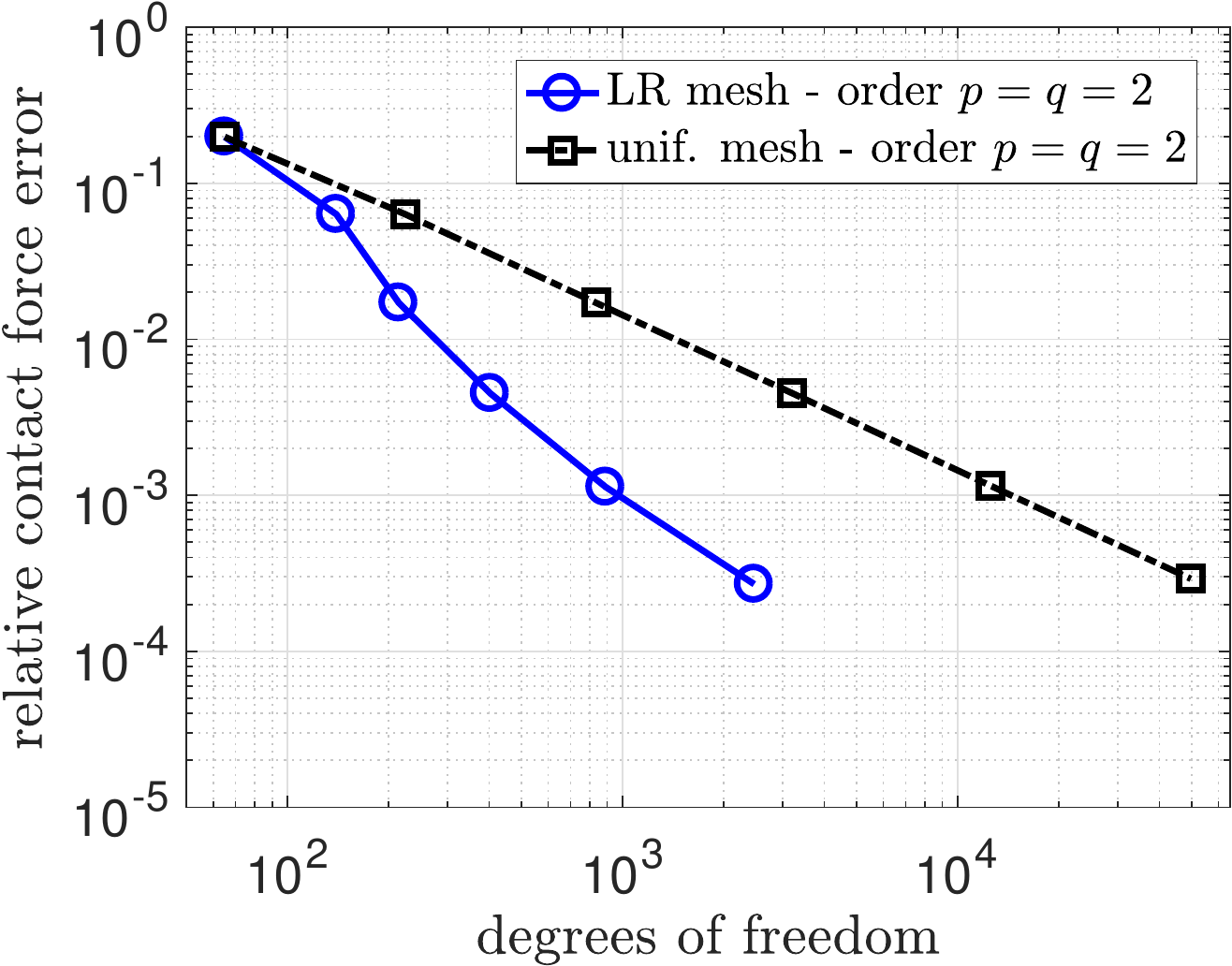}}
{\includegraphics[width=0.5\linewidth]{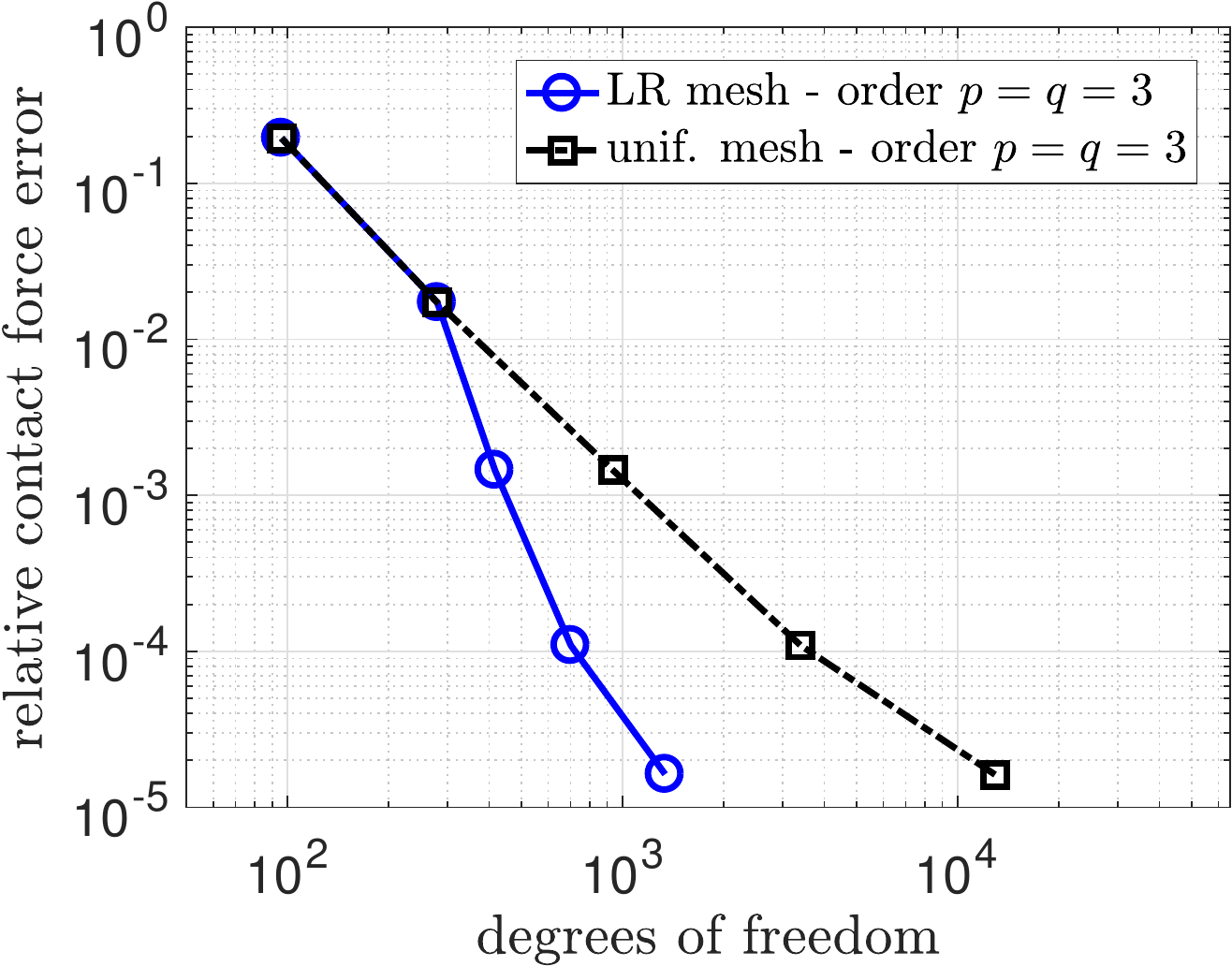}}
\caption{Rigid sphere in contact with a square membrane sheet: Relative error of the normal contact force for quadratic discretizations~(left) and cubic discretizations~(right). The error is defined w.r.t.~a uniformly discretized mesh with cubic NURBS elements that is taken as the reference.}
\label{fig:RSM_3}
\end{figure}

The second example considers a square membrane sheet with dimension $2\,L_0 \times 2\,L_0$ that is initially pre-stretched by $\lambda = 1.1$. The pre-stretching is applied to avoid the membrane instability. A rigid sphere, located at $\bX=[0,0,L_0]$ with radius $R = L_0$, is pushed gradually downwards until the bottom of the sphere reaches $z=-L_0/2$ as shown in Fig.~\ref{fig:RSM_1}. The initial mesh consists of $4\times 4$ LR NURBS elements with $5\times 5$ Gaussian quadrature points each. The material behavior is the same as in the previous example. The boundaries at $X = 2\,L_0$ and $Y = 2\,L_0$ are fixed in all directions. At $X = 0$ and $Y = 0$ the deformation is constrained to be zero in $x$- and $y$-direction, respectively. Contact between the rigid sphere and the membrane is treated by the penalty method with the penalty parameter
\begin{equation}
\label{eq:pen_elem_2}
\varepsilon_\mathrm{n}^{el}=\varepsilon_\mathrm{n}^0\cdot \left( \frac{l_{0_x}\,l_{0_y}}{l_{el_x}\,l_{el_y}}\right)^{p-1}~,
\end{equation}
depending on the individual element size. Here, $l_{el_x}$ and $l_{el_y}$ denote the current element lengths, and $l_{0_x}$ and $l_{0_y}$ the initial element lengths. $p$ is the polynomial order\footnote{the order in both parametric directions is set equally, i.e. $q=p$} and $\varepsilon_\mathrm{n}^0 = 10\,E_0/L_0$ is the constant penalty parameter. A mesh and order depending penalty parameter was studied in \citet{composite} and \citet{Sauer2014-2} giving a good balance between accuracy and convergence behavior. 
The contact domain is locally refined by the procedure described in Sec.~\ref{sec:ALR}. A series of resulting meshes for quadratic LR NURBS elements is illustrated in Fig.~\ref{fig:RSM_2} from refinement depth 1 to 5. A strong local aggregation of elements is obtained in the contact domain. In this example, the parameter $d_\mathrm{ref}^d$ is taken as zero.
The performance of LR NURBS meshes and uniform meshes using NURBS discretizations is investigated by examining the relative error
\begin{equation}
e_\mathrm{n}^\mathrm{rel}= \frac{\mid f_\mathrm{n}^\mathrm{ref}-f_\mathrm{n} \mid}{\mid f_\mathrm{n}^\mathrm{ref}\mid}~,
\label{eq:rel_err}
\end{equation}
of the normal contact force $f_\mathrm{n}$ w.r.t.~a reference solution. For the reference solution, a uniform mesh with cubic NURBS discretizations is taken. This mesh has more than $5 \cdot 10^4$ dofs and $\varepsilon_\mathrm{n}^{el} > 10^6\,E_0/L_0$, which is highly accurate. The relative contact force error of LR NURBS and homogeneous meshes is shown in Fig.~\ref{fig:RSM_3}. The proposed local refinement technique using LR NURBS elements show very good convergence behavior, both for quadratic and cubic elements. 
It turns out that for cubic elements the obtained relative error decreases to $\approx 1.63\, \cdot 10^{-5}$, while for quadratic elements $\approx 2.1\, \cdot 10^{-4}$ is achieved. Since the penalty parameter is increased by Eq.~\eqref{eq:pen_elem_2} and since it is depending on the polynomial order, higher accuracy for cubic elements can be expected. The dofs are reduced by a factor of more than $16$ for LR NURBS meshes with quadratic elements at the $5^{\mathrm{th}}$ refinement depth. For LR NURBS meshes with cubic elements at the $4^{\mathrm{th}}$ refinement depth the dofs are reduced by a factor of more than $37$. This is a huge decrease of the computational cost. In Tab.~\ref{tab:RSM_1} the mesh properties and numerical results for cubic elements are listed. This example shows that with the local refinement procedure, a high accuracy is achieved while decreasing the computational cost. 
\begin{table}[h]
\begin{center}
\begin{tabular}{| p{2.6cm} | r | r | r |}
    \hline
    LR mesh & dofs & $\sharp$ el. & rel. error \\ \hline
    depth 0 & $96$ & $16$ & $1.95\cdot10^{-1}$  \\ \hline
    depth 1 & $280$ & $64$ & $1.72\cdot10^{-2}$ \\ \hline
    depth 2 & $416$ & $112$ & $1.45\cdot10^{-3}$ \\ \hline
    depth 3 & $701$ & $211$ & $1.09\cdot10^{-4}$ \\
    \hline
    depth 4 & $1340$ & $430$ & $1.63\cdot10^{-5}$ \\
    \hline
\end{tabular}
\hspace{4mm}
\begin{tabular}{| p{2.6cm} | r | r | r |}
    \hline
    uniform mesh & dofs & $\sharp$ el. & rel. error \\ \hline
    depth 0  & $96$ & $16$ & $1.95\cdot10^{-1}$  \\ \hline
    depth 1 & $280$ & $64$ & $1.72\cdot10^{-2}$ \\ \hline
    depth 2 & $936$ & $256$ & $1.45\cdot10^{-3}$ \\ \hline
    depth 3 & $3400$ & $1024$ & $1.09\cdot10^{-4}$ \\
    \hline
    depth 4 & $12936$ & $4096$ & $1.62\cdot10^{-5}$ \\
    \hline
\end{tabular}
\end{center}
\caption{Rigid sphere in contact with a square membrane sheet: Comparison of the mesh parameters and the computational error between the LR mesh and the uniformly refined mesh. Cubic discretization is considered. Corresponding refinement depths have almost equal error but hugely different number of dofs.}
\label{tab:RSM_1}
\end{table}

\newpage

\begin{figure}[h]
\centering
	\includegraphics[width=0.5\linewidth]{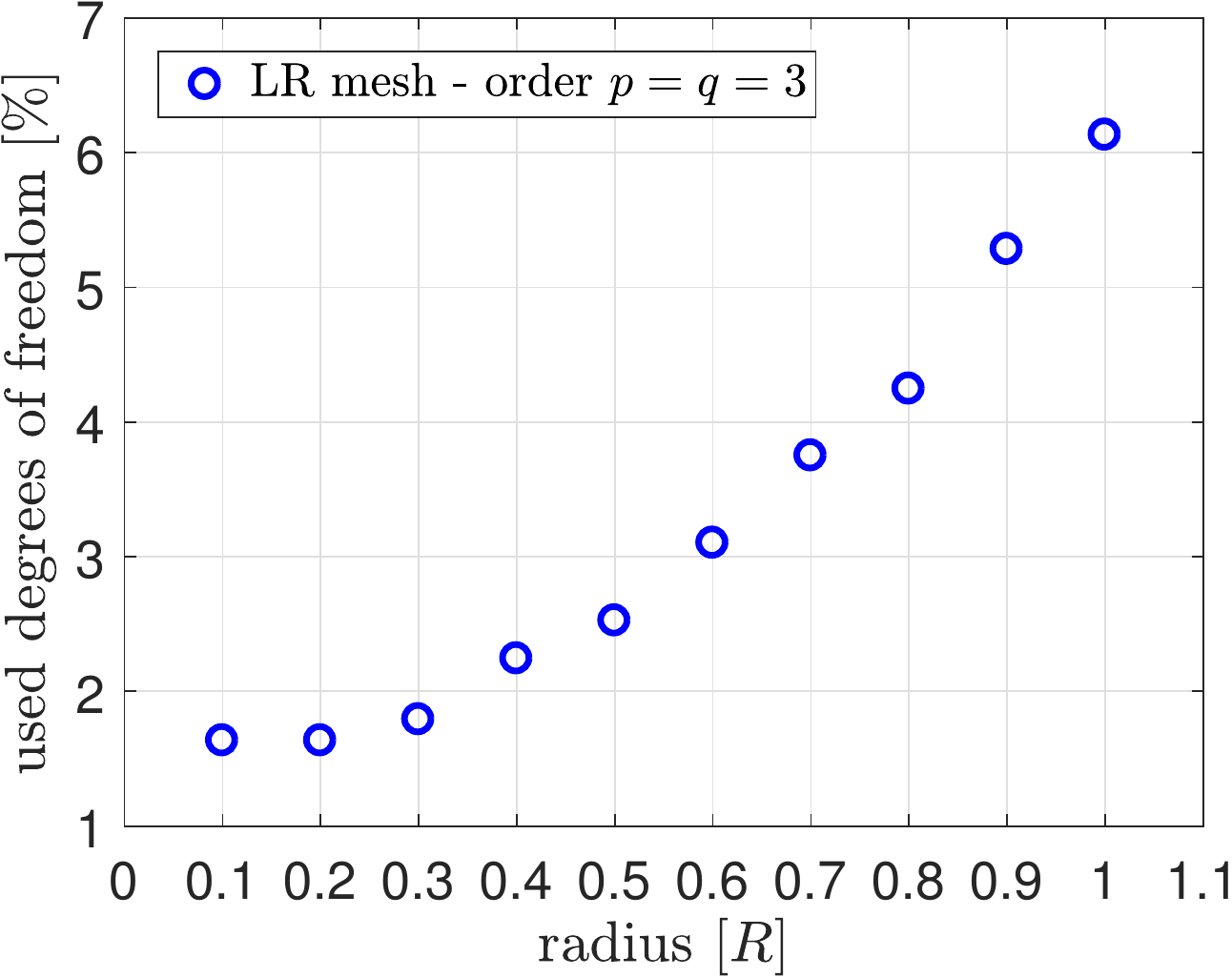}
	\caption{Rigid sphere in contact with a square membrane sheet: Percentage of the used dofs with respect to the radius $R$. The $y$-axis shows the ratio of the used dofs with respect to the reference model.}
	\label{fig:RSM_rad} 
\end{figure}

In a second part of this example, the performance of LR NURBS elements for different sizes of the contact domain is investigated. Computations for different radii of the rigid sphere are performed. The initial radius is set to $R=1/10\,L_0$ and is increased step-wise to $R=L_0$. The sphere is positioned at $X=Y=0$ and lowered gradually to $z=-R/2$. The performance of LR NURBS meshes using cubic elements is investigated. Local refinement of depth 5 is applied. The relative error of the normal contact force is defined w.r.t.~uniformly discretized meshes using cubic NURBS elements. The smallest element size of each mesh is identical, leading to the same value for $\varepsilon_\mathrm{n}^{el}$. The percentage of the used dofs with respect to the radius $R$ is shown in Fig.~\ref{fig:RSM_rad}. The relative contact force error for all radii $R$ is in the range of $5\cdot 10^{-8}$ and $3\cdot 10^{-7}$. The LR NURBS meshes capture the reference solutions nicely. A high accuracy is achieved while only using few dofs. Consequently, for $R=1/10L_0$ less than $2\%$ of the time is used in comparison to the reference model. The time for the local refinement procedure is negligible. This reduction of the computational effort that still achieves high accuracy demonstrates the benefit of the local refinement technique using LR NURBS.

\FloatBarrier

\subsection{Frictionless sliding contact}
\label{Sec:RSS}

\begin{figure}[h]
\centering
\includegraphics[width=0.49\linewidth, trim = 230 560 200 480,clip]{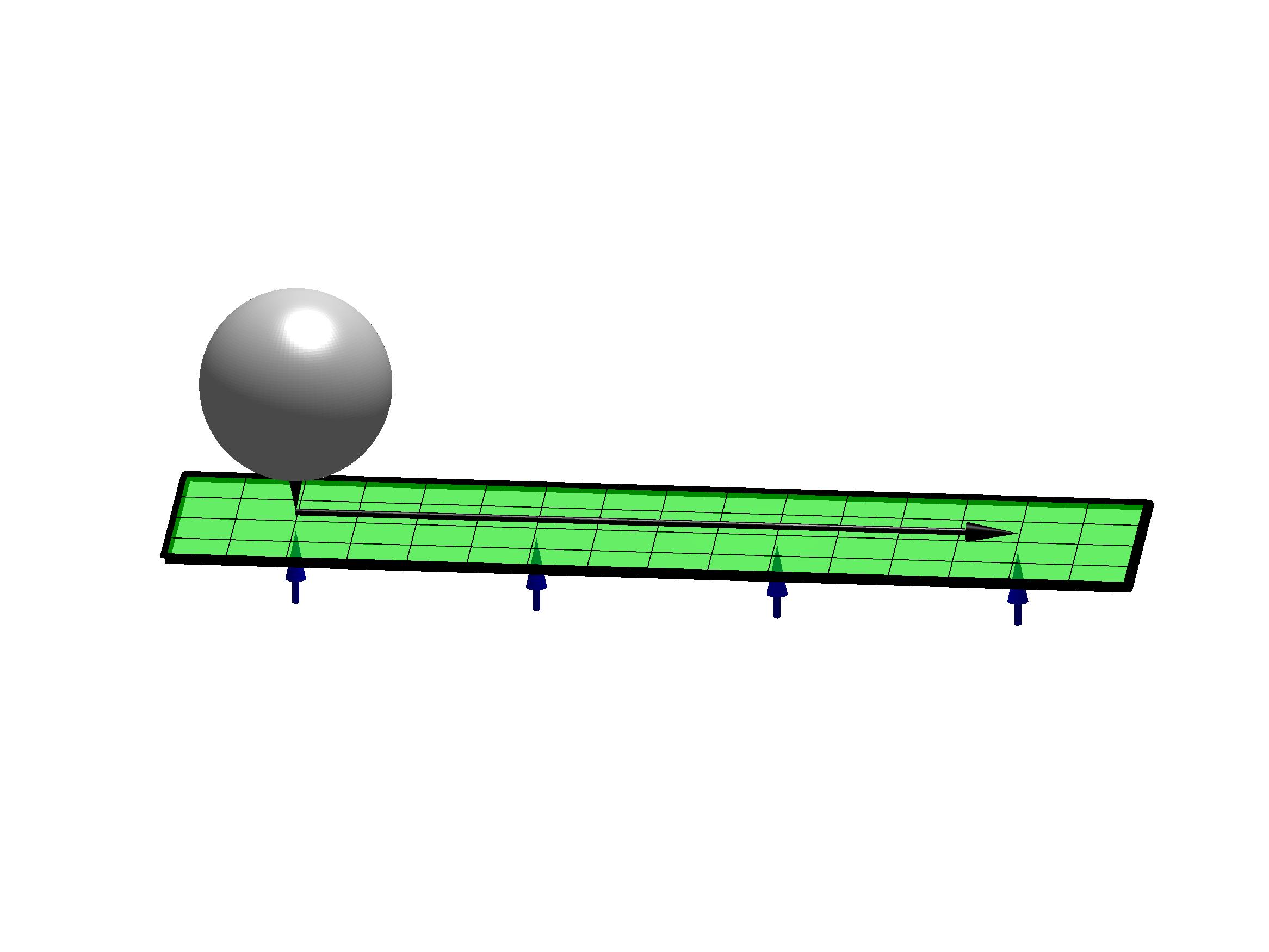}
\includegraphics[width=0.49\linewidth, trim = 230 620 220 530,clip]{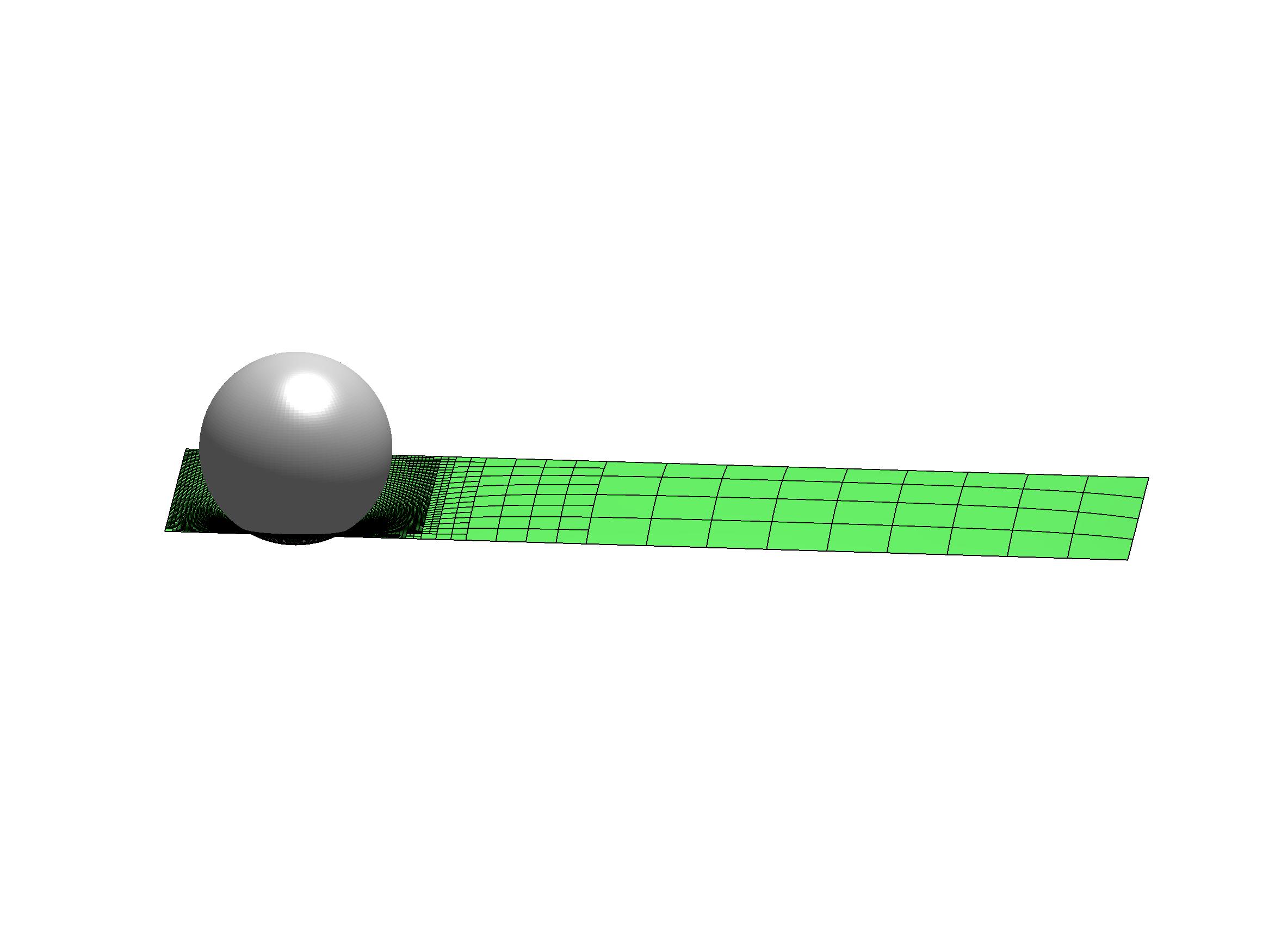}
\caption{Frictionless sliding contact: Initial problem setup~(left). The membrane is initially discretized by $4\times 16$ elements. The LR NURBS mesh, after the sphere is moved downward, is shown on the right figure. Local refinement of depth 4 is applied leading to a highly resolved mesh in the contact domain.}
\label{fig:RSS_1}
\end{figure}

In the third example the performance of the adaptive local refinement and coarsening technique from Sec.~\ref{sec:ALR} is investigated. For this, frictionless sliding contact is considered. The problem setup consists of a rectangular membrane sheet having dimension $8\,\lambda\,L_0 \times 2\,\lambda\,L_0$ with isotropic pre-stretch $\lambda = 1.25$. The boundaries are clamped. A rigid sphere with radius $L_0$ is initially located at $\boldsymbol{X}_0=\lambda\cdot[L_0,L_0,L_0]$ and first moved downwards to $\boldsymbol{x}_0=\lambda\cdot[L_0,L_0,L_0/2]$. This is followed by horizontal motion until the sphere reaches $\boldsymbol{x}_0=\lambda\cdot[7\,L_0,L_0,L_0/2]$, see left side of Fig.~\ref{fig:RSS_1}. The volume enclosed by the membrane is constrained to be constant so that it behaves like a cushion \citep{membrane}. The material and the computational parameters are the same as in the previous example. During frictionless sliding the contact domain changes, and the adaptive local refinement and coarsening procedure from Sec.~\ref{sec:ALR} is applied to obtain highly resolved meshes in the contact domain.

\begin{figure}[h]
\centering
\includegraphics[width=0.49\linewidth, trim = 240 600 210 600,clip]{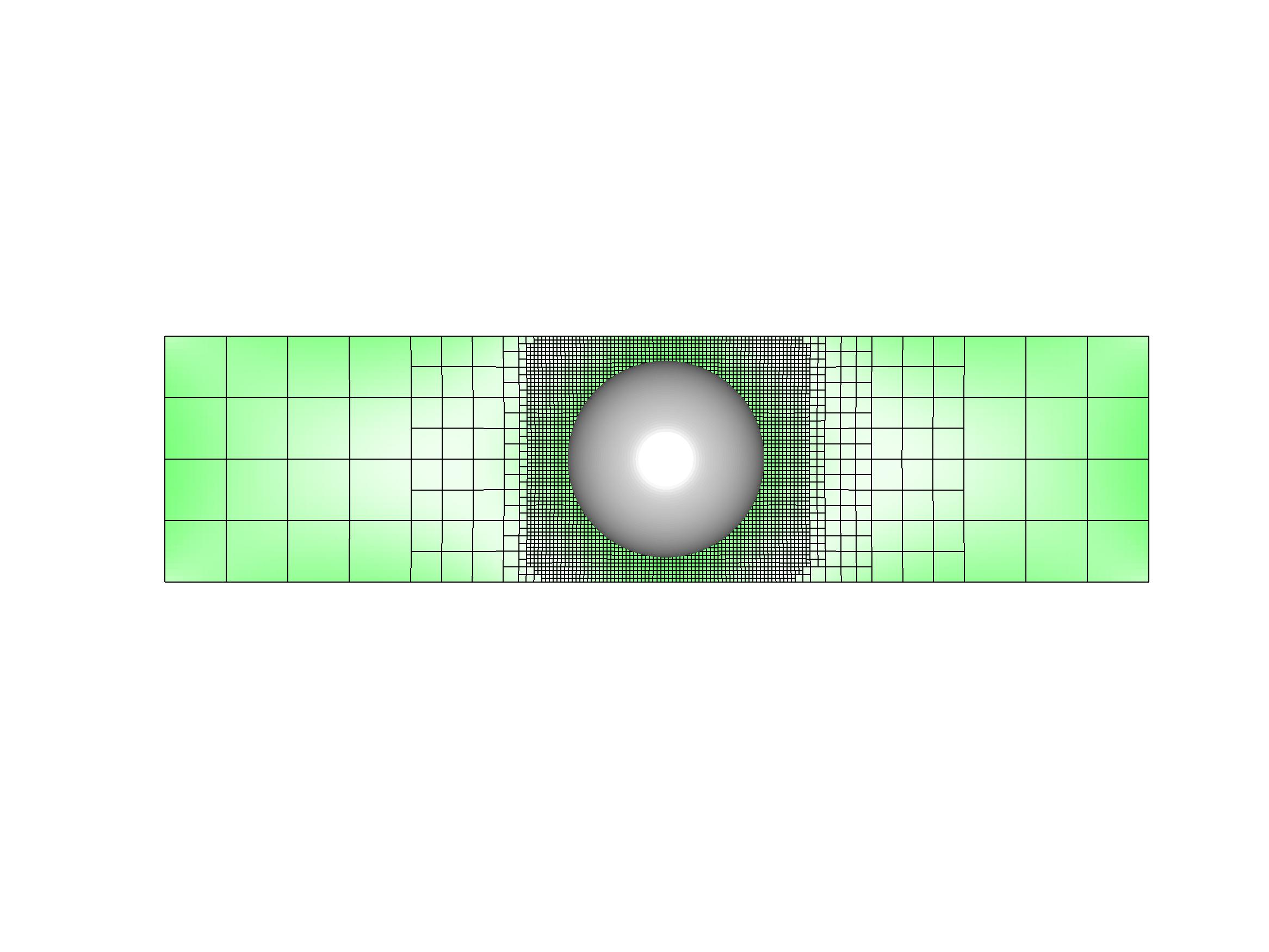}
\includegraphics[width=0.49\linewidth, trim = 240 600 210 600,clip]{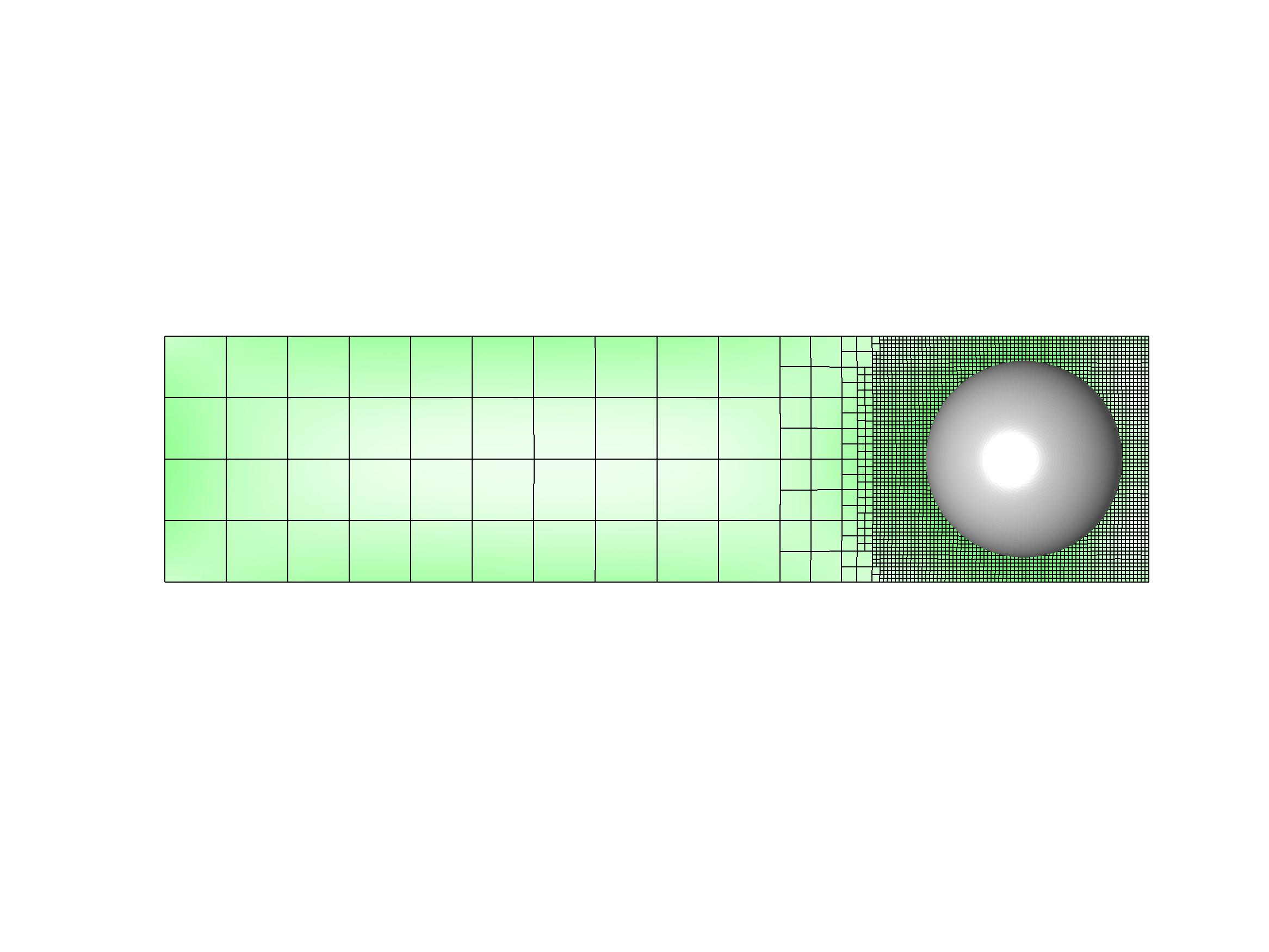}\\
\includegraphics[width=0.49\linewidth, trim = 240 720 210 650,clip]{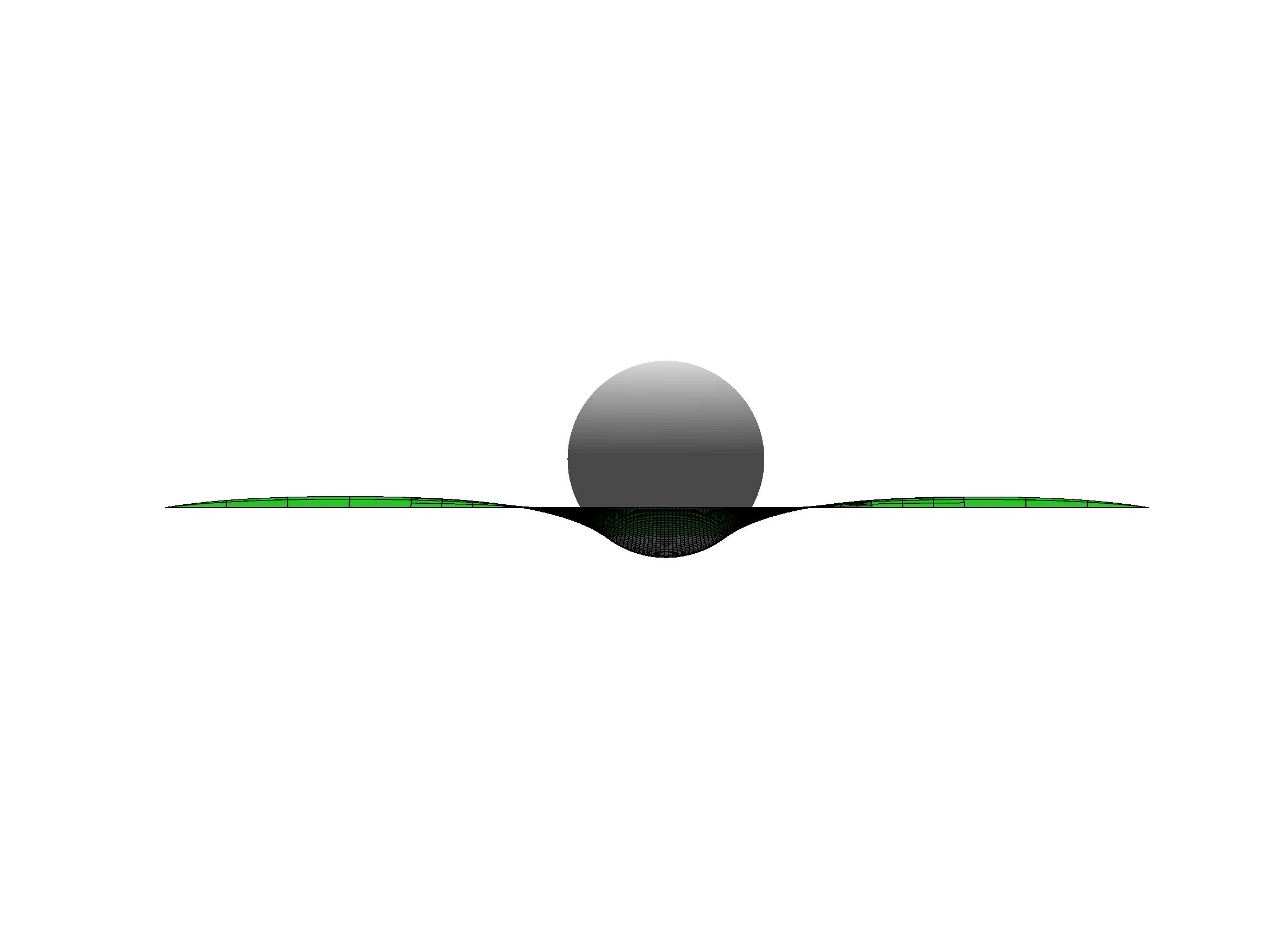}
\includegraphics[width=0.49\linewidth, trim = 240 720 210 650,clip]{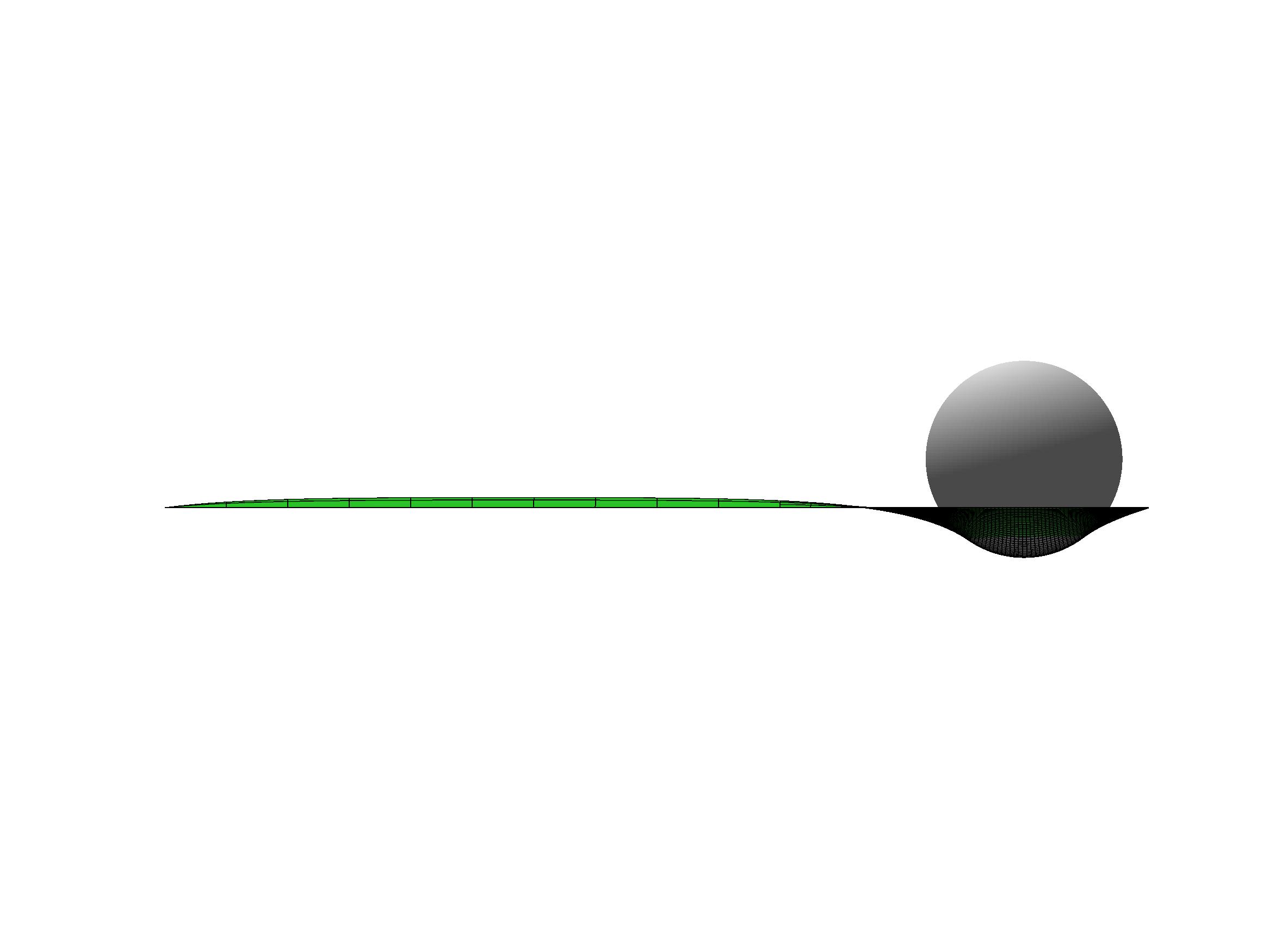}
\caption{Frictionless sliding contact: Adaptive local refinement and coarsening of the membrane during frictionless sliding in top and side view. Highly resolved meshes within the local contact domain are obtained while the periphery is still represented with the coarse, initial mesh.}
\label{fig:RSS_2}
\end{figure}

The right side of Fig.~\ref{fig:RSS_1} shows the deformed configuration after the downward displacement of the rigid sphere has been applied. 
The parameters for the automatic control from Sec.~\ref{sec:ALR} are $d_{\mathrm{ref}}^d=3\,d_\mre^{1-d}$, $d_{\mathrm{safe}}^d=2\,d_\mre^{1-d}$ and $d_{\mathrm{crs}}^d=4\,d_\mre^{0}$.
In Fig.~\ref{fig:RSS_2} two meshes resulting from the adaptive local refinement and coarsening technique are illustrated. It can be observed that during sliding only the contact domain has a high aggregation of LR NURBS elements.

\begin{figure}
\centering
\includegraphics[width=0.49\linewidth]{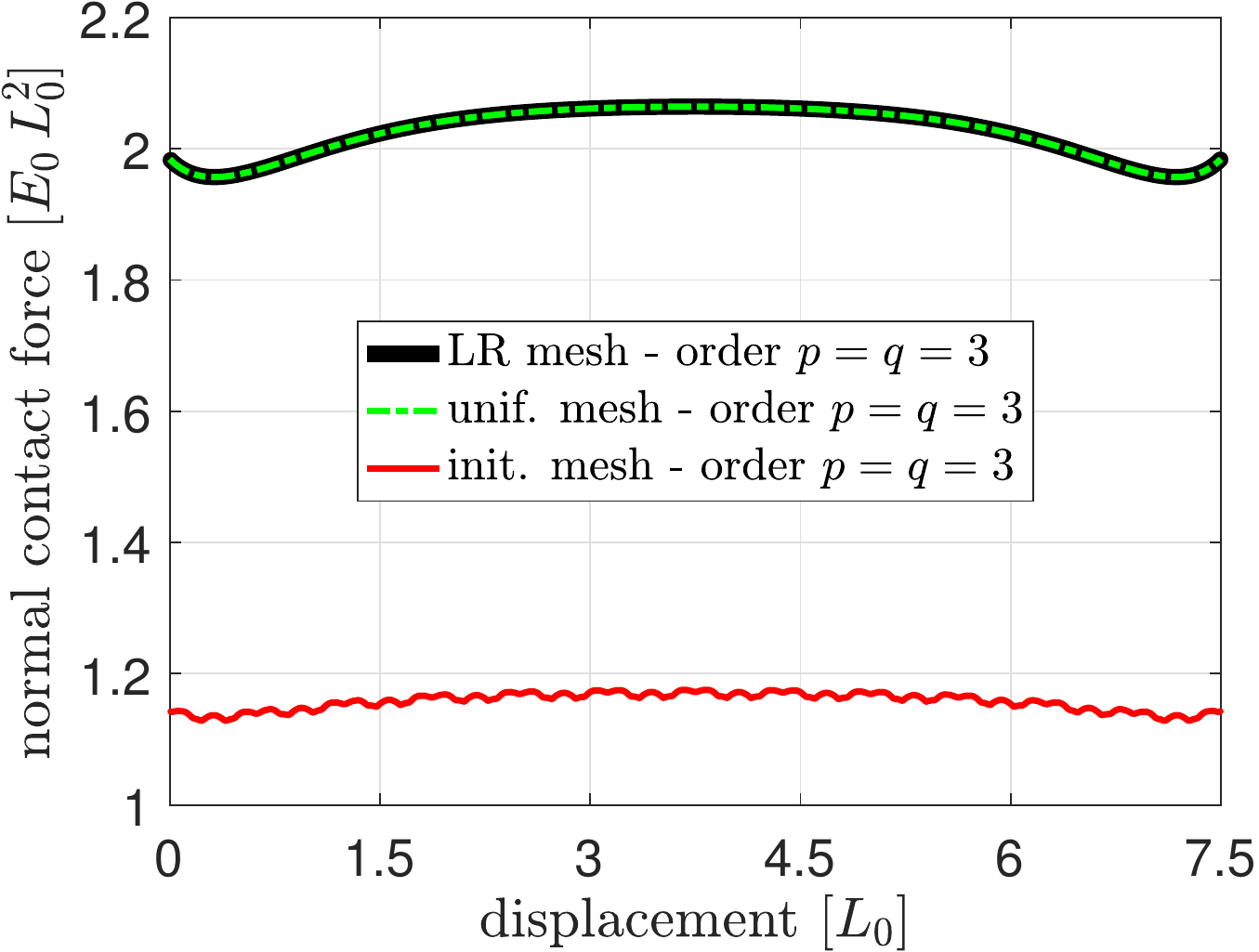}
\includegraphics[width=0.49\linewidth]{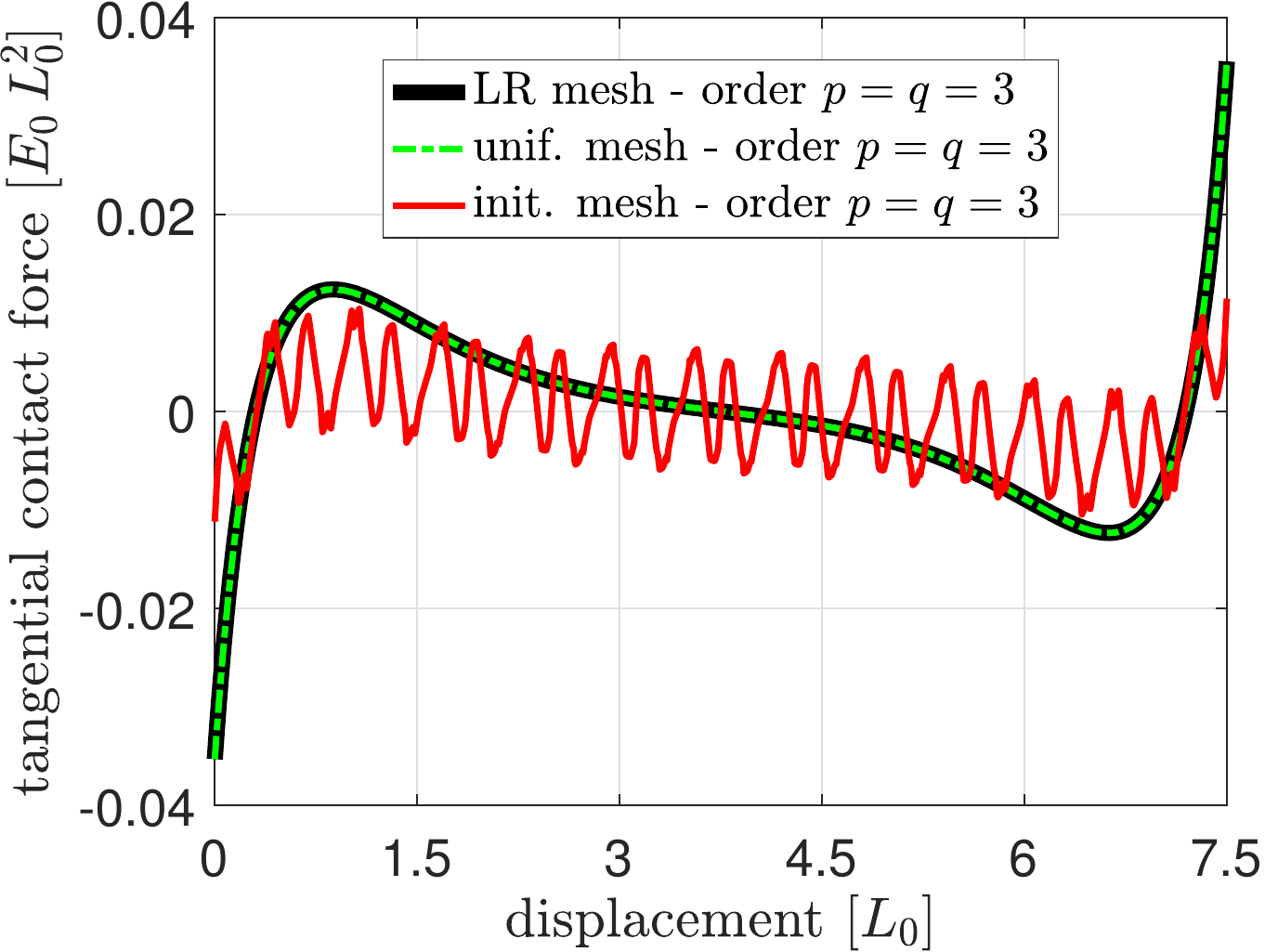}
\caption{Frictionless sliding contact: Normal~(left) and tangential~(right) contact forces for LR NURBS, uniformly refined and initial, coarse meshes. The LR NURBS mesh uses less than $30\%$ of the dofs in comparison with the uniformly refined mesh.}
\label{fig:RSS_3}
\end{figure}

In the following, the performance of adaptive local refinement and coarsening is investigated. For this, LR NURBS meshes are compared to uniformly discretized meshes using standard NURBS. The normal and tangential contact forces for the LR NURBS, the uniformly refined and the initial, coarse mesh are illustrated in Fig.~\ref{fig:RSS_3}. The normal and tangential contact forces for the LR NURBS and the uniformly refined mesh match each other nicely. The contact forces of the initial, coarse mesh are oscillating. The oscillations arise from the coarse discretization. The normal contact force of the initial, coarse mesh is much smaller than the normal contact forces of the refined meshes. The reason is the mesh dependent penalty parameter, see Eq.\eqref{eq:pen_elem_2}. The net tangential contact force $f_\mathrm{t}$ is small in comparison to the net normal contact force $f_\mathrm{n}$, as the sliding is considered to be frictionless. 
\begin{figure}[h]
\centering
\includegraphics[width=0.49\linewidth]{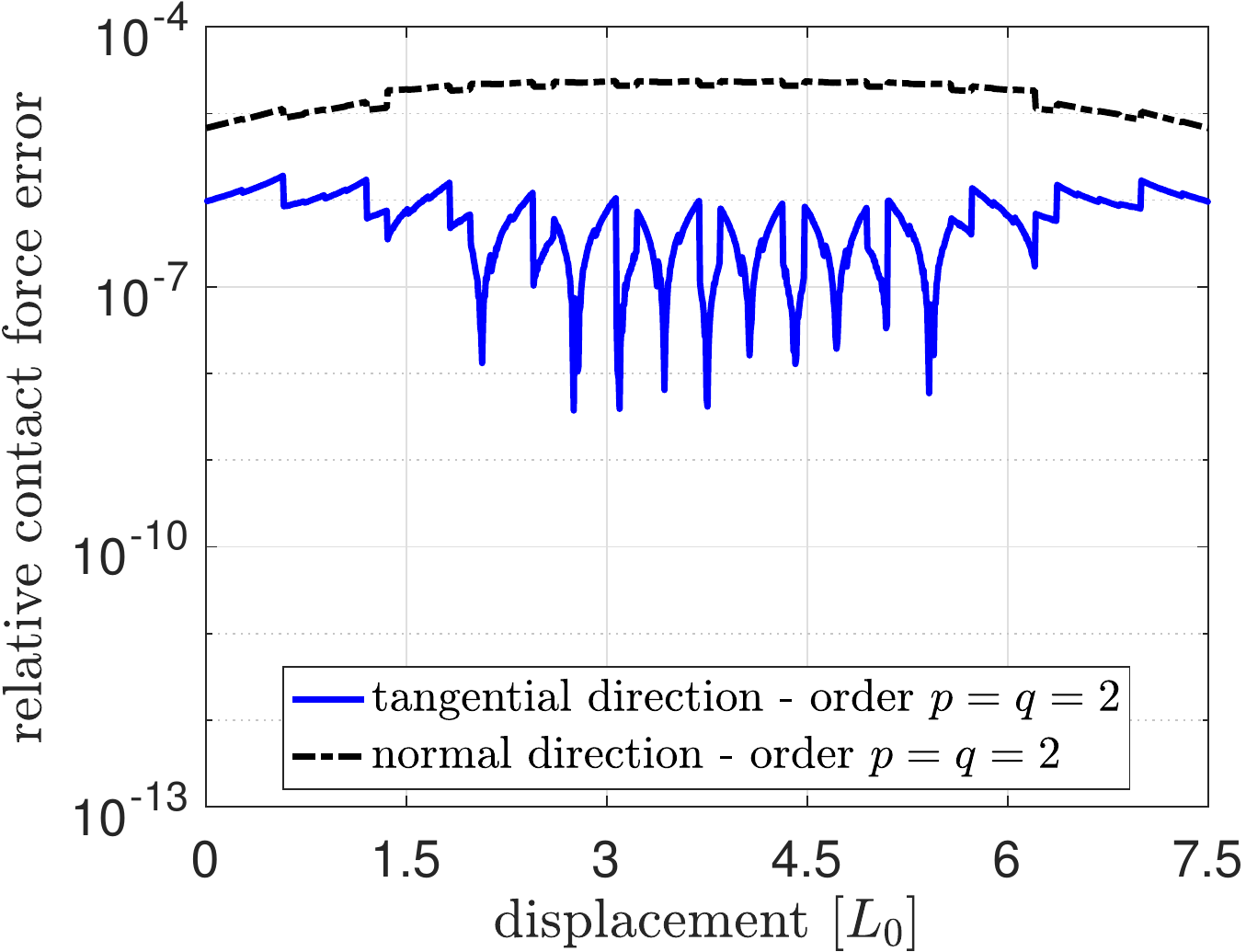}
\includegraphics[width=0.49\linewidth]{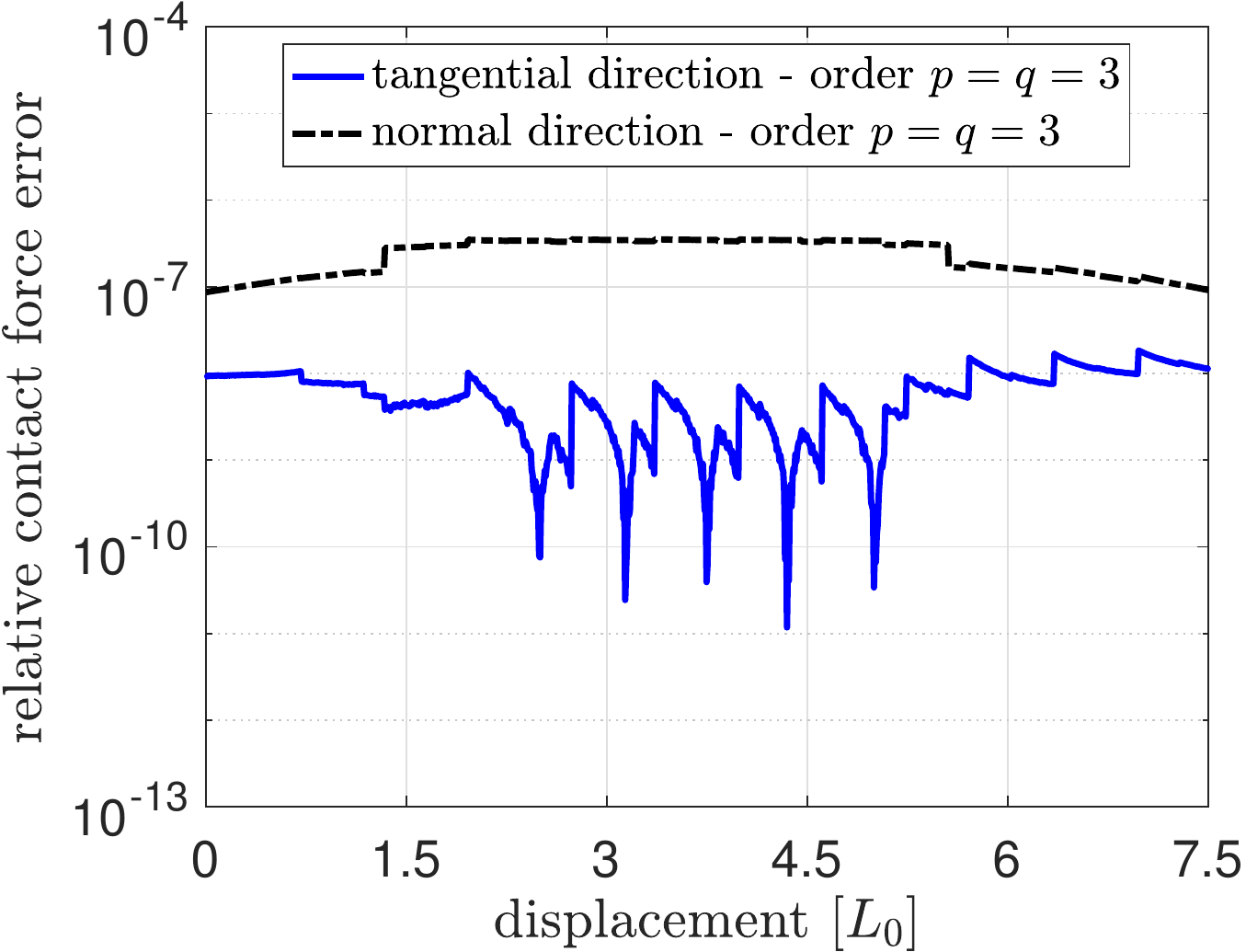}
\caption{Frictionless sliding contact: Relative contact force error of LR NURBS. Quadratic LR NURBS elements are compared to quadratic, uniformly refined NURBS elements~(left). Cubic LR NURBS elements are compared to cubic, uniformly refined NURBS elements~(right). The LR NURBS mesh uses less than $30\%$ of the dofs in comparison with the uniform mesh.}
\label{fig:RSS_4}
\end{figure}
In Fig.~\ref{fig:RSS_4} the relative contact force error of $f_\mathrm{t}$ and $f_\mathrm{n}$ is illustrated. It can be seen that the error is below $10^{-4}$ for quadratic LR NURBS and below $10^{-6}$ for cubic LR NURBS. The LR NURBS meshes capture the reference solution nicely with less than $30\%$ dofs in both examples. The high relative error of quadratic LR NURBS elements in comparison to cubic elements is caused by the coarse periphery. The evolution of the relative error in tangential direction shows abrupt changes. These result from the geometric approximation error of the coarsening steps. 
The setting of the parameters for the automatic control leads to an almost constant evolution of the relative normal contact force error. This example shows that the adaptive local refinement and coarsening technique using LR NURBS elements leads to highly accurate results while decreasing the computational cost. 


\subsection{Frictional contact of two deformable membranes}
\begin{figure}[h]
\centering
{\includegraphics[width=0.49\linewidth, trim = 250 680 200 500,clip]{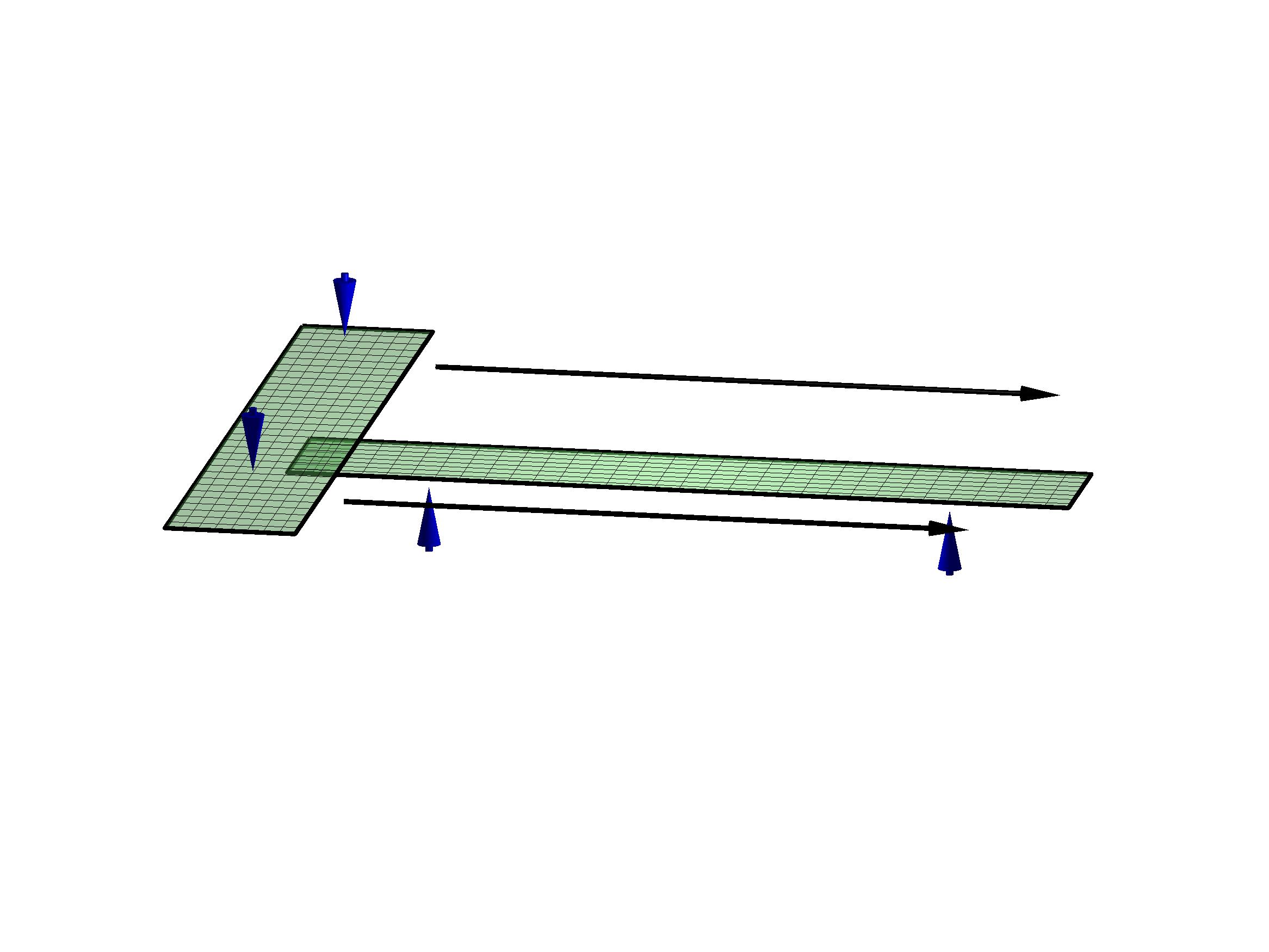}}
{\includegraphics[width=0.49\linewidth, trim = 0 -30 0 0,clip]{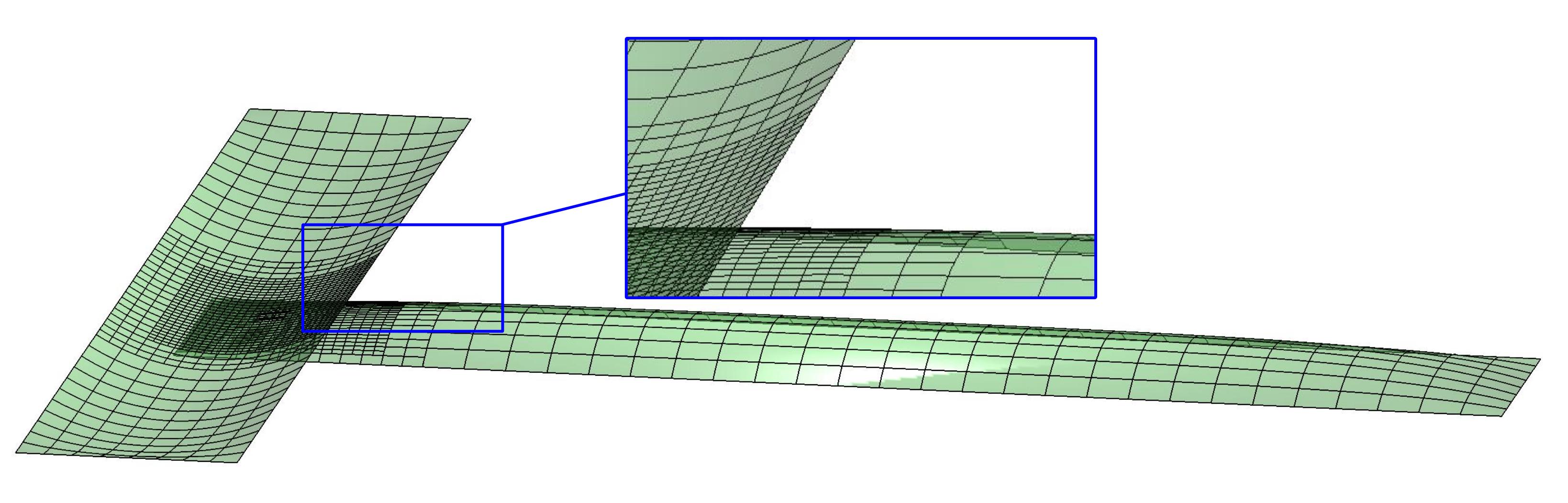}}
\caption{Frictional contact of two deformable membranes: Initial model setup~(left). The blue arrows indicate the pressure associated with the inflation. The black arrows indicate the sliding direction. Initially, each membrane surface is discretized by $8\times 32$ elements. LR NURBS mesh at the time-step of first contact detection~(right). Local refinement of depth 2 is applied.}
\label{fig:SHF_1}
\end{figure}

\begin{figure}[h]
\centering
{\includegraphics[width=0.99\linewidth, trim = 0 0 0 10,clip]{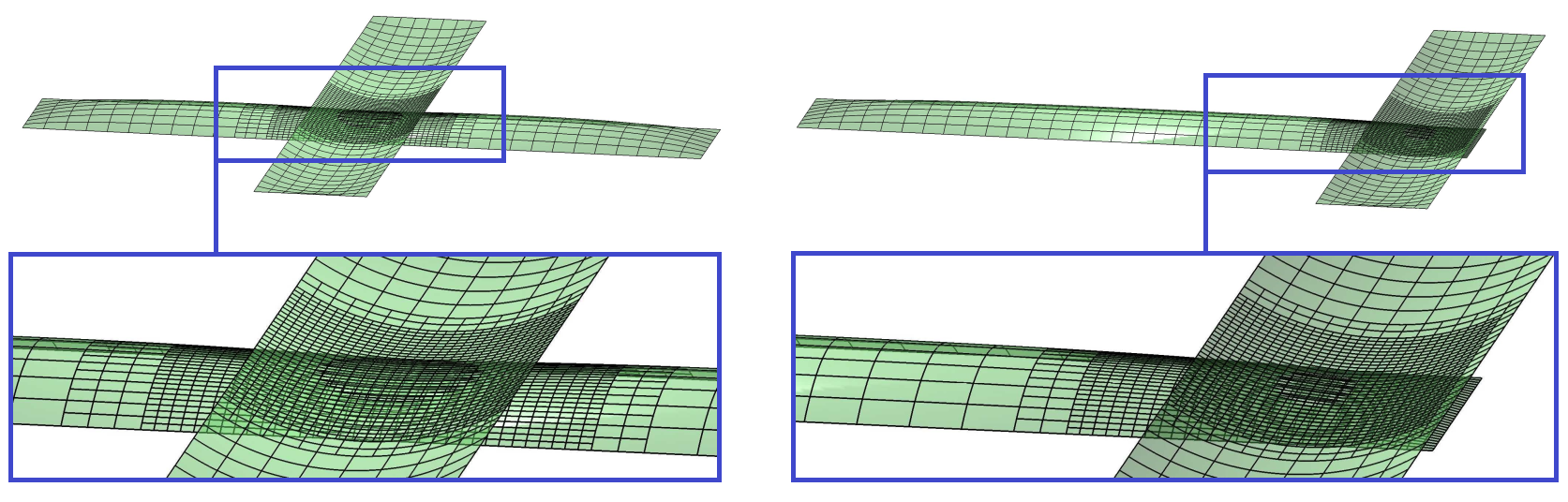}}
\caption{Frictional contact of two deformable membranes: Adaptive local refinement and coarsening of the membranes during frictional sliding. Additional zoom into the locally refined contact domain. Highly resolved meshes within the local contact domain are obtained while the periphery is still represented with the coarse, initial mesh.}
\label{fig:SHF_1b}
\end{figure}

In the next example, the performance of LR NURBS is investigated for frictional contact considering two deformable membranes. This example addresses two new aspects. First, the preservation of sliding variables during coarsening and refinement. Second, the local refinement and coarsening of two objects. We consider two deformable rectangular solid membranes with dimension $2\,L_0^u \times 0.5\,L_0^u$ and $0.5\,L_0^l \times 2\,L_0^l$. They are fixed at their boundaries in all parametric directions. An isotropic pre-stretch of $\lambda=1.5$ in longitudinal direction is applied to avoid membrane instabilities (like wrinkling) during frictional sliding. The left side of Fig.~\ref{fig:SHF_1} shows the problem setup. Both surfaces are inflated until they almost touch. This is followed by sliding the upper membrane along the lower one. During sliding, the inflated volume of each membrane is constrained to be constant. The adaptive locally refined surfaces at initial contact detection are illustrated on the right side of Fig.~\ref{fig:SHF_1}. The friction coefficient is set to $\mu_\mrf=0.25$ and the enclosed volume of both surfaces is constrained to be constant after inflation. The material behavior is the same as in the previous examples. The number of Gaussian quadrature points is $5\times 5$ for each element. Contact is treated by the penalty method. The penalty parameter is increased by Eq.~\eqref{eq:pen_elem_2} with $\varepsilon_0=80\,E_0/L_0$ and it is taken equal in tangential and normal direction. The two-half-pass algorithm \citep{Sauer2014-2} is used to evaluate the contact forces. Due to frictional sliding the contact domain changes and the presented adaptive local refinement and coarsening procedure is applied on both surfaces. In Fig.~\ref{fig:SHF_1b} two meshes resulting from the adaptive local refinement and coarsening technique are illustrated.

\begin{figure}[H]
\centering
\includegraphics[width=0.49\linewidth]{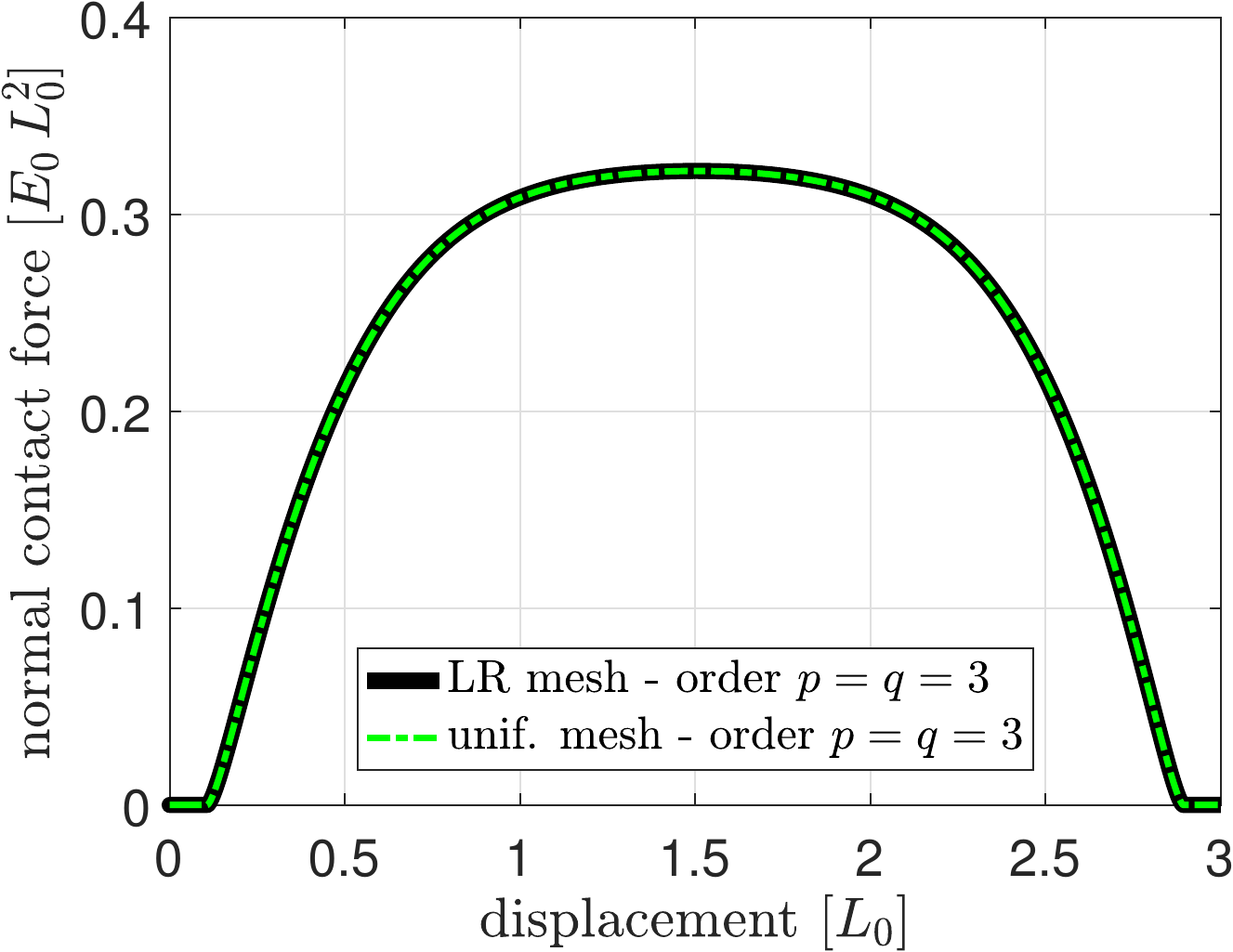}
\includegraphics[width=0.49\linewidth]{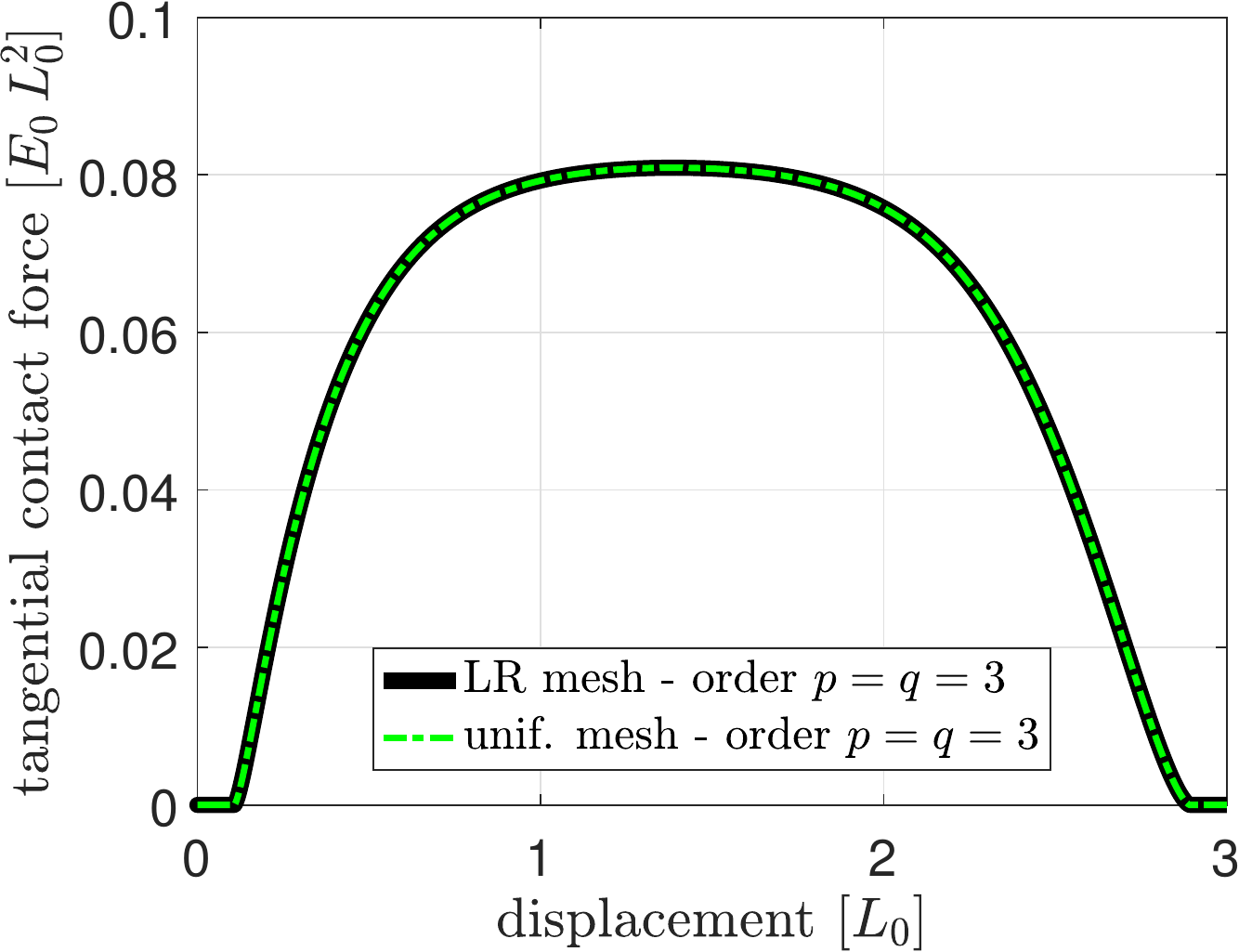}
\caption{Frictional contact of two deformable membranes: Normal~(left) and tangential~(right) contact forces for both uniform and LR NURBS mesh. The LR NURBS mesh uses less than $30\%$ of the dofs in comparison with the uniform mesh.}
\label{fig:SHF_3}
\end{figure}

\begin{figure}[H]
\centering
\includegraphics[width=0.49\linewidth]{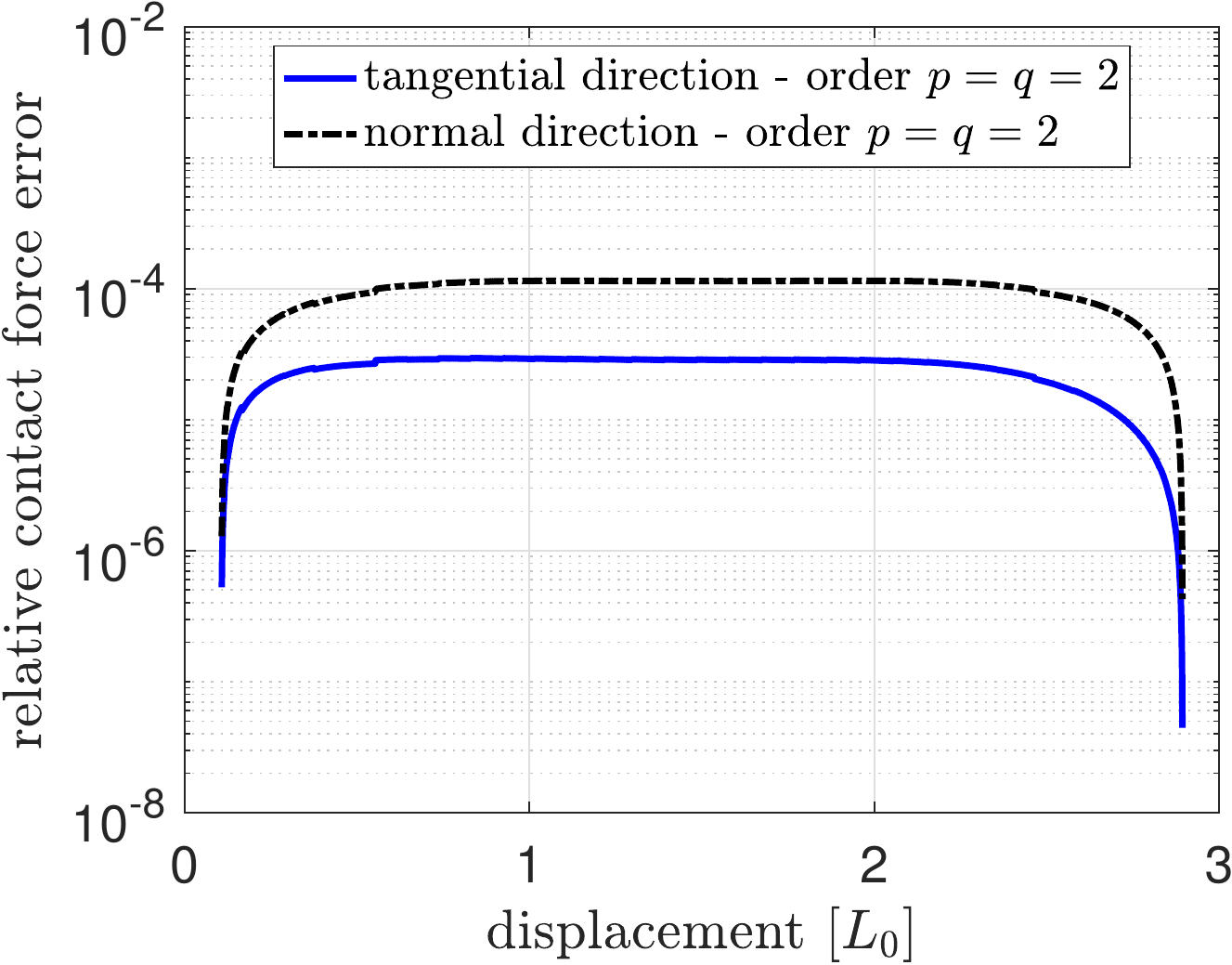}
\includegraphics[width=0.49\linewidth]{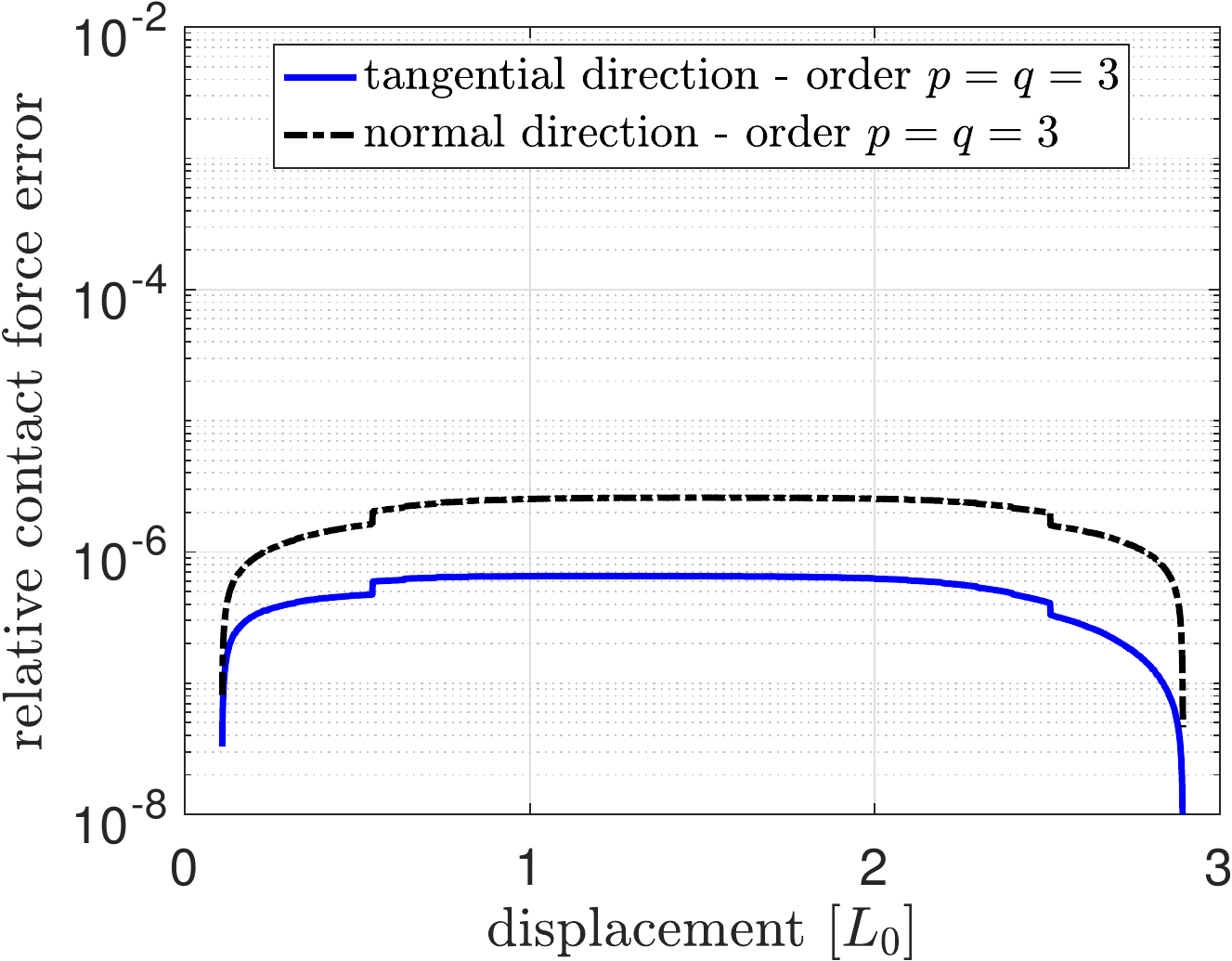}
\caption{Frictional contact of two deformable membranes: Relative contact force error of LR NURBS. Quadratic LR NURBS elements are compared to quadratic, uniformly refined NURBS elements~(left). Cubic LR NURBS elements are compared to cubic, uniformly refined NURBS elements~(right). The LR NURBS mesh uses less than $30\%$ of the dofs in comparison with the uniform mesh.}
\label{fig:SHF_4}
\end{figure}

The performance of quadratic and cubic LR NURBS is investigated. The LR NURBS meshes are compared to their uniform counterparts using quadratic and cubic NURBS discretizations separately. The normal contact force $f_\mathrm{n}$ and the tangential contact force $f_\mathrm{t}$ acting on the lower surface are shown in Fig.~\ref{fig:SHF_3}. Their behavior is as expected, and both discretizations match each other nicely.
Next, the relative contact force errors
\begin{equation}
e^\mathrm{rel}_\mathrm{t}= \frac{\mid f_\mathrm{t}^{\mathrm{ref}}-f_\mathrm{t} \mid}{\mid f^{\mathrm{max}}_\mathrm{n}\mid}~,\quad \text{and}\quad e_\mathrm{n}^\mathrm{rel}= \frac{\mid f_\mathrm{n}^{\mathrm{ref}}-f_\mathrm{n} \mid}{\mid f^{\mathrm{max}}_\mathrm{n}\mid}~,
\label{eq:rel_erry}
\end{equation}
are investigated. They are illustrated in Fig.~\ref{fig:SHF_4}. The evolution of the relative error in normal direction shows a similar behavior as in the previous example. The cubic LR NURBS elements perform better than quadratic elements due to the coarse periphery. The evolution of the relative error in tangential direction is more smooth than in the previous example, see Sec.~\ref{Sec:RSS}. The reason is that the geometric approximation error shows no remarkable influence in this example. 
A maximum relative error of $\approx 3 \cdot 10^{-4}$ for quadratic LR NURBS discretizations is achieved. The relative error for cubic LR NURBS elements is much smaller and a maximum of $\approx 3 \cdot 10^{-6}$ is achieved. The parameters for the automatic control from Sec.~\ref{sec:ALR} are $d_{\mathrm{ref}}^d=3\,d_\mre^{1-d}$, $d_{\mathrm{safe}}^d=2\,d_\mre^{1-d}$ and $d_{\mathrm{crs}}^d=4\,d_\mre^{0}$. With this setting less than $30\%$ of the dofs of the reference meshes are used for both quadratic and cubic LR meshes. This example shows that the technique of adaptive local refinement and coarsening can be successfully applied to frictional contact of membranes. The computational cost is reduced while still achieving high accuracy. In the next example the adaptive local refinement and coarsening technique is applied to LR NURBS-enriched volume elements considering frictional contact.
\subsection{Frictional ironing contact between two deformable solids}
\label{sec:soli}

Local refinement for general 3D isogeometrical models is challenging and the linear independency of LR meshes in 3D has not been developed yet. To overcome this issue, the NURBS-enrichment technique by \citet{Corbett2014-1} is used to extend LR NURBS surface discretizations to 3D.

\subsubsection{LR NURBS-enriched finite elements}

The core idea of this technique is to use standard finite Lagrange elements in the bulk domain and enrich the surface by isogeometric finite elements. The local refinement of NURBS-enriched elements is performed in two main steps
\begin{enumerate}
\item Local refinement of the isogeometric surface mesh using LR NURBS. The adpative local refinement technique from Sec.~\ref{sec:ALR} is applied.
\item Extend the local refinement through all the Lagrange elements in the thickness direction
\end{enumerate}
This work considers the local refinement of NURBS-enriched and Lagrange elements in two dimensions only. The refinement of the thickness direction is not considered here, but it is possible to use standard refinement procedures.
The local refinement of the Lagrange elements typically involve the generation of hanging nodes. For the treatment of these hanging nodes the approach of \cite{Demkowicz1989-1} is used.

\subsubsection{Numerical results}

\begin{figure}[h]
\centering
\includegraphics[width=1\linewidth, trim = 550 1230 400 1130,clip]{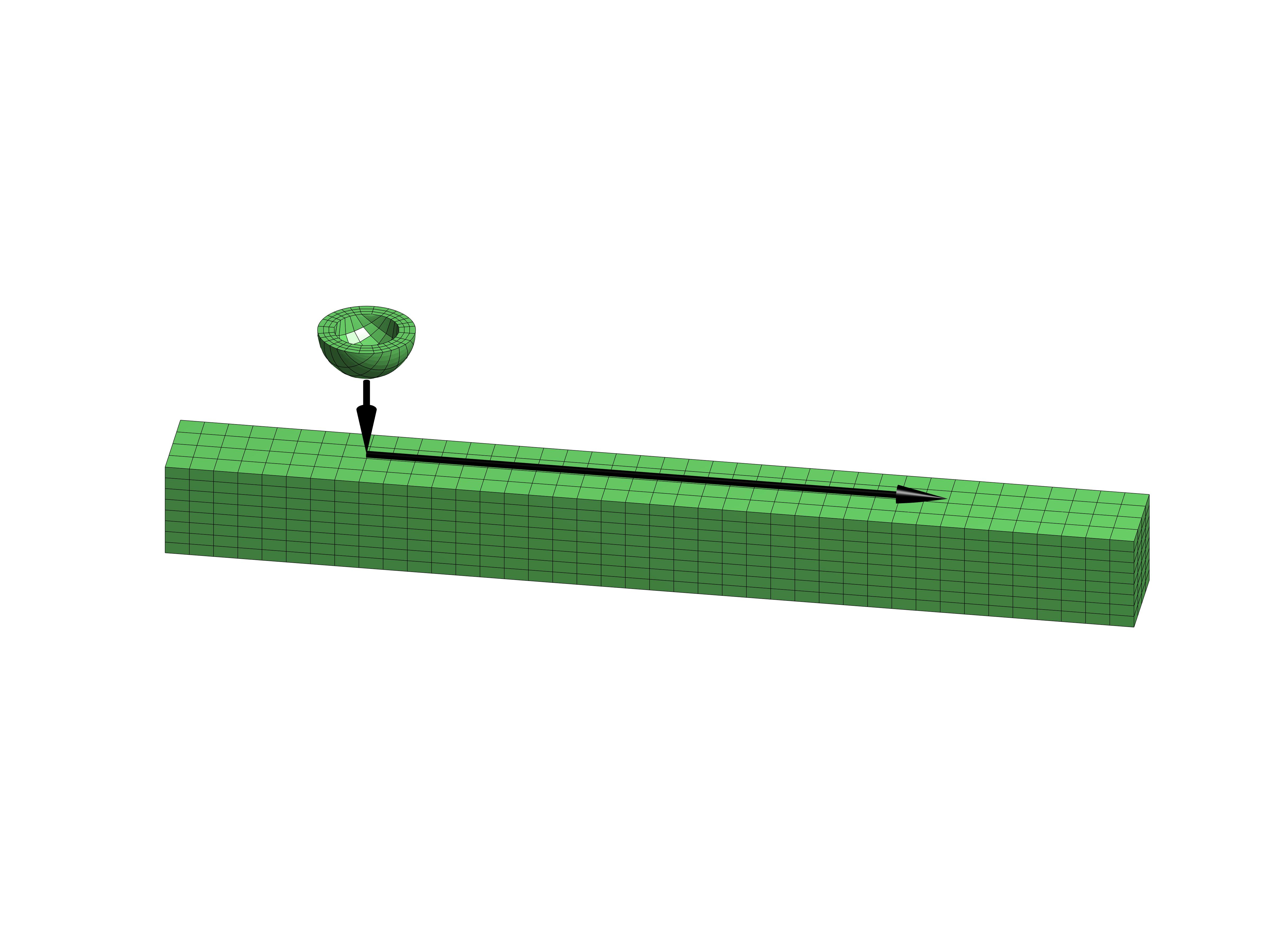}
\caption{Frictional ironing: Initial problem setup of the hemisphere and the block.}
\label{fig:sol1}
\end{figure}

The last example considers a deformable hollow hemisphere in frictional contact with a deformable block. Both solids have a quadratic LR NURBS-enriched contact surface. The hemisphere is discretized with three elements in its thickness direction and $7\times 7$ elements on the contact surface. The block is discretized with eight elements in its thickness direction and $4\times 40$ elements on the contact surface. Each contact element has $5\times 5$ Gaussian quadrature points. The bulk domain is discretized by linear Lagrange finite elements. Initially, the block has dimension $L_0 \times L_0 \times 10\,L_0$ and the hollow hemisphere has the inner radius $R_0^\mathrm{in}=L_0/6$ and the outer radius $R_0^\mathrm{out} = L_0/2$, see Fig.~\ref{fig:sol1}. The base of the block is fixed in all directions. Periodic boundary conditions are used on the left and right boundary of the block. First, a downward displacement of $2/3 R_0^\mathrm{out}$ is applied to the top of the hollow hemisphere, which is followed by a displacement of $6\,L_0$ in tangential direction. The downward displacement is considered to be frictionless while the tangential displacement is considered to be frictional. The solids follow the isotropic, non-linear Neo-Hookean material model (e.g. see \citet{taylor1})
\begin{equation}
\bsig = \frac{\Lambda}{\text{det}(\bF)}\,\text{ln} \left( \text{det}(\bF)\right)\bI+\frac{\mu}{\text{det}(\bF)}\left(\bF\,\bF^{\bT}-\bI \right)~,
\end{equation}
with the deformation gradient $\bF$ and the identity tensor $\bI$. The Lam\'e constants $\Lambda$ and $\mu$ can be expressed in terms of the Poisson's ratio $\nu$ and Young's modulus $E$ by
\begin{equation}
\mu = \frac{E}{2\left(1+\nu \right)}~, \qquad \text{and} \qquad \Lambda=\frac{2\,\mu\,\nu}{1-2\,\nu}~.
\end{equation}
Considered here are $\nu_1=\nu_2=0.3$ for both solids and $E_1=E_0$ for the block and $E_2=10\,E_0$ for the hollow hemisphere. Contact is computed by the two-half-pass algorithm using the penalty method with $\varepsilon_0=80\,E_0/L_0$ and increasing $\varepsilon^{el}_\mathrm{n}=\varepsilon^{el}_\mathrm{t}$ by Eq.~\eqref{eq:pen_elem_2}. The friction coefficient is set to $\mu_\mrf=0.25$.

\begin{figure}[h]
\centering
\includegraphics[width=.99\linewidth, trim = 417 1340 305 680,clip]{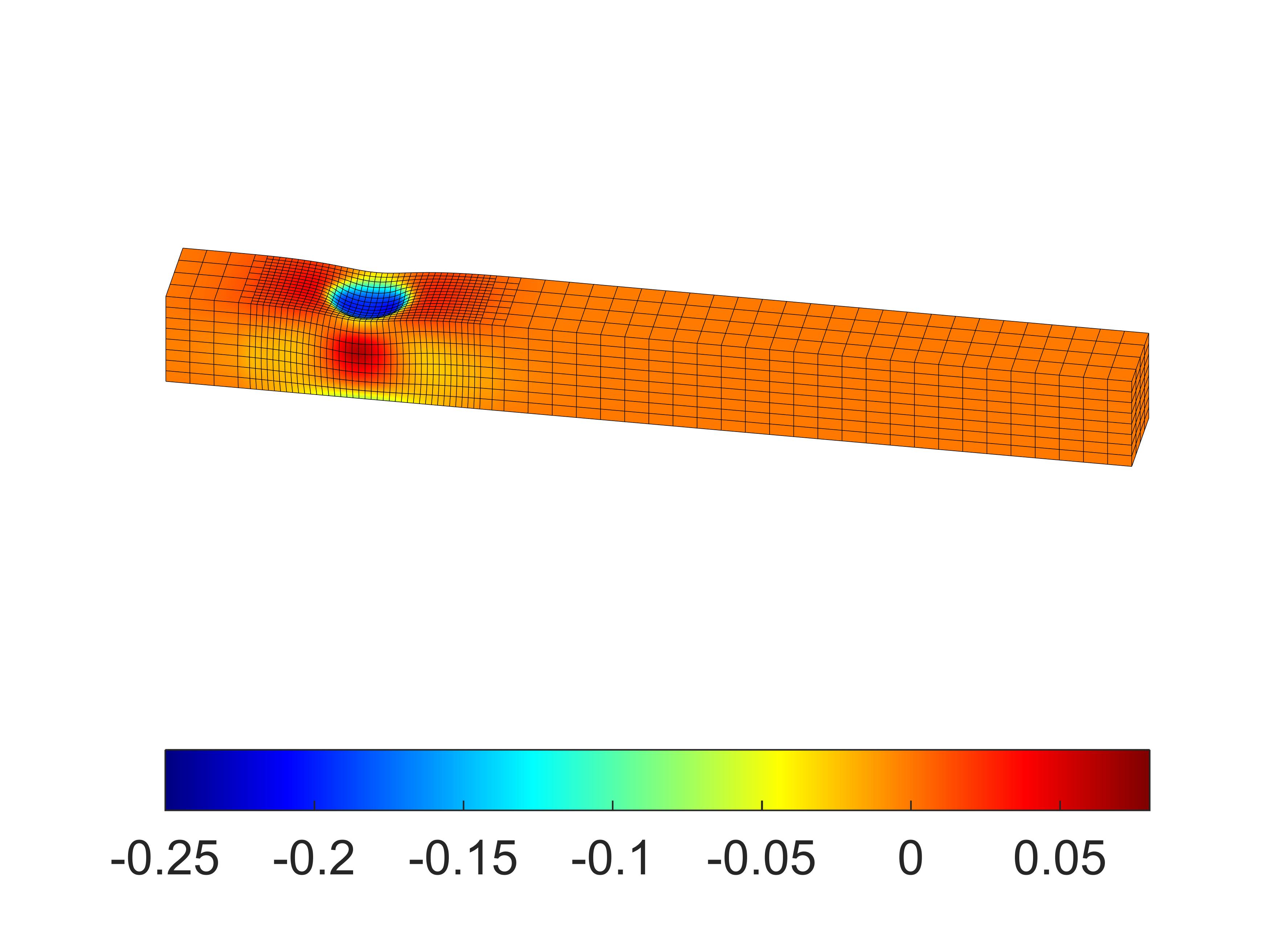}\\
\includegraphics[width=0.4\linewidth, trim = 300 190 290 2060,clip]{pics/finished/solid/new_B_adapt_p2_150_r600_cb}
\caption{Frictional ironing: Deformed, LR NURBS mesh of the block. Refinement of depth 2 is applied. The coloring shows the normalized stress invariant $I_1=\mathrm{tr}\,\bsig /E_0$.}
\label{fig:sol2}
\end{figure}

\begin{figure}[h]
\centering
\includegraphics[width=0.45\linewidth, trim = 620 570 500 190,clip]{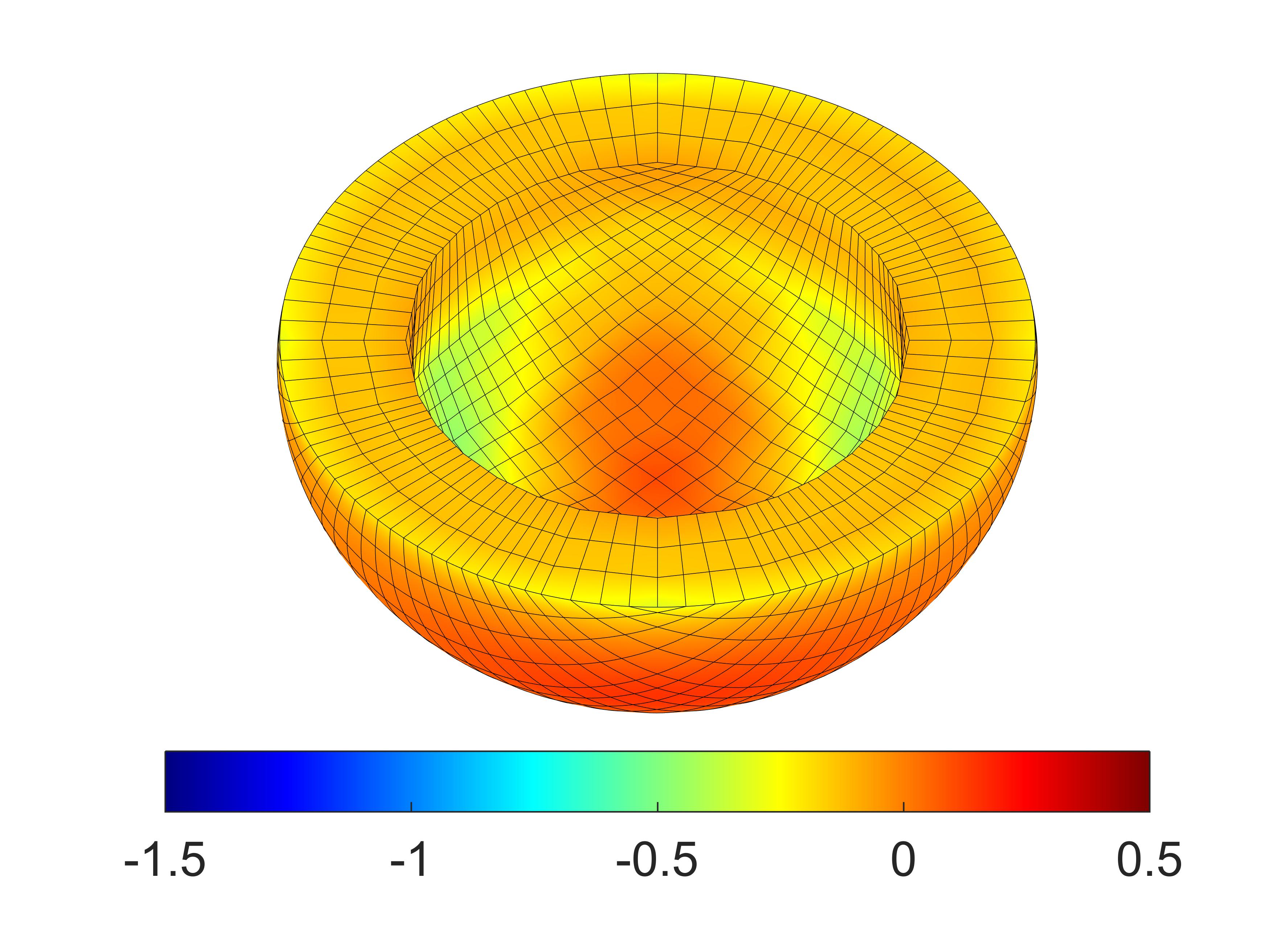}
\hspace{5mm}
\includegraphics[width=0.45\linewidth, trim = 620 570 500 190,clip]{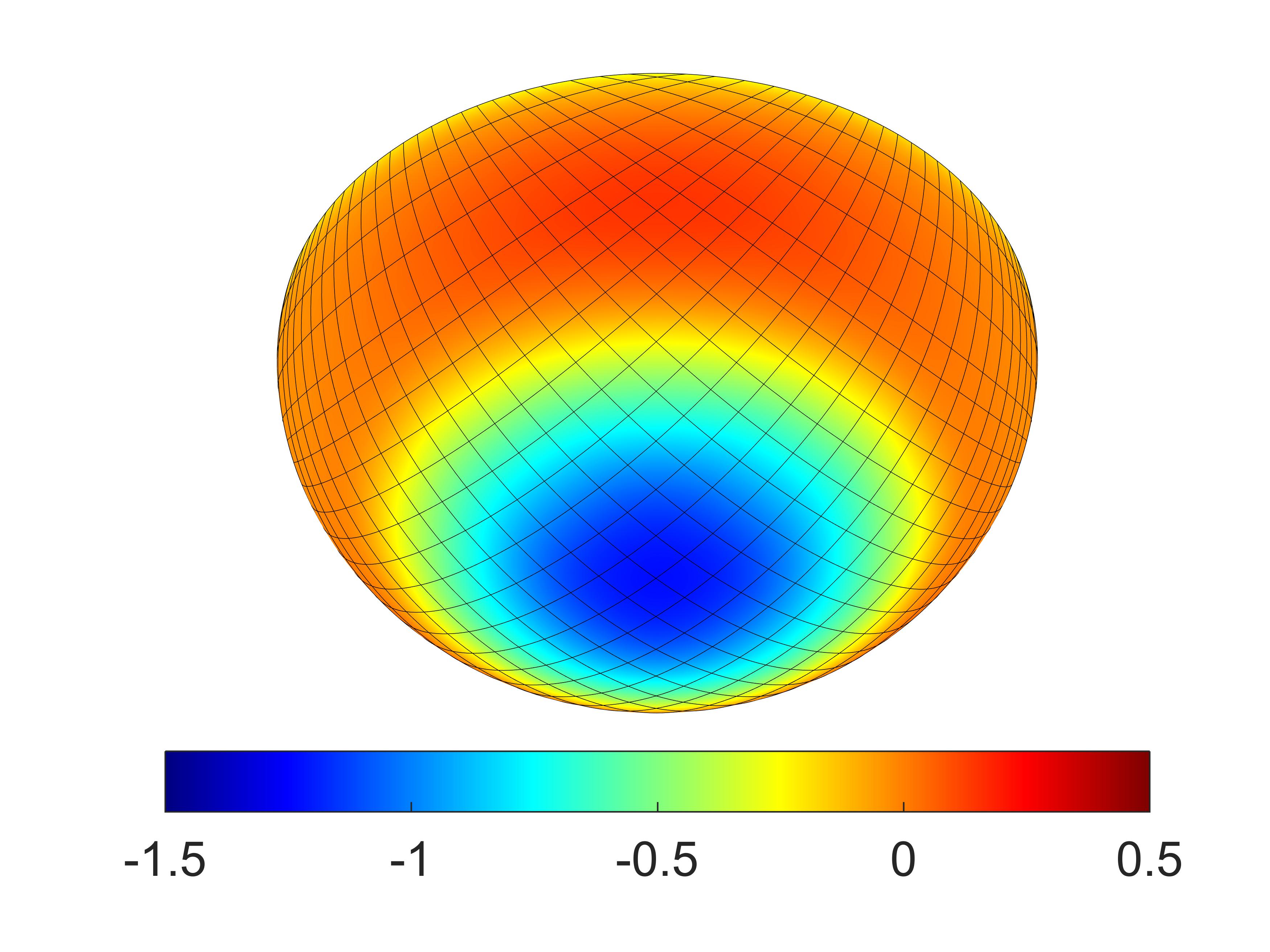}\\
\includegraphics[width=0.4\linewidth, trim = 320 200 200 2000,clip]{pics/finished/solid/new_S_top_150_r600nax}
\hspace{14mm}
\includegraphics[width=0.4\linewidth, trim = 320 200 200 2000,clip]{pics/finished/solid/new_S_bot_150_r600nax}
\caption{Frictional ironing: Refined model of the hollow hemisphere. Top view~(left) and bottom view~(right). The coloring shows $I_1=\mathrm{tr}\,\bsig /E_0$.}
\label{fig:sol3}
\end{figure}

Next, the performance of adaptive local refinement and coarsening using quadratic LR NURBS-enriched finite elements is investigated. For this, LR NURBS meshes are compared to uniformly discretized meshes using quadratic NURBS-enriched finite elements. Local refinement of depth 2 is applied. Fig.~\ref{fig:sol2} shows the locally refined and deformed block after the hollow hemisphere has been moved downward. The coloring shows the normalized stress invariant $I_1=\mathrm{tr}\,\bsig /E_0$. It can be observed that the area of major interest is locally refined while the periphery remains coarse. As the surface of the hemisphere is almost entirely in contact with the block the local refinement leads to a uniformly refined hemisphere, see Fig.~\ref{fig:sol3}. 

\begin{figure}[h]
\centering
{\includegraphics[width=0.48\linewidth, trim =0 0 30 0,clip]{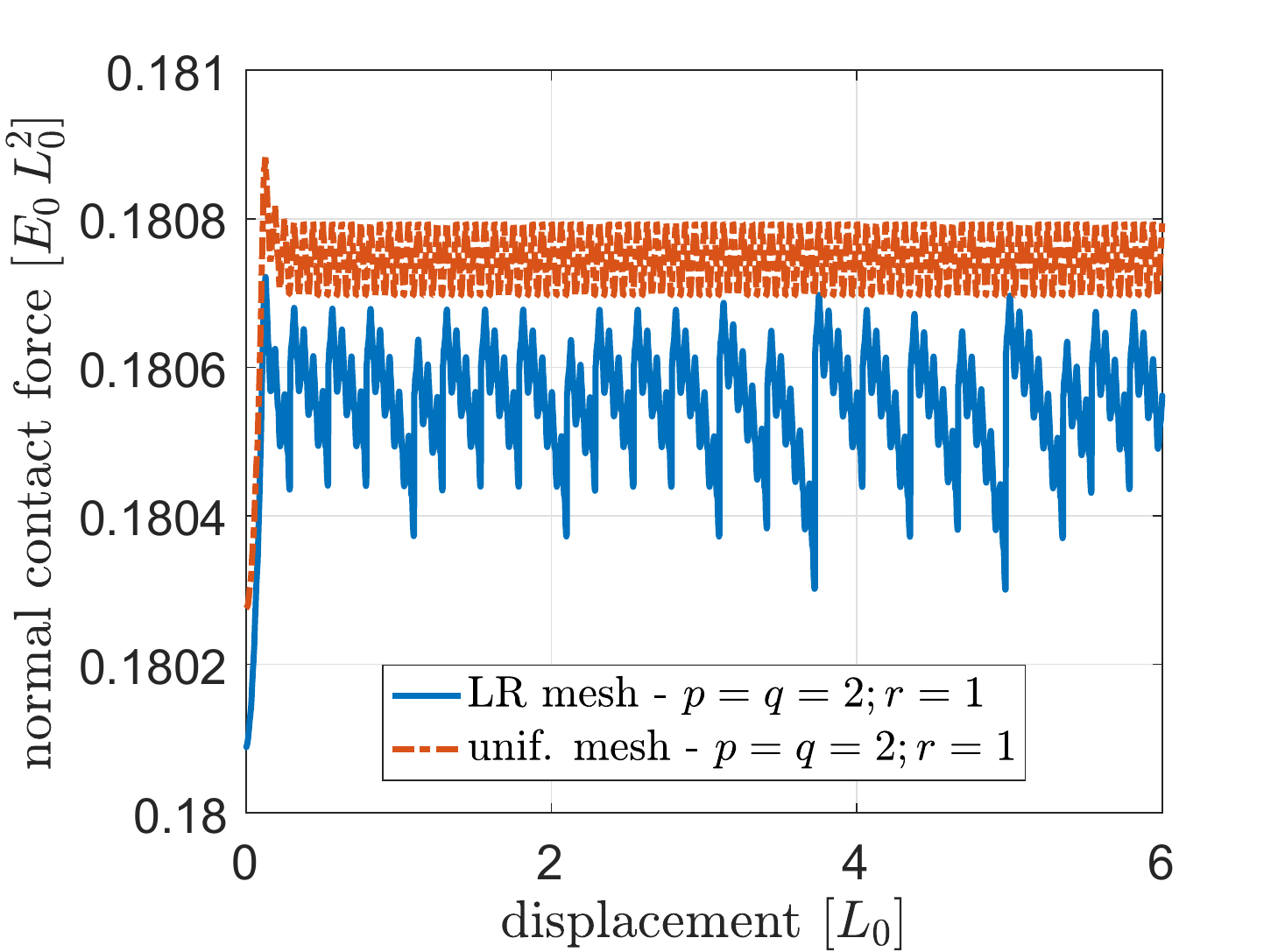}}
\hspace{2mm}
{\includegraphics[width=0.48\linewidth, trim =0 0 30 0,clip]{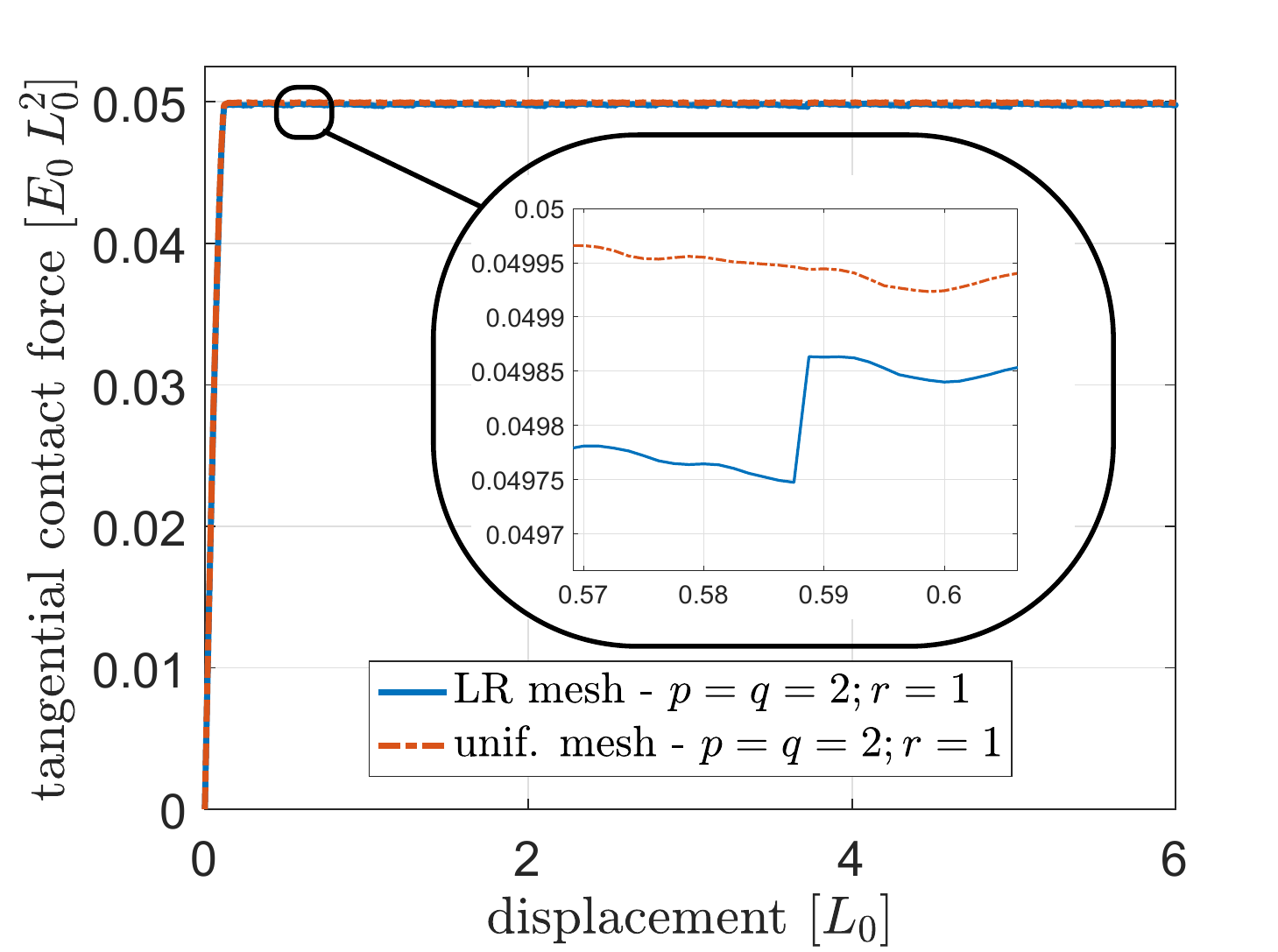}}
\caption{Frictional ironing: Normal~(left) and tangential~(right) contact forces of the LR NURBS and uniform meshes. The LR NURBS mesh uses $\approx 40\%$ of the dofs in comparison with the uniform mesh.}
\label{fig:sol4}
\end{figure}

\begin{figure}[h]
\centering
\includegraphics[width=0.48\linewidth]{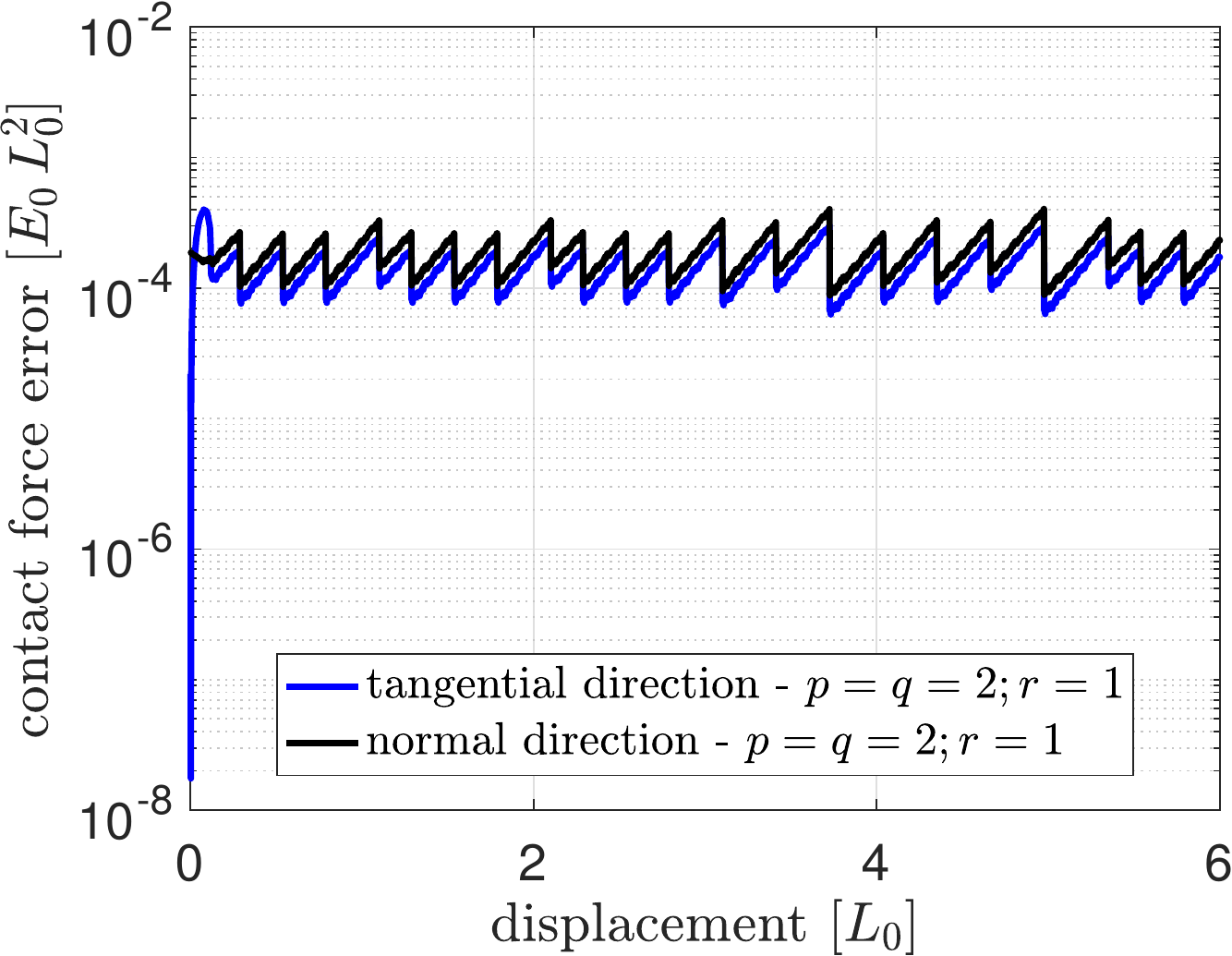}
\hspace{2mm}
\includegraphics[width=0.48\linewidth]{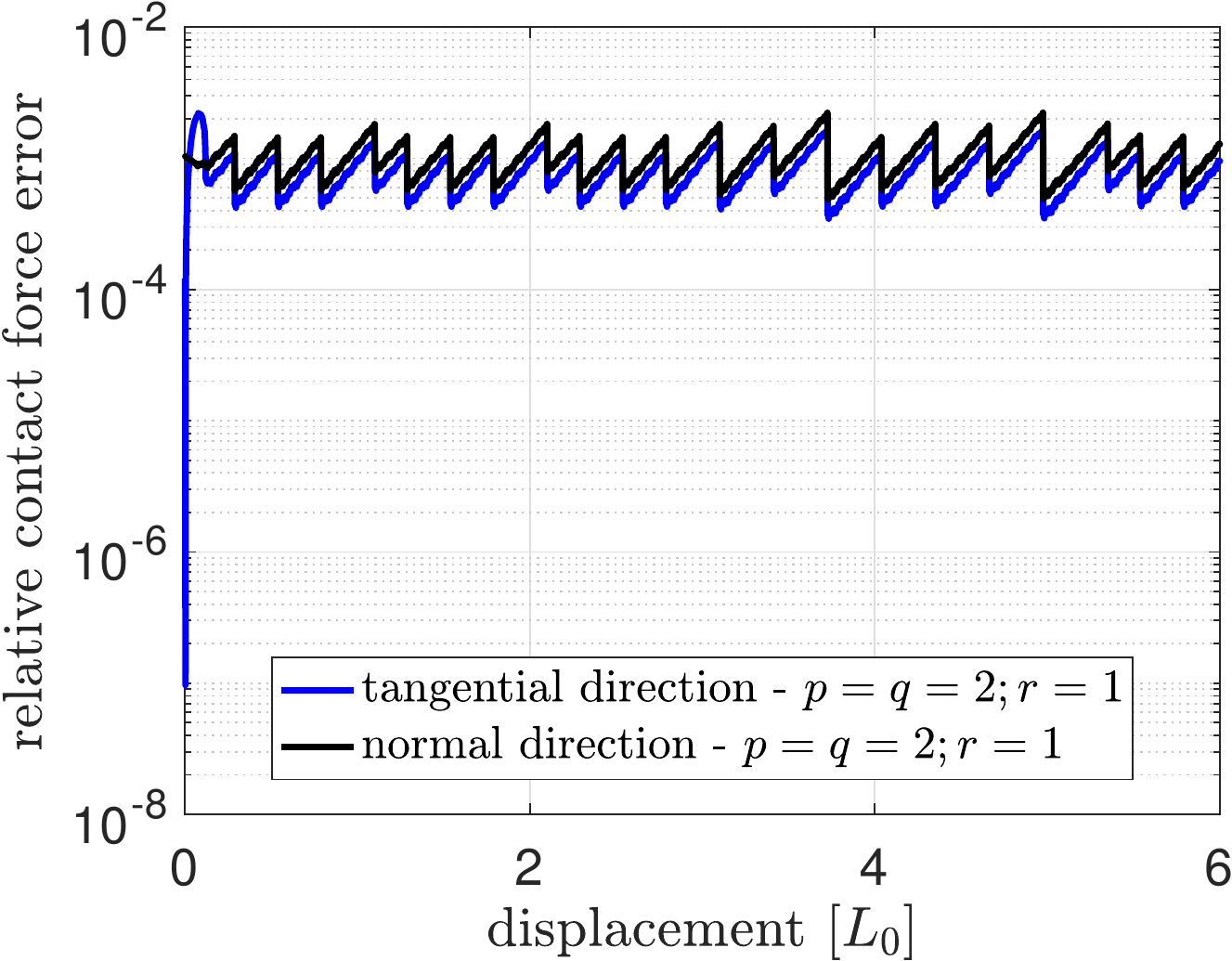}
\caption{Frictional ironing: Absolute~(left) and relative~(right) error of the tangential and normal contact forces of the LR NURBS mesh w.r.t.~the reference solution. The LR NURBS mesh uses $\approx 40\%$ of the dofs in comparison with the uniform mesh.}
\label{fig:sol5}
\end{figure}

The normal and tangential contact forces for LR NURBS and uniform meshes are shown in Fig.~\ref{fig:sol4}. These are evaluated at the block's contact surface. The contact forces for the two meshes only differ slightly. By first investigating the normal contact force an oscillatory and periodic behavior can be observed. The oscillations can be reduced by further mesh refinement. The normal contact force of the LR NURBS mesh shows abrupt increases at several points. This is caused by an adaptive local refinement step. The same can be observed for the tangential contact force, see right side of Fig.~\ref{fig:sol4}. The absolute and relative contact force errors in normal and tangential direction are illustrated in Fig.~\ref{fig:sol5}. The relative contact force errors are approximately in the range of $4\cdot 10^{-4}$ to $2\cdot 10^{-3}$. The errors show a continuous increase followed by an abrupt decrease. The errors reflect the influence of the refinement and coarsening technique also seen in Fig.~\ref{fig:sol4}. Comparing the relative contact force error to the results of the previous example, one order of magnitude is lost (in case of quadratic LR NURBS elements). The third dimension of the solids affect the numerical results strongly. For further improvements one could take local error measures into account. It is expected that by using cubic LR NURBS-enrichment the error would decrease but this would not reduce the influence of the solid domain. The parameters for the automatic control from Sec.~\ref{sec:ALR} are $d_{\mathrm{ref}}^d=5\,d_\mre^{1-d}$, $d_{\mathrm{safe}}^d=4\,d_\mre^{1-d}$ and $d_{\mathrm{crs}}^d=6\,d_\mre^{0}$. With this setting the LR NURBS mesh has still $\approx 40\%$ of the dofs of the reference mesh. Reducing the size of the locally refined domain would lead to an increase of the error of the contact forces. Still, the adaptive local refinement and coarsening of LR NURBS-enriched elements show the benefits of this procedure. 
\section{Conclusion}
\label{sec:conclusion}

This work presents a novel concept of adaptive local surface refinement using LR NURBS elements in the framework of computational contact mechanics. The B\'ezier extraction of LR NURBS elements is obtained by an additional mapping of the B\'ezier extraction operator. With this a convenient embedding of LR NURBS elements into general finite element codes is achieved. 

The adaptive local refinement and coarsening technique is automatically controlled by the presented refinement indicators. The numerical results show that the geometric approximation error arising from the coarsening is negligible. The technique is applied to pure isogeometric surface elements formulation and to isogeometrically-enriched volume elements. The numerical examples show the benefit of using LR NURBS. Due to the local refinement, the computational cost decreases, while still achieving high accuracy. A good convergence behavior for LR NURBS elements is also achieved. 

The numerical examples in Sec.~\ref{sec:infl} and Sec.~\ref{sec:soli} consider curved surfaces that truly use NURBS instead of B-splines. So they show that the extension from LR B-splines to LR NURBS is implemented successfully. The automatic control of the adaptive local refinement and coarsening technique works robustly for frictionless and frictional contact. As seen in the examples, quadratic LR NURBS achieve less accuracy than cubic LR NURBS. The reason for this is that the latter capture the solution in the coarse (i.e. unrefined) mesh more accurately. This motivates the incorporation of local error measures that indicate elements for refinement. Especially, the computations performed with quadratic LR NURBS-enriched elements show the demand for such an error measure. The refinement depth is prescribed in the presented examples. By using error measures the refinement depth can be specified automatically to achieve a desired accuracy. This work considers the local refinement of LR NURBS-enriched and Lagrange elements in two dimensions. More complex 3D objects call for an enhanced adaptive local refinement technique, which includes local refinement in the third dimension.

In this work, only bivariate LR NURBS are considered. Their extension to trivariate LR NURBS would be interesting, since they allow local refinement of 3D LR NURBS elements. This is a topic of its own because of the complex structure of trivariate LR NURBS meshes. Ensuring that the resulting LR mesh is linearly independent has not been proven yet and is challenging for arbitrary meshline extensions.

\section*{Acknowledgements}
The authors are grateful to the German Research Foundation (DFG) for supporting this research through project GSC 111.

\bigskip

\bibliographystyle{apalike}
\bibliography{bibliography}
\end{document}